\documentclass[iop]{emulateapj}
\slugcomment{{\sc Accepted to ApJ:} June 24, 2015}
\usepackage{graphicx}
\usepackage{latexsym}
\usepackage{amssymb}
\usepackage{longtable}
\usepackage{amsmath}
\usepackage{url}
\def\hei{He I $\lambda$10830 }
\def\kms{km s$^{-1}$ }
\def\Msun{\ifmmode {\rm\,\it{M_\odot}}\else ${\rm\,M_\odot}$\fi}
\def\Rsun{\ifmmode {\rm\,\it{R_\odot}}\else ${\rm\,R_\odot}$\fi}
\def\simlt{\mathrel{\spose{\lower 3pt\hbox{$\mathchar"218$}}
     \raise 2.0pt\hbox{$\mathchar"13C$}}}
\def\simgt{\mathrel{\spose{\lower 3pt\hbox{$\mathchar"218$}}
\     \raise 2.0pt\hbox{$\mathchar"13E$}}}

\usepackage[pdftex,backref,breaklinks,colorlinks,citecolor=blue]{hyperref}
\usepackage[all]{hypcap}

\begin{document}

\title{Optical Mass Flow Diagnostics in Herbig Ae/Be Stars}
\author{P. Wilson Cauley \altaffilmark{1}}
\email{pcauley@wesleyan.edu}
\affil{Wesleyan University}
\affil{Department of Astronomy and Van Vleck Observatory, 96 Foss Hill Dr., Middletown, CT 06459}
\author{Christopher M. Johns--Krull \altaffilmark{1}}
\email{cmj@rice.edu}
\affil{Rice University}
\affil{Department of Physics and Astronomy, 6100 Main St., MS 108, Houston, TX 77005}

\altaffiltext{1}{Visiting Astronomer, McDonald Observatory, The University of Texas, Austin, TX 78712, USA}

\begin{abstract}

We examine a broad range of mass flow diagnostics in a large sample of Herbig Ae/Be stars (HAEBES)
using high resolution optical spectra. The H$\beta$ and He I 5876 \AA\ lines show the
highest incidence of P--Cygni (30\%) and inverse P--Cygni (14\%) morphologies, respectively. The Fe II 4924 \AA\
line also shows a large incidence of P--Cygni profiles (11\%).
We find support for many of the conclusions reached in
a study based on the analysis of the \hei line in a large sample of HAEBES.
Namely, HAEBES exhibit smaller fractions of both blue--shifted absorption (i.e. mass outflow) and
red--shifted absorption (i.e. mass infall or accretion) than their lower mass cousins, the classical
T Tauri stars (CTTSs). In particular, the optical data supports the conclusion that HAEBES 
displaying red--shifted absorption, in general, show maximum
red--shifted absorption velocities that are smaller fractions of their stellar escape velocities
than is found for CTTSs.  This suggests that HAEBE accretion flows are originating deeper in the
gravitational potentials of their stars than in CTTS systems. In addition, we find a lack of inner
disk wind signatures in the blue--shifted absorption objects; only stellar wind signatures are
clearly observed. These findings, along with the lack of detected magnetic fields around HAEBES,
support the idea that large magnetospheres are not prevalent around HAEBES and that accretion flows
are instead mediated by significantly smaller magnetospheres with relatively smaller truncation
radii (e.g. 1--2 $R_*$). Red--shifted absorption is much more common around Herbig Ae stars than Be
stars, suggesting that Herbig Be stars may accrete via a boundary layer rather than along magnetic
field lines.    

\end{abstract}

\newpage

\section{INTRODUCTION}
\label{sec:sec1}

The interplay between mass loss and accretion in pre-main sequence systems is important in
determining both the final stellar mass, the mass in the disk available to form planets, and the
local physical conditions (e.g. incident UV flux, temperature) affecting planet formation and
evolution. While many details of the physical mechanisms involved in mass accretion and mass outflow
in the immediate circumstellar environments ($r$$\lesssim$0.1 AU) for low--mass Class II pre--main
sequence stars, or classical T Tauri stars (CTTSs), are now understood \citep[e.g.][]{muzz01,roma09},
the picture is less clear concerning these processes around the intermediate mass Herbig Ae/Be
stars (HAEBES). 

Optical spectroscopic investigations into the nature of the circumstellar environments around HAEBES
have been numerous. A small number of these studies have been performed on medium or large samples
of HAEBES. \citet{fink84} presented medium resolution ($\Delta$v$\sim$16 km s$^{-1}$) H$\alpha$ and
Na I D lines for a sample of 43 HAEBES that focused mainly on the P--Cygni profiles of a small
number (8) of objects. The investigations of \citet{hp92}, \citet{bc94}, and \citet{bc95} all
examined samples of $\sim$30 HAEBES and a small or medium number of spectral line diagnostics.
\citet{hp92} and \citet{bc94} compare their samples to known TTS characteristics but these studies
focus mainly on pure emission line morphologies and not on red and blue--shifted absorption.
\citet{viera03}, in a spectroscopic HAEBE study covering the largest number of objects to date,
classified the H$\alpha$ profiles of 131 stars using low and medium resolution spectra.
\citet{viera03} also examined forbidden line emission in their sample. However, no detailed
interpretation of the H$\alpha$ profile morphologies or forbidden line strengths are performed,
yielding little information about the kinematics of the inner circumstellar material.  Finally, the
HAEBE variability study of \citet{mend11a} examined H$\alpha$, [\ion{O}{1}]6300, He I 5876, and the
Na I D lines in a medium sized sample using medium resolution spectra. High spectral resolution
observations of a wide range of accretion and outflow diagnostics in a large sample of HAEBES is
clearly needed. 

Besides the studies listed above, most optical spectroscopic HAEBE studies have focused on single
objects or small sample sizes and a small number of spectral line diagnostics
\citep[e.g.][]{catala93,kraus08,grady10}. While this is useful for providing details about specific
objects, studies of this sort provide little information concerning the physical characteristics of
mass flows around HAEBES as a group, which are important for illuminating the general differences
between the early evolution of low and intermediate mass pre--main sequence stars. Comparing large
sample of HAEBES and CTTSs can also provide insight into the important physical mechanisms
responsible for generating mass flows in young stars.   

While both HAEBES and CTTSs host gaseous accretion disks, the lack of strong detected magnetic
fields on HAEBES \citep[e.g.][]{wade07,alecian13} and their large $v$sin$i$ values compared to CTTSs
\citet{jk07} provide hints that accretion and outflow processes do not proceed identically in both
groups of objects. Magnetospheric accretion, widely accepted to operate in most CTTS systems,
funnels material from a truncated inner disk onto the surface of the star \citep{konigl91}.
Accretion along magnetic field lines is not possible from outside of a narrow region around the
corotation radius in the disk \citep{shu94}. \citet{muzz04} pointed out that in order for
magnetospheric accretion to operate around HAEBES, the large $v$sin$i$ values of HAEBES require the
corotation radii, and thus the magnetic truncation radii, to be closer to the stellar surface than
in CTTSs. This has implications for the ability of HAEBES to launch accretion--generated outflows in
the same way as CTTSs \citep[see \S of][for an expanded discussion]{cauley14}. Spectral line
profile morphologies can provide insight into how these mass flows are generated.      
 
In this paper we present a high resolution optical study of a large sample of HAEBES using a large
number of spectral line diagnostics with the goal of investigating differences in the accretion and
outflow mechanisms operating around HAEBES compared to CTTSs. This work is the second part of a
multi--wavelength study of accretion and outflows in HAEBES. The first part is presented in
\citet[][hereafter Paper I]{cauley14}\defcitealias{cauley14}{Paper I}, and focused on the study of
\hei in a sample of 56 HAEBES. Our comprehensive coverage of optical mass flow indicators enables us
to derive reliable estimates and confidence intervals of the incidence of accretion and outflow
signatures in HAEBES. We describe our observations and data reduction in \S 2. In \S 3 we discuss
the calculation of stellar radial velocities and rotational velocities. Section 4 covers the
classification of the line profiles. We present the individual line and overall mass flow statistics
in \S 5, as well as a comparison to the \hei statistics from \citetalias{cauley14}. A comparison of
the entire sample statistics to those of CTTSs is also be given in \S 5. Section 6 investigates the
physical cause of the differences in morphology statistics. Our conclusions are summarized in \S 7.    

\section{Observations and Data Reduction}
\label{sec:sec2}

Our sample consists of 87 HAEBES covering a wide range of spectral types and was chosen from the
catalogues of \citet{viera03}; \citet{the94}; and \citet{fink84}. Classification as a HAEBE in a
previous survey was the only selection criteria. It is important to note that our sample likely
contains objects at various phases in their pre--main sequence evolution. However, the small number
of HAEBES currently identified in the literature weakens the significance of a more specific
comparison between distinct evolutionary groups. Thus we include all objects in our central analysis
regardless of spectral type and estimated age, though presumably all are pre--main sequence in
nature.

\subsection{Observations}
\label{subsec:sec21}

Our observations were performed using two cross--dispersed echelle spectrographs on two telescopes:
the Sandiford Echelle Spectrometer \citep[SES][]{mccarthy93} on the McDonald Observatory 2.1 m Otto
Struve telescope and the Tull 2dCoud\'{e} echelle spectrometer \citep[TS2][]{tull95} on the 2.7 m
Harlan J. Smith telescope. The specifics of each observation are given in \autoref{tab:tab1}. Some
objects were observed multiple times during different observing runs. The use of observations from 
different epochs in the analysis is described in \autoref{sec:sec5}.

\capstartfalse
\LongTables
\begin{deluxetable*}{lccccc}
%\rotate
%\linespacing{1}
\tablecaption{Log of echelle observations \label{tab:tab1}}
%\tablewidth{0pt}
\tablehead{\colhead{Object ID}&\colhead{Instrument}&\colhead{Telescope}&\colhead{UT Date}&
\colhead{Integration time (s)}&\colhead{S/N @ 6400 \AA$^a$}\\
\colhead{(1)}&\colhead{(2)}&\colhead{(3)}&\colhead{(4)}&\colhead{(5)}&\colhead{(6)}}
%\tabletypesize{\footnotesize}
\startdata
AB Aur          & TS2 & 2.7 m & 18--Sep--2011 & 120  & 65\\
AE Lep          & SES & 2.1 m & 01--Nov--2012 & 2400 & 35\\
BD+61 154       & TS2 & 2.7 m & 20--Sep--2011 & 2500 & 45\\
    "            & SES & 2.1 m & 30--Oct--2012 & 1200 & 30\\
    "            & SES & 2.1 m & 31--Oct--2012 & 2000 & 40\\
BF Ori          & TS2 & 2.7 m & 21--Jan--2013 & 3000 & 65\\
BH Cep          & TS2 & 2.7 m & 20--Sep--2011 & 2500 & 45\\
   "             & SES & 2.1 m & 01--Nov--2012 & 3600 & 30\\
   "             & SES & 2.1 m & 02--Nov--2012 & 3600 & 30\\
CQ Tau          & TS2 & 2.7 m & 19--Sep--2011 & 2100 & 65\\
DW CMa          & TS2 & 2.7 m & 20--Jan--2013 & 3600 & 20\\
GSC 04794--00827& TS2 & 2.7 m & 21--Jan--2013 & 3600 & 50\\
HD 141569       & TS2 & 2.7 m & 20--Sep--2011 & 180  & 90\\
    "            & TS2 & 2.7 m & 20--Jan--2013 & 75   & 60\\
HD 142666       & TS2 & 2.7 m & 20--Jan--2013 & 600  & 65\\
HD 163296       & TS2 & 2.7 m & 18--Sep--2011 & 300  & 55\\
HD 169142       & TS2 & 2.7 m & 19--Sep--2011 & 200  & 50\\
HD 190073       & TS2 & 2.7 m & 18--Sep--2011 & 180  & 55\\
    "            & TS2 & 2.7 m & 26--Sep--2012 & 350  & 60\\
HD 203024       & TS2 & 2.7 m & 20--Sep--2011 & 390  & 60\\
    "            & SES & 2.1 m & 30--Oct--2012 & 600  & 30\\
    "            & SES & 2.1 m & 31--Oct--2012 & 1500 & 50\\
HD 244314       & TS2 & 2.7 m & 20--Sep--2011 & 2200 & 65\\
    "            & TS2 & 2.7 m & 20--Jan--2013 & 3000 & 80\\
HD 244604       & TS2 & 2.7 m & 20--Sep--2011 & 1300 & 75\\
HD 245185       & TS2 & 2.7 m & 19--Sep--2011 & 2300 & 65\\
HD 249879       & TS2 & 2.7 m & 26--Sep--2012 & 3200 & 60\\
    "            & SES & 2.1 m & 01--Nov--2012 & 3600 & 35\\
    "            & SES & 2.1 m & 02--Nov--2012 & 3600 & 45\\
HD 250550       & SES & 2.1 m & 01--Nov--2012 & 2000 & 40\\
    "            & SES & 2.1 m & 02--Nov--2012 & 3000 & 100\\
HD 287823       & TS2 & 2.7 m & 22--Sep--2011 & 3000 & 55\\
HD 290409       & TS2 & 2.7 m & 22--Sep--2011 & 3000 & 70\\
HD 290500       & TS2 & 2.7 m & 23--Sep--2011 & 3600 & 60\\
HD 290764       & TS2 & 2.7 m & 23--Sep--2011 & 2400 & 70\\
HD 290770       & TS2 & 2.7 m & 21--Sep--2011 & 1200 & 90\\
HD 34282        & TS2 & 2.7 m & 21--Sep--2011 & 1500 & 70\\
HD 35187        & TS2 & 2.7 m & 18--Sep--2011 & 240  & 70\\
    "            & SES & 2.1 m & 01--Nov--2012 & 1200 & 35\\
    "            & SES & 2.1 m & 02--Nov--2012 & 2400 & 40\\
HD 35929        & TS2 & 2.7 m & 18--Sep--2011 & 400  & 50\\
HD 36408        & TS2 & 2.7 m & 18--Sep--2011 & 120  & 40\\
HD 37357        & TS2 & 2.7 m & 18--Sep--2011 & 500  & 55\\
HD 37411        & SES & 2.1 m & 30--Oct--2012 & 2400 & 30\\
HD 38120        & SES & 2.1 m & 30--Oct--2012 & 600  & 30\\
    "            & SES & 2.1 m & 31--Oct--2012 & 2000 & 55\\
HD 50083        & TS2 & 2.7 m & 18--Sep--2011 & 180  & 50\\
HK Ori          & TS2 & 2.7 m & 21--Sep--2011 & 3600 & 30\\
IL Cep          & TS2 & 2.7 m & 20--Sep--2011 & 1300 & 80\\
    "            & TS2 & 2.7 m & 22--Sep--2011 & 1200 & 75\\
    "            & TS2 & 2.7 m & 23--Sep--2011 & 1300 & 80\\
IP Per          & TS2 & 2.7 m & 19--Sep--2011 & 2400 & 70\\
IRAS 05044--0325& TS2 & 2.7 m & 20--Jan--2013 & 3200 & 35\\
IRAS 06071+2925 & TS2 & 2.7 m & 20--Jan--2013 & 3600 & 30\\
IRAS 07061--0414& TS2 & 2.7 m & 21--Jan--2013 & 3600 & 60\\
IRAS 17481--1415& TS2 & 2.7 m & 20--Sep--2011 & 3600 & 30\\
IRAS 18306--0500& TS2 & 2.7 m & 20--Sep--2011 & 3600 & 30\\
IRAS 18454+0250 & TS2 & 2.7 m & 19--Sep--2011 & 3000 & 20\\
IRAS 19343+2926 & TS2 & 2.7 m & 19--Sep--2011 & 3000 & 20\\
    "            & TS2 & 2.7 m & 26--Sep--2012 & 3000 & 15\\
    "            & SES & 2.1 m & 01--Nov--2012 & 3600 & 10\\
    "            & SES & 2.1 m & 02--Nov--2012 & 3600 & 10\\
LkH$\alpha$ 134 & TS2 & 2.7 m & 20--Sep--2011 & 2700 & 60\\
    "            & SES & 2.1 m & 01--Nov--2012 & 3600 & 35\\
    "            & SES & 2.1 m & 02--Nov--2012 & 3600 & 40\\
LkH$\alpha$ 208 & SES & 2.1 m & 01--nov--2012 & 3600 & 35\\
    "            & SES & 2.1 m & 02--Nov--2012 & 3600 & 35\\
    "            & TS2 & 2.7 m & 22--Jan--2013 & 3000 & 65\\
LkH$\alpha$ 233 & TS2 & 2.7 m & 22--Sep--2011 & 3600 & 20\\
    "            & TS2 & 2.7 m & 26--Sep--2012 & 3600 & 20\\
LkH$\alpha$ 257 & TS2 & 2.7 m & 21--Sep--2011 & 3800 & 35\\
LkH$\alpha$ 324 & TS2 & 2.7 m & 20--Sep--2011 & 3600 & 40\\
LkH$\alpha$ 339 & TS2 & 2.7 m & 22--Jan--2013 & 1960 & 20\\
MWC 1080        & TS2 & 2.7 m & 21--Sep--2011 & 2500 & 20\\
    "            & TS2 & 2.7 m & 22--Sep--2011 & 2500 & 20\\
    "            & TS2 & 2.7 m & 23--Sep--2011 & 2500 & 20\\
    "            & SES & 2.1 m & 30--Oct--2012 & 2000 & 20\\
    "            & SES & 2.1 m & 31--Oct--2012 & 3000 & 10\\
MWC 120         & TS2 & 2.7 m & 18--Sep--2011 & 400  & 80\\
MWC 137         & TS2 & 2.7 m & 20--Jan--2013 & 1800 & 15\\
MWC 147         & TS2 & 2.7 m & 26--Sep--2012 & 600  & 20\\
MWC 300         & TS2 & 2.7 m & 18--Sep--2011 & 2700 & 10\\
MWC 361         & TS2 & 2.7 m & 18--Sep--2011 & 180  & 70\\
MWC 480         & TS2 & 2.7 m & 18--Sep--2011 & 180  & 50\\
MWC 610         & TS2 & 2.7 m & 18--Sep--2011 & 360  & 50\\
    "            & TS2 & 2.7 m & 26--Sep--2012 & 900  & 60\\
MWC 614         & TS2 & 2.7 m & 18--Sep--2011 & 120  & 75\\
MWC 758         & TS2 & 2.7 m & 18--Sep--2011 & 400  & 55\\
MWC 778         & SES & 2.1 m & 30--Oct--2012 & 3600 & 10\\
    "            & SES & 2.1 m & 31--Oct--2012 & 4000 & 10\\
MWC 863         & TS2 & 2.7 m & 23--Sep--2011 & 500  & 50\\
NZ Ser          & TS2 & 2.7 m & 19--Sep--2011 & 3000 & 40\\
R Mon           & TS2 & 2.7 m & 26--Sep--2012 & 2000 & 20\\
RR Tau          & SES & 2.1 m & 30--Oct--2012 & 2500 & 35\\
    "            & SES & 2.1 m & 31--Oct--2012 & 3600 & 35\\
    "            & TS2 & 2.7 m & 20--Jan--2013 & 3600 & 55\\
SV Cep          & TS2 & 2.7 m & 20--Sep--2011 & 2500 & 65\\
T Ori           & TS2 & 2.7 m & 23--Sep--2011 & 3000 & 70\\
    "            & SES & 2.1 m & 30--Oct--2012 & 1200 & 20\\
    "            & SES & 2.2 m & 31--Oct--2012 & 3600 & 50\\
UX Ori          & TS2 & 2.7 m & 22--Sep--2011 & 3600 & 30\\
UY Ori          & SES & 2.1 m & 30--Oct--2012 & 3600 & 20\\
V1185 Tau       & TS2 & 2.7 m & 19--Sep--2011 & 2400 & 70\\
    "            & SES & 2.1 m & 30--Oct--2012 & 2000 & 30\\
    "            & SES & 2.1 m & 31--Oct--2012 & 3000 & 40\\
V1578 Cyg       & TS2 & 2.7 m & 20--Sep--2011 & 2500 & 90\\
    "            & TS2 & 2.7 m & 22--Sep--2011 & 2400 & 90\\
    "            & TS2 & 2.7 m & 23--Sep--2011 & 2400 & 80\\
V1685 Cyg       & TS2 & 2.7 m & 18--Sep--2011 & 2500 & 25\\
    "            & TS2 & 2.7 m & 26--Sep--2012 & 3000 & 30\\
    "            & SES & 2.1 m & 01--Nov--2012 & 2000 & 25\\
    "            & SES & 2.1 m & 02--Nov--2011 & 3000 & 30\\
V1686 Cyg       & TS2 & 2.7 m & 19--Sep--2011 & 3600 & 20\\
V1787 Ori       & TS2 & 2.7 m & 22--Jan--2013 & 3600 & 35\\
V1818 Ori       & TS2 & 2.7 m & 22--Jan--2013 & 3600 & 40\\
V346 Ori        & TS2 & 2.7 m & 21--Sep--2011 & 2400 & 75\\
V351 Ori        & TS2 & 2.7 m & 18--Sep--2011 & 650  & 50\\
V361 Cep        & TS2 & 2.7 m & 18--Sep--2011 & 2200 & 50\\
V373 Cep        & TS2 & 2.7 m & 21--Sep--2011 & 3500 & 40\\
V374 Cep        & TS2 & 2.7 m & 21--Sep--2011 & 2800 & 30\\
    "            & TS2 & 2.7 m & 26--Sep--2012 & 3000 & 50\\
    "            & SES & 2.1 m & 30--Oct--2012 & 2000 & 30\\
    "            & SES & 2.1 m & 31--Oct--2012 & 2000 & 60\\
V380 Ori        & SES & 2.1 m & 30--Oct--2012 & 2400 & 15\\
    "            & SES & 2.1 m & 31--Oct--2012 & 3600 & 15\\
V586 Ori        & TS2 & 2.7 m & 20--Sep--2011 & 1500 & 75\\
V590 Mon        & TS2 & 2.7 m & 21--Jan--2013 & 3600 & 40\\
V599 Ori        & TS2 & 2.7 m & 22--Jan--2013 & 3200 & 30\\
V718 Sco        & TS2 & 2.7 m & 23--Sep--2011 & 600  & 50\\
V791 Mon        & SES & 2.1 m & 01--Nov--2012 & 2400 & 30\\
    "            & SES & 2.1 m & 02--Nov--2012 & 3600 & 45\\
VV Ser          & TS2 & 2.7 m & 18--Sep--2011 & 3000 & 45\\
    "            & TS2 & 2.7 m & 21--Sep--2011 & 3000 & 40\\
    "            & TS2 & 2.7 m & 22--Sep--2011 & 3000 & 40\\
WW Vul          & TS2 & 2.7 m & 18--Sep--2011 & 1600 & 40\\
    "            & TS2 & 2.7 m & 26--Sep--2012 & 2000 & 70\\
    "            & SES & 2.1 m & 30--Oct--2012 & 600  & 25\\
    "            & SES & 2.1 m & 31--Oct--2012 & 1500 & 45\\
XY Per          & TS2 & 2.7 m & 19--Sep--2011 & 2000 & 65\\
Z CMa           & TS2 & 2.7 m & 21--Jan--2013 & 1800 & 20\\
%\enddata
\tablenotetext{a}{Continuum S/N for SES spectra taken on 31--Oct--2012 and 02--Nov--2012 is
estimated at 4800 \AA.}
\end{deluxetable*}
\capstarttrue

The TS2 data were obtained during three separate runs in 2011 September (7 nights), 2012 September
(3 nights), and 2013 January (3 nights). For the TS2 observations, we employed the 1.2'' slit with
the E2 grating to achieve a resolving power of $R$$\sim$60,000, or a velocity resolution of $\sim$5
km s$^{-1}$. The grating configuration provided wavelength coverage from approximately 3700--10000
\AA\hspace{0pt}. Exposure times were limited to 1--hour in order to maximize the number of objects
observed with at least moderate signal--to--noise. A S/N of $\sim$50 per pixel is typical for the
TS2 data, although this varies from S/N$\sim$10 up to $\sim$90 depending on a combination of the
object brightness and weather conditions during the exposure. ThAr comparison lamp spectra were
taken multiple times each night to be used for wavelength calibrations. Flat field exposures were
collected at the beginning of each night. Telluric and spectroscopic standards were observed
periodically throughout the course of each observing run.

The SES data were collected during a single 4--night observing run in 2012 October. We used two
different grating settings, each one on two separate nights, to record both "red" (5430--6620 \AA)
and "blue" (4375--4870 \AA) wavelengths. This was necessary due to the SES's smaller wavelength
coverage ($\sim$500--1000 \AA) per exposure than that of TS2. Even using two wavelength settings, we
were not able to observe all of the spectral diagnostics that are available in the TS2 data. Thus
objects that are only observed with the SES are lacking data for some diagnostics. The resolving
power for each grating setting is $R$$\sim$60,000. ThAr lamp spectra were taken throughout the night
and flat field exposures were obtained at the beginning of each night. Telluric and spectroscopic
standards were observed, although the spectroscopic standards obtained using the TS2 are used
interchangeably due to the almost identical velocity resolutions of the two instruments.

\subsection{Data reduction}
\label{subsec:sec22}

All observations were reduced using custom IDL routines optimized for extracting echelle spectra.
Each object exposure is bias subtracted and divided by a normalized median flat to remove pixel to
pixel variations. The spectral orders are located and then fit using a 7$^{th}$ order polynomial.
Each order is optimally extracted which also removes most hot pixels and cosmic rays. Wavelength
calibrations are performed on the ThAr exposures using a two dimensional polynomial fit to all
spectral orders simultaneously. Hundreds of individual lamp lines in the exposures are matched to
known atlas wavelengths. This procedure results in dispersion solutions that are accurate to
$\sim$0.10 pixels, or 0.005 \AA, across the entire spectral range.  Telluric absorption was
subtracted out of the spectral region near 6300 \AA\hspace{0pt} in order to make identification of
the O I $\lambda$6300 line unambiguous. Due to the prominence of the H$\alpha$ and Na I D lines,
telluric absorption in these regions does not affect the classification of the line morphologies
since these absorption are small perturbations of the true line profile shape.  Equivalent width
calculations of H$\alpha$ lines include the telluric features which results in a maximum additional
uncertainty of $\sim$0.40 \AA.    

Some exposures contain strong telluric O I emission near 5577, 6300, and 6363 \AA. We have modified
our echelle reduction program to locate the night sky emission above and below the spectral trace of
the star and subtract it from the actual object spectrum. However, this routine was not always
successful, especially for objects with profiles that were not centered on the slit or for
irregularly shaped profiles (e.g. binaries). Thus some profiles contain residual night sky emission
(e.g. HD 244314). This extra emission can be identified by its narrow ($\sim$10 km s$^{-1}$) width
and is easily distinguished from the stellar line profile.

Approximate continuum normalization in each order was performed by dividing each order by the blaze
function of the same order, derived from the reduced spectrum of the median flat for that particular
grating setting. A linear fit is then performed to remove any residual slope. Some objects display
strong photospheric Balmer features which have wings that often extend to the ends of the spectral
order. The continuum normalization in these cases is likely not exactly the true stellar continuum.
However, identification of non--photospheric features in these lines is not problematic with the use
of a comparison standard since the chosen continuum level does not affect the line shape. The Fe II
4352 \AA\ and N II 6583 \AA\ features often lie on the wings of H$\gamma$ and H$\alpha$, respectively.
These profiles are normalized to the continuum in the Balmer lines.

\section{Physical Parameters}
\label{sec:sec3}

In order to correctly separate circumstellar from photospheric features, it is important to have
estimates of both the stellar rotational velocity ($v$sin$i$) and the radial velocity (RV) of the
object relative to the barycenter. For many of the objects in our sample these values have been
estimated in the literature. For objects that do not have previous estimates, we fit rotationally
broadened and velocity--shifted synthetic spectra to a single spectral order to obtain RV and
$v$sin$i$ estimates. The order selected is chosen based on the strongest available photospheric lines.
This method works well for A--type and late B--type objects but fails for the
earliest spectral types. Typically the strong He I
$\lambda$4471 and Mg II $\lambda$4481 lines are used to measure $v$sin$i$ for rapidly rotating
B--type stars \citep{halbedel96}; for more slowly rotating B--type stars, metal lines near 6400 \AA\hspace{0pt}
and 4500 \AA\hspace{0pt} are used \citep{fekel03}. Many of our early type objects, however, show no absorption
features at any of these wavelengths. For a
typical S/N spectrum in our sample, this method can also fail for stars with $v$sin$i\gtrsim$ 200 km s$^{-1}$
 because the features become too weak to reliably measure.  Lower resolving powers of
$R$$\sim$10,000-30,000 are typically used to measure the large $v$sin$i$ values of B--type stars
\citep[e.g.][]{halbedel96}. We do not obtain $v$sin$i$ estimates for these objects.

For objects that do not have literature or synthetic spectrum fit RV estimates, we use the
interstellar Na I D lines as an RV proxy. \citet{finkjank84} were the first to show, using a sample
of 10 HAEBES, that the velocities of the molecular clouds in which HAEBES are embedded are very
similar (mean residual of -3.1 $\pm$3 km s$^{-1}$) to the velocities of observed interstellar Ca II
and Na I D velocities in the same objects. They also showed that independently measured stellar RVs
were similar to the molecular cloud velocities, although with a larger mean residual of 8.5 $\pm$ 8
km s$^{-1}$. Thus the Na I D velocities can, in general, provide a rough estimate of the stellar RV
to within $\sim$10 km s$^{-1}$. As the mass flow diagnostics we are interested in are typically
several hundred \kms wide, this is adequate for our purposes.

\begin{figure}\label{fig:fig1a}
  \begin{center}
     \includegraphics[scale=.55,trim= 55mm 30mm 55mm 40mm]{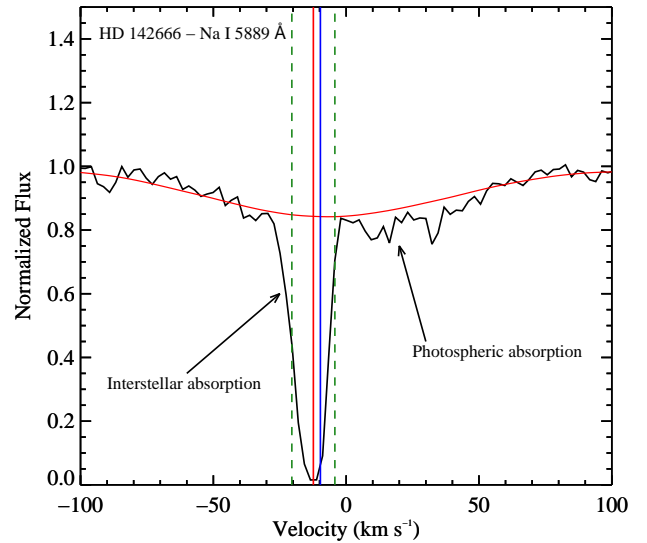}
     \caption[Interstellar Na I D velocity vs. stellar RV]{Comparison of the interstellar Na I RV determination
     with the value calculated from fitting a synthetic spectrum to photospheric absorption lines for
     HD 142666. The observed profile is shown in black; the broadened spectroscopic standard 
     HR 3221 is over plotted in red. The RV value estimated from the ISM feature is shown with the
     vertical red line. The photospheric model fit RV is shown with the vertical blue line. The $\pm$8.5 \kms
     uncertainty for the Na I value is shown with the vertical green dashed lines. It is clear that manually
     selecting the center of the ISM absorption results in negligible errors compared with the systematic
     uncertainty of using the Na I to estimate the system RV.} 
  \end{center}
\end{figure}

For objects in our sample for which we determine RVs based on synthetic spectrum fits, we compare
these measurements to the RV determined from the object's interstellar Na I D lines. RVs determined
by the synthetic spectral fits from \citet{alecian13} are also used. The interstellar Na I RV estimate is
made by manually pointing at the approximate center of the interstellar absorption. This is appropriate since the
estimates are only accurate to $\sim$8 \kms. An example of the Na I interstellar medium (ISM)
absorption is shown in \autoref{fig:fig1a} for HD 142666. The Na I ISM absorption is very strong and
narrow and is clearly distinguished from the photospheric profile. This is a particularly good example
of the strength of using the Na I value as an estimate for the system RV: the Na I and synthetic model
values differ by only $\sim$2.5 \kms. Further examples of the Na I ISM features can be seen in
the individual object plots in the Appendix.

\begin{figure}\label{fig:fig1}
  \begin{center}
     \includegraphics[scale=.55,trim= 25mm 0mm 0mm 140mm]{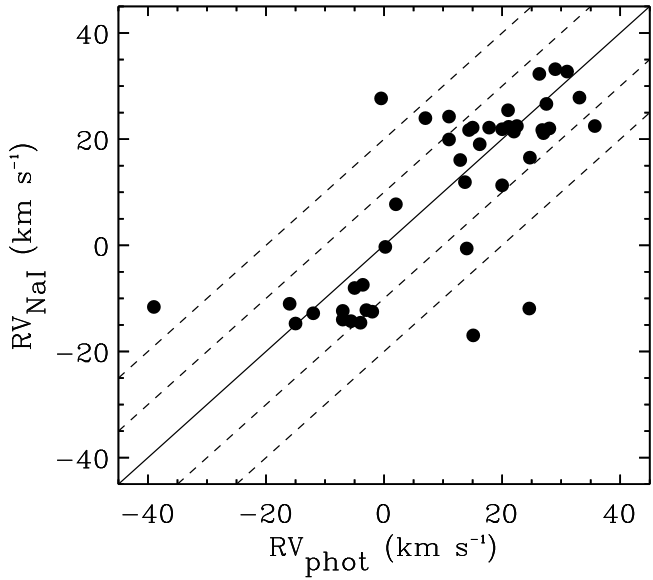}
     \caption[Interstellar Na I D velocity vs. stellar RV]{Radial velocities calculated from
synthetic
model photosphere fits versus Na I D interstellar absorption velocities. The solid line represents
equal velocities. The dashed lines moving outward from the solid line represent differences between
the velocities of 10 and 20 \kms. The mean deviation is 8.1 \kms. We note that the
four objects (HD 50083, IL Cep, MWC 614, and SV Cep) lying outside of the 20 \kms boundary 
are all known binaries. The large velocity discrepancies in these objects may be due to their
orbital motion.} 
  \end{center}
\end{figure}

The result of the comparison is plotted in \autoref{fig:fig1}. Most objects show a velocity discrepancy of $<$10 \kms. The four
objects that lie outside of the 20 \kms boundaries are all confirmed binaries and the large velocity
differences may be due in part to their orbital motion. We find a mean velocity difference of
8.1$\pm$3.6 \kms for the objects in \autoref{fig:fig1}, where the uncertainty is the
95\% confidence interval given by the \textit{t}--distribution \citep{feigelson12}. The
mean difference drops to 7.9 \kms if the four outliers are excluded, a negligible difference. {The
standard deviation of the absolute value of the difference between the interstellar RV values and
those determined by the model fits is 8.5 \kms. We take this to be the uncertainty associated with 
the values determined from the Na I lines.}

Our sample thus confirms the findings of \citet{finkjank84}. For future HAEBE studies we suggest that
the interstellar Na I D line velocities be used for objects that, (1) do not allow for reliable
photospheric fits either due to lack of absorption features or very high $v$sin$i$, or (2) only
$\sim$10 \kms accuracy is needed in the radial velocity.  All other physical information about the
objects examined in this study (e.g. $M_*$, $R_*$, $\dot{M}$) are taken from the literature. The
relevant references are given in \autoref{tab:tab2}.

\capstartfalse
\begin{deluxetable*}{lccccccccc}
%\linespacing{1}
\tablecaption{HAEBE optical sample physical parameters$^a$ \label{tab:tab2}}
\tablehead{\colhead{} & \colhead{} & \colhead{$v_{rad}$} &
\colhead{$M_*$}&\colhead{$R_*$}&\colhead{\textit{v}sin\textit{i}}&\colhead{log($\dot{M}$)} &
\colhead{Disk}&\colhead{\textit{i}}&\colhead{}\\
\colhead{Object ID}&\colhead{Spectral Type}&\colhead{(km s$^{-1}$)}&
\colhead{($\Msun$)}&\colhead{($\Rsun$)}&\colhead{(km s$^{-1}$)}&
\colhead{(M$_\odot$ yr$^{-1}$)}& \colhead{Detected$^b$} & \colhead{($^\circ$)} & \colhead{References}\\
\colhead{(1)} & \colhead{(2)} & \colhead{(3)} & \colhead{(4)} & \colhead{(5)} &
\colhead{(6)} & \colhead{(7)} & \colhead{(8)} & \colhead{(9)} & \colhead{(10)}}
%\tabletypesize{\scriptsize}
\startdata
AB Aur     & A0& 24.7 & 2.50 & 2.62 & 116 &-6.85 & Y & 40 &1,2\\
AE Lep     & B6& \textbf{28.0} &\nodata & \nodata & \nodata&\nodata &\nodata &\nodata &39\\
BD+61 154  & B8& -16.0& 3.40 & 2.42 & 112 &\nodata & Y &   70  &1,6,9\\
BF Ori     & A2& 22.0 & 2.58 & 3.26 & 39 &$<$-8.00& Y &\nodata&1,2,6,8\\
BH Cep     & F5& -5.6 & 1.70 & 2.40 & 98 &$<$-8.30& Y & 84 &2,38,40\\
CQ Tau     & F2& 35.7 & 2.93 & 5.10 & 98 &$<$-8.30& Y &   29  &1,2,12,37\\
DW CMa     & B3&\nodata&\nodata&\nodata&\nodata&\nodata& Y &\nodata&4,13\\
GSC 04794-00827 & A? & \textbf{25.0} & \nodata&\nodata&\nodata&\nodata&\nodata&\nodata&43\\
HD 141569  & A0& 35.7 & 2.33 & 1.94 & 228 & -6.90  & Y &   55  &1,2,14\\
HD 142666  & A5& \textbf{-9.7} & 2.15 & 2.82 & 65 & -7.22  & Y &\nodata&1,3,15\\
HD 163296  & A1& -9.0 & 2.23 & 2.28 & 129& -7.16  & Y &\nodata &1,3,6,18\\
HD 169142  & A7& -0.4 & 1.69 & 1.64 & 48 &\nodata& Y & 0 &1,46\\
HD 190073  & A1&  0.2 & 2.85 & 3.60 &  4 &\nodata & Y &   45  &1,19\\
HD 203024  & A1& -14.0& 2.80 & 3.40 & 162 &\nodata& Y & 74 & 1,44 \\
HD 244314  & A1& 22.5 & 2.33 & 2.07 & 52 & -6.90  &\nodata &\nodata & 1,3\\
HD 244604  & A4& 26.8 & 2.66 & 3.69 & 98& -7.20  & Y &\nodata&1,3,21\\
HD 245185  & A1& 16.0 & 2.19 & 1.85 & 118& -7.20 & Y & 81 &1,3,6,44\\
HD 249879  & A2& 11.0 & 4.00 & 5.90 & 249& -8.00 &\nodata&\nodata&1,3\\
HD 250550  & B8& -22.0& 4.80 & 3.50 & 79& -7.80  & Y &\nodata&1,3,13\\
HD 287823  & A0& -0.3 & 2.50 & 2.60 & 10&\nodata & ? &\nodata&1\\
HD 290409  & A2& 80.3 & 2.04 & 1.75 & 250&\nodata &\nodata&\nodata&1\\
HD 290500  & A2& 29.0 & 1.96 & 1.68 & 85& \nodata &\nodata &\nodata &1\\
HD 290764  & A8& 26.3 & 1.86 & 2.30 & 60&\nodata& Y & 32 &41\\
HD 290770  & B9& \textbf{24.6} & 2.86 & 2.49 & 240&\nodata &\nodata &\nodata &1\\
HD 34282   & A3& 16.2 & 1.59 & 1.66 & 105&$<$-8.30& Y &\nodata&1,2,18,22\\
HD 35187   & A2& 27.0 & 1.93 & 1.58 & 93& -7.60  & Y &\nodata&1,3,24\\
HD 35929   & F1& 21.1 & 4.13 & 8.10 & 61.8&\nodata &Y &56 &1,44\\
HD 36408   & B8& 15.0 & 4.10 & 3.50 &\nodata&$<$-8.00& ? &\nodata&2\\
HD 37357   & A1& 21.4 & 2.47 & 2.83 & 124& \nodata & Y &\nodata&1,47\\
HD 37411   & B9& \textbf{14.1} & \nodata & \nodata & \nodata&\nodata&Y &63 &33,44\\
HD 38120   & B9& 28.0 & 2.49 & 1.91 & 97&-6.90  & Y?&  $<$8 &1,3,35\\
HD 50083   & B4& -0.5 & 12.10 & 10.0 & 233&\nodata & ? &\nodata&1\\
HK Ori     & A3& \textbf{22.9} & 3.00 & 4.10 & \textbf{20} & -5.24  & Y &\nodata&2,6\\
IL Cep     & B4& -39.0&\nodata&\nodata& 179 &\nodata & ? &\nodata&1\\
IP Per     & A3& 13.7 & 1.86 & 2.10 & 80&\nodata & ? &\nodata&1\\
IRAS 05044-0325 & B3& \textbf{25.2} &\nodata&\nodata&\nodata&\nodata&\nodata &\nodata &43\\
IRAS 06071+2925 & B9& \textbf{12.5} &\nodata&\nodata&\nodata&\nodata&\nodata &\nodata &43\\
IRAS 07061-0414 & B0& \textbf{43.0} &\nodata&\nodata&\nodata&\nodata&\nodata&\nodata&42\\
IRAS 17481-1415 & A0& \textbf{-5.1} & 2.50 &\nodata&\nodata&\nodata&\nodata&\nodata&10\\
IRAS 18306-0500 & B2& 33.1& 12.00 & 6.80 &\nodata&\nodata& Y & 87 &40,44\\
IRAS 18454+0250 & B1& \textbf{$\sim$0} & 19.00 & 8.50 &\nodata&\nodata&\nodata&\nodata&42,13\\
IRAS 19343+2926 & B1& \textbf{-10.7} &\nodata&\nodata&\nodata&\nodata&\nodata&\nodata&43 \\
LkH$\alpha$ 134 & B3 & \textbf{-13.0} & 6.00 & 5.05 &\nodata&\nodata& Y & 58 &40,44\\
LkH$\alpha$ 208 & A4& \textbf{13.6} & 2.40 & 3.20 &\nodata&\nodata& Y & 56 &40,44\\
LkH$\alpha$ 233 & A3& \textbf{-9.10} & 3.20 & 4.45 &\nodata&\nodata& Y & 18 &40,44\\
LkH$\alpha$ 257 & B5& \textbf{-15.2} & 3.61 & 2.68 &\nodata &\nodata& Y & 57 & 44\\
LkH$\alpha$ 324 & B8& \textbf{-10.5} & 5.10 & 6.30 &\nodata&\nodata& Y & 42 &40,44\\
LkH$\alpha$ 339 & A1& \textbf{26.4} & 3.18 & 3.92 &\nodata&\nodata& Y & 40 &40,44\\
MWC 1080   & B1&\nodata& 17.4 & 7.3 &\nodata&\nodata & Y &   83  &1,6,10\\
MWC 120    & B9& 47.0 & 3.94 & 4.60 & 120&-6.85  & Y &\nodata& 1,3\\
MWC 137    & B1 &\nodata &\nodata &\nodata &\nodata&\nodata & Y &   80  &6,10,13\\
MWC 147    & B2& \textbf{18.6} & 6.60 & 4.80 & 83 & -6.12 & Y &\nodata&1,3,13,48\\
MWC 300    & B1& \textbf{-4.5} & 7.37 & 3.33 &\nodata &\nodata& Y & 62 &13,44\\
MWC 361    & B4& \textbf{-7.0} & 10.70 & 10.40 & 26 &\nodata& Y & 26 & 1,44\\
MWC 480    & A4& 12.9 & 1.93 & 1.93 & 98&$<$-7.23& Y &   37  &1,2,12\\
MWC 610    & B3& 14.0 & 8.00 & 4.70 & 219&\nodata & ? &\nodata&1\\
MWC 614    & A0& 15.1 &\nodata &\nodata & 69 &-6.59   & Y &\nodata&1,3,28\\
MWC 758    & A5& 17.8 & 2.90 & 4.40 & 54&-6.05   & Y &   21  &1,3,12\\
MWC 778    & B2& \textbf{10.3} & 10.00 & 3.50  &\nodata &\nodata & Y & 75 &13,42,49\\
MWC 863    & A1& -5.0 & 2.56 & 2.89 & 108&-6.12   & Y &   38  &1,2,29\\
NZ Ser     & B3& \textbf{-17.0} & 10.10 & 5.80 &\nodata&\nodata & Y & 5 &10,13\\
R Mon      & B8& \textbf{22.3} & 5.10 & 12.00 &\nodata & \nodata & Y & 60 &10,13\\
RR Tau     & A0& \textbf{15.3} & 5.80 & 9.30  & \textbf{225} & -6.86 & Y & 44 &2,32,44\\
SV Cep     & A1& \textbf{24.6} & 2.62 & 3.00 &180& \nodata & Y & 49 &1,44\\
T Ori      & A3& 56.1 & 3.13 & 4.47 &147 &-6.60   & Y &\nodata&1,2,30\\
UX Ori     & A1& 12.0 & 6.72 & 12.1 &221&-7.18   & Y & $<$8  &1,3,6,35\\
UY Ori     & B9& \textbf{19.8} &\nodata &\nodata&\nodata & \nodata & \nodata & \nodata &4\\
V1185 Tau  & A2& \textbf{20.0} & \nodata & \nodata &\textbf{250} &\nodata & \nodata &\nodata&\nodata\\
V1578 Cyg  & A1& \textbf{-3.0} & 5.90 & 9.70 &199 &\nodata & Y &\nodata&1,6\\
V1685 Cyg  & B4& \textbf{-16.0} &\nodata &\nodata &\nodata & \nodata & Y & 41? &30,34\\
V1686 Cyg  & F8& \textbf{-12.2} & 3.50 & 14.00 &\nodata &\nodata& Y & \nodata &6,40\\
V1787 Ori  & A5& \textbf{26.4} & 1.88 & 1.84 &\nodata & -7.43 &\nodata & \nodata &45\\
V1818 Ori  & B7& \textbf{26.4} & \nodata &\nodata& 46 & \nodata & \nodata & \nodata &43\\
V346 Ori   & A7& 20.0 & 1.72 & 1.96 &116&-6.90   & N? & \nodata &1,3,35\\
V351 Ori   & A6& 15.0 & 2.88 & 4.38 & 100&\nodata & ? &\nodata&1\\
V361 Cep   & B4& \textbf{-20.0} & 8.11 & 6.70 & 278 &\nodata & N? & \nodata &1,6\\
V373 Cep   & B8& \textbf{-18.6} & 5.30 & 14.00 &\nodata&\nodata & Y & 45 &39,40\\
V374 Cep   & B5& \textbf{-13.6} &\nodata& \nodata&\nodata & \nodata & \nodata &\nodata&51\\
V380 Ori   & B9& 27.5 & 2.87 & 3.00 & 7&-5.60   & Y &\nodata&1,3,6\\
V586 Ori   & A1& 31.0  & 2.28 & 1.94 & 200& \nodata & Y & 81 &1,44\\
V590 Mon   & B7& \textbf{17.7} & 4.71 & 5.86 &\nodata & \nodata & \nodata & 84 &40,44\\
V599 Ori   & F0& \textbf{7.0} & 1.83 & 5.29 & \textbf{96} & \nodata & Y & 42 &43,44\\
V718 Sco   & A4& -3.6 & 1.93 & 2.25 &113&\nodata & Y?&   32? &1,35\\
V791 Mon   & B5& \textbf{-2.6} &\nodata &\nodata &\nodata &\nodata & ? &\nodata&4\\
VV Ser     & B7& \textbf{-2.0} & 4.00 & 3.10 & 124 & -7.50 & Y & 46 &1,3,44\\
WW Vul     & A2& -4.0 & 3.70 & 5.40 & 196 & -6.38 & \nodata & \nodata &1,2\\
XY Per     & A2& 2.0 & 1.95 & 1.65 & 224&-7.02   & ? &\nodata&1,3\\
Z CMa      & B9& \textbf{-27.0} & 3.80 & 3.20 &\nodata&-6.72   & Y &\nodata&3,31\\
\enddata
%References: 1=A13,2=Lee & Chen 2009,3=Manoj 2006,4=M11,5=DB11,6=Kraus et al 2013
%7=Malfait et al 1998, 8=Viera 03, 9=Reed 2003, 10=Sartori et al 2010, 11=Verhoeff et al 2012,
%12=Liu et al 2011, 13=Garcia-Lopez et al 2006, 14=Carrati o Garrati et al 2012
\tablenotetext{a}{All spectral types, values of $M_*$, $R_*$, $\dot{M}$, $i$, and disk detections are taken from
the literature. Values of the radial velocity and $v$sin$i$ determined in this study are given in bold. Typical
radial velocity uncertainties are $\sim$5--10 km s$^{-1}$; typical $v$sin$i$ uncertainties are
$\sim$5--10 km s$^{-1}$ although large values of $v$sin$i$ can have uncertainties as large as
30--100 km s$^{-1}$.}
\tablenotetext{b}{Disk detections are indicated to provide context for the line profiles presented in
this study. Question marks (?) indicate that the measurements are uncertain as to whether or not
a disk is present. The detections are taken from the literature and are based mainly on millimeter and near--IR
flux measurements (see given references).}

\tablecomments{References: 1=\citet{alecian13}, 2=\citet{mend11}, 3=\citet{db11}, 4=\citet{viera03},
5=\citet{carmona10}, 6=\citet{hillenbrand92}, 7=\citet{tang12}, 8=\citet{fink84}, 9=\citet{boss11},
10=\citet{aa09}, 11=\citet{guillo11}, 12=\citet{guillo13}, 13=\citet{verhoeff12}, 14=\citet{thi14},
15=\citet{schegerer13}, 16=\citet{eisner09}, 17=\citet{chen12}, 18=\citet{marinas11},
19=\citet{ragland12}, 20=\citet{okamoto09}, 21=\citet{vink02}, 22=\citet{natta04},
23=\citet{ackeancker04}, 24=\citet{oud92}, 25=\citet{fuente02}, 26=\citet{mg01},
27=\citet{perrin06}, 28=\citet{liu07}, 29=\citet{fukagawa03}, 30=\citet{eisner04},
31=\citet{schutz05}, 32=\citet{garcia06}, 33=\citet{malfait98}, 34=\citet{hernandez04},
35=\citet{dent05}, 36=\citet{corporon99}, 37=\citet{mend11a}, 38=\citet{liu11},
39=\citet{lee09}, 40=\citet{manoj06}, 41=\citet{kraus13}, 42=\citet{reed03}, 43=\citet{sartori10},
44=\citet{liu11}, 45=\citet{caratti12}, 46=\citet{osorio14}, 47=\citet{juhasz10},
48=\citet{bagnoli10}, 49=\citet{perrin09}, 50=\citet{acke05}, 51=\citet{the94}}

\end{deluxetable*}
\capstarttrue

\newpage

\section{Profile Classification}
\label{sec:sec4}

\begin{figure*}\label{fig:fig2}
     \includegraphics[scale=.93,trim= 20mm 35mm 5mm 5mm]{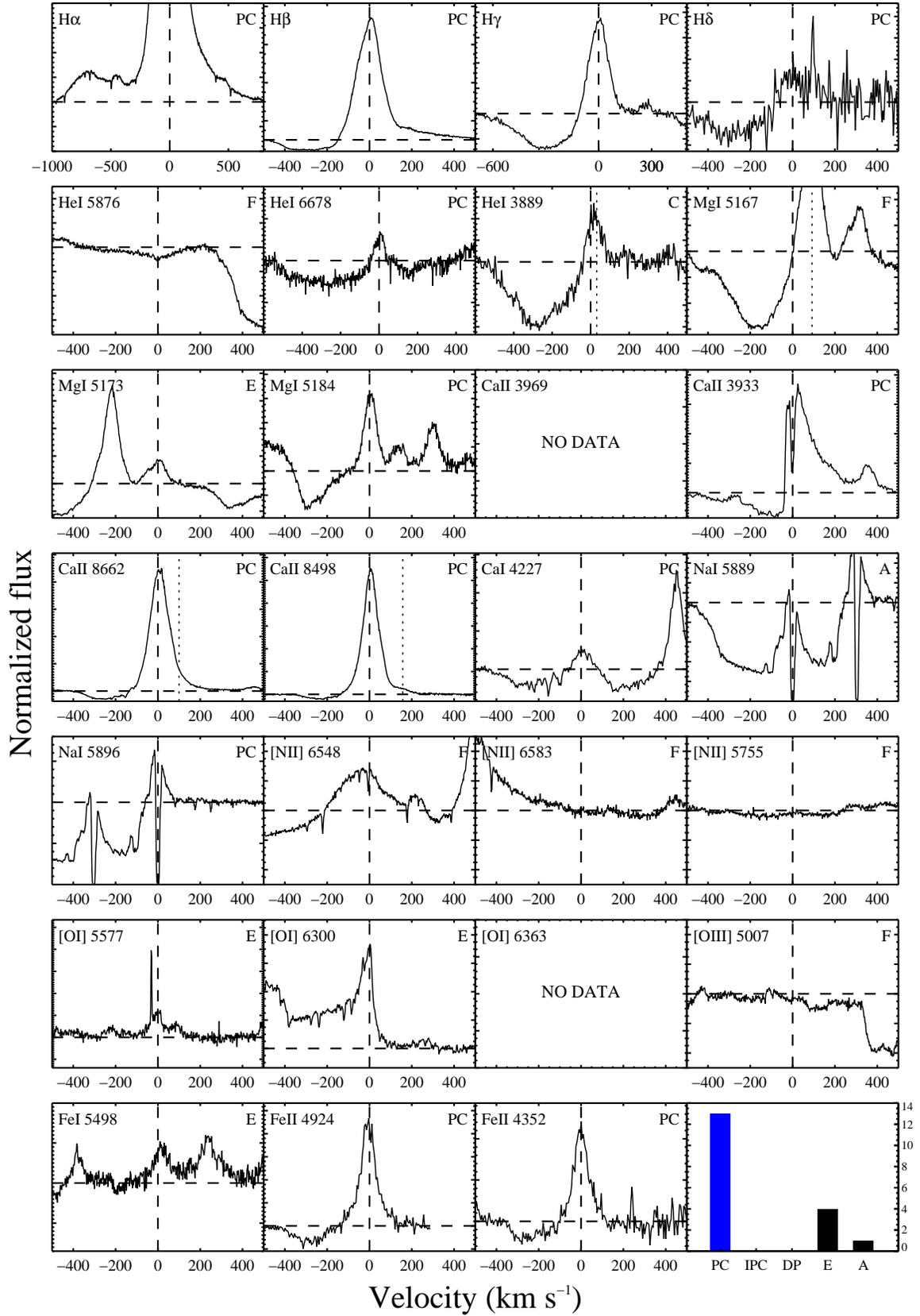}
     \caption{Extracted line profiles for Z CMa. The continuum, which is normalized to 1.0,  is
marked with a horizontal dashed line. The stellar rest velocity is marked with a vertical dashed
line. The profile classification is noted in the upper--right of each plot window. The plotted
spectral line is listed in the upper--left. The bottom right--most plot window is a bar plot of the
profile types for the object. The featureless profile count is excluded from the bar plot. 
Z CMa displays strong outflow signatures in numerous lines. The H$\alpha$ vertical plot
range is abbreviated to show the blue--shifted absorption. The [NII] 6548 $\AA$\hspace{0pt}
appears to show emission but is actually showing the far blue wing of the H$\alpha$ line. 
Although the \ion{Na}{1} doublet lines should show scaled versions of the same profile morphology, 
the blue--shifted absorption from the 5896 \AA\hspace{0pt} profile decreases the emission
peak of the 5889 \AA\hspace{0pt} profile to below the local continuum, resulting in a classification 
of `A' for the 5889 \AA\hspace{0pt} line.
Note the vertical dotted lines denoting the rest wavelengths of lines that potentially 
contaminate the line of interest (e.g., H$\zeta$ in the He I 3889 \AA\ panel).}
\end{figure*}

Line profiles are classified into six morphological groups: P--Cygni (PC), inverse P--Cygni (IPC),
double--peaked emission (DP), single--peaked emission (E), absorption (A), and featureless (F).
For the purpose of generating mass flow statistics, pure absorption profiles (A) are further grouped as red or blue
absorption based on the maximum absorption velocity: $v_{red}$ $>$ 0 km s$^{-1}$ and $v_{blue}$ $<$
0 km s$^{-1}$. The objects counted as showing blue(red)--shifted absorption either have
a PC(IPC) profile \textit{or} an A profile that is blue(red)--shifted. Objects are classified as E or DP if the profile is in
emission (single and double peaked, respectively) and there is no clear
evidence of absorption below the local continuum. Profiles are labeled as F if they show no detectable
departure from the estimated photospheric line profile. We note that a small number of the PC profiles, most
of them H$\alpha$ profiles, show
blue--shifted absorption that does not extend below the stellar continuum, i.e., the absorption is superposed
on the wing of the emission profile (see, for example, the H$\alpha$ profile in \autoref{fig:fig2}). 
While these are not strictly P--Cygni profiles, we have chosen to extend the PC classification to these
objects since it is the closest morphology group that describes the profile. 

\begin{figure*}\label{fig:fig3}
  %\begin{center}
     \includegraphics[scale=.93,trim= 20mm 35mm 5mm 5mm]{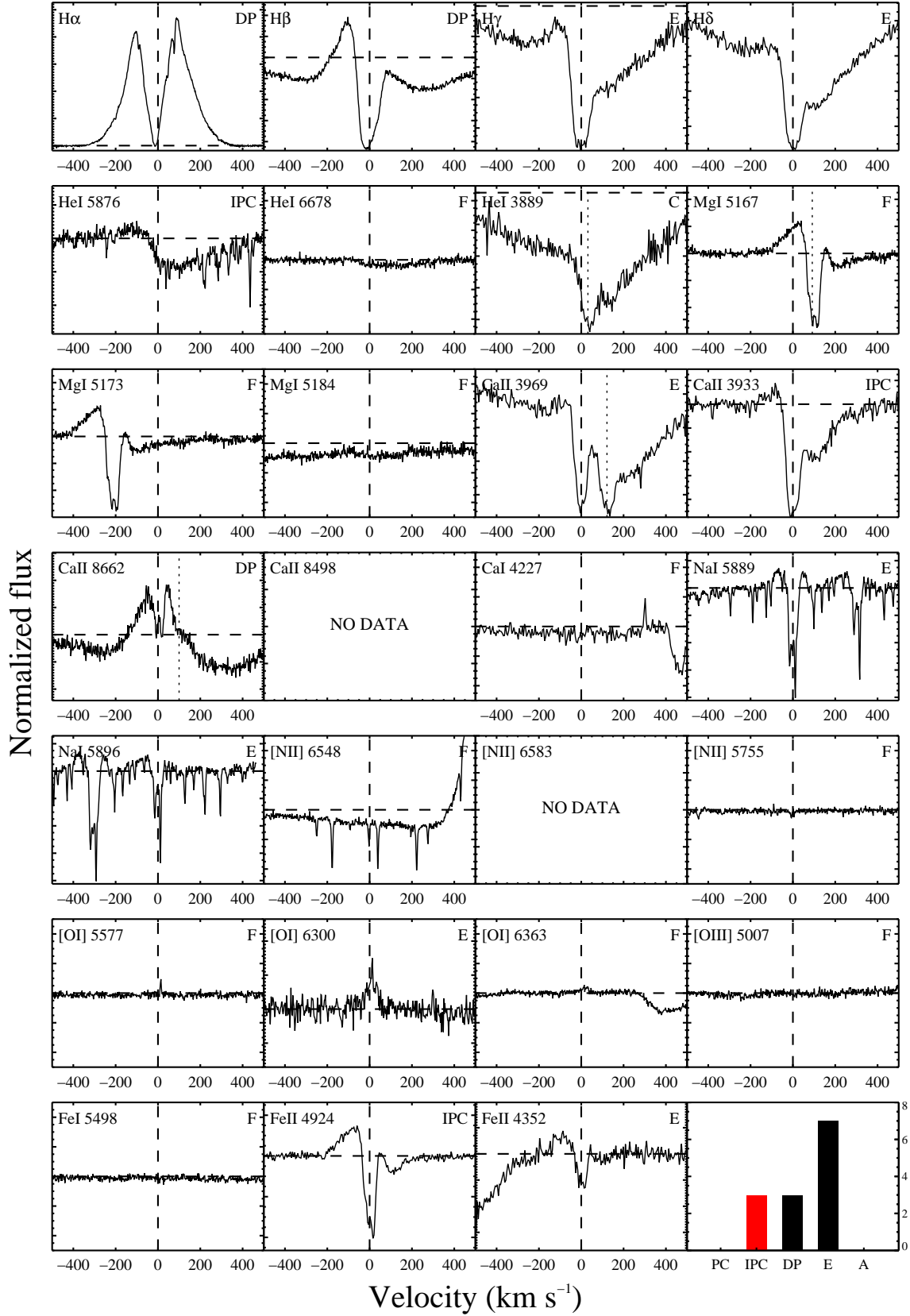}
     \caption{Extracted line profiles for MWC 120. The format is the same as \autoref{fig:fig2}. Note the
     strong IPC morphology in the \ion{He}{1} 5876 \AA\hspace{0pt} line.}
  %\end{center}
\end{figure*}

\begin{figure}\label{fig:fig4}
  %\begin{center}
     \includegraphics[scale=.50,trim=50mm 10mm 10mm 10mm]{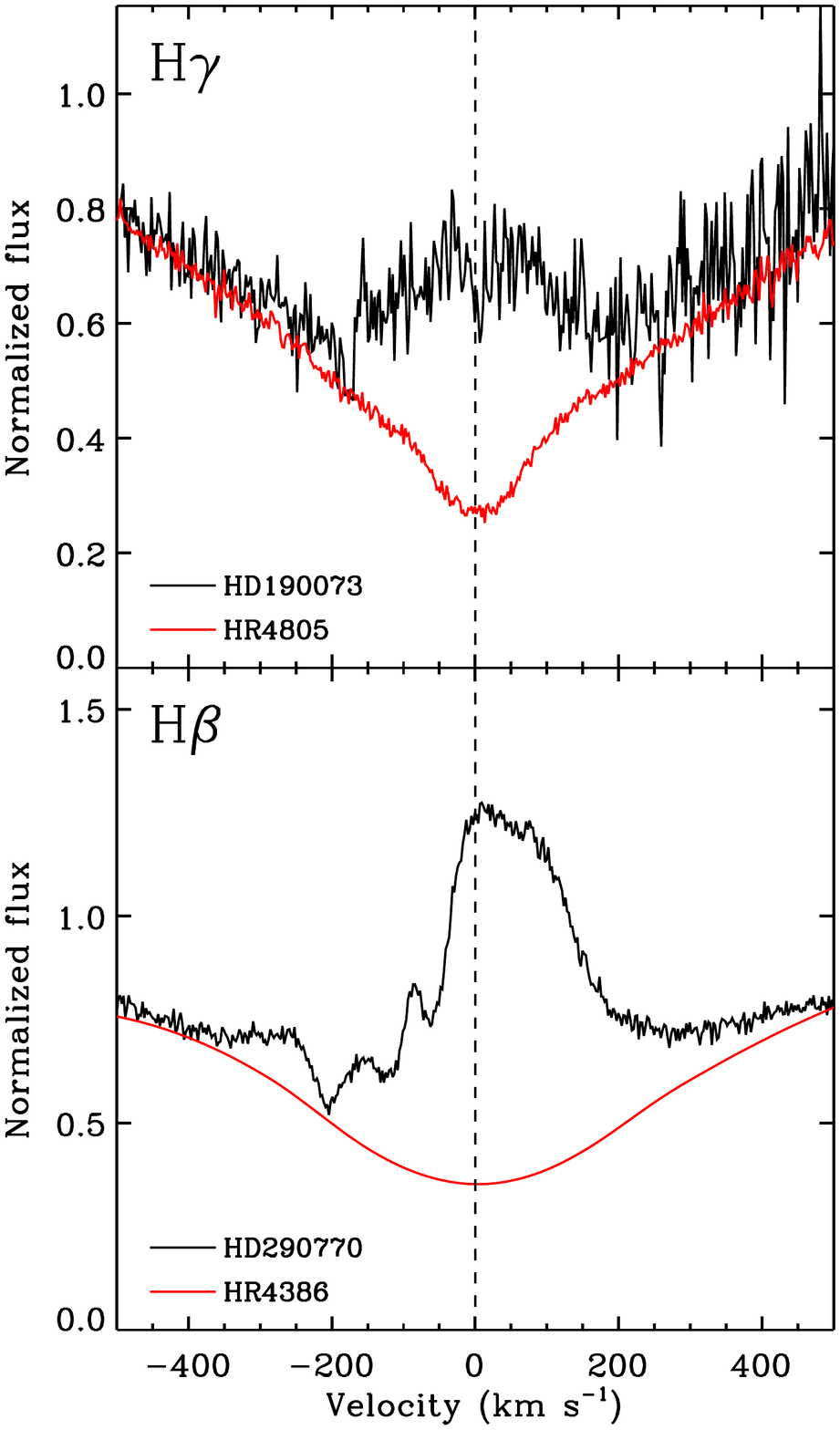}
     \caption{A comparison of the
H$\gamma$ profile of HD 190073 with a spectroscopic standard of similar T$_{eff}$ (top panel).
The object profile is plotted in black and the standard photospheric profile is plotted in red. The
core of the line is clearly filled in by emission, placing this profile in the E morphology group. A similar comparison is shown for HD 290770 
(bottom panel) at H$\beta$. The standard, HR4386, has been rotationally broadened to match the 
$v$sin$i$ of HD 290770. The blue--shifted absorption can be seen superimposed on the emission profile, resulting
in a profile classification of PC.}
  %\end{center}
\end{figure}

\begin{figure*}\label{fig:fig5a}
  %\begin{center}
     \includegraphics[scale=.93,trim= 20mm 35mm 5mm 5mm]{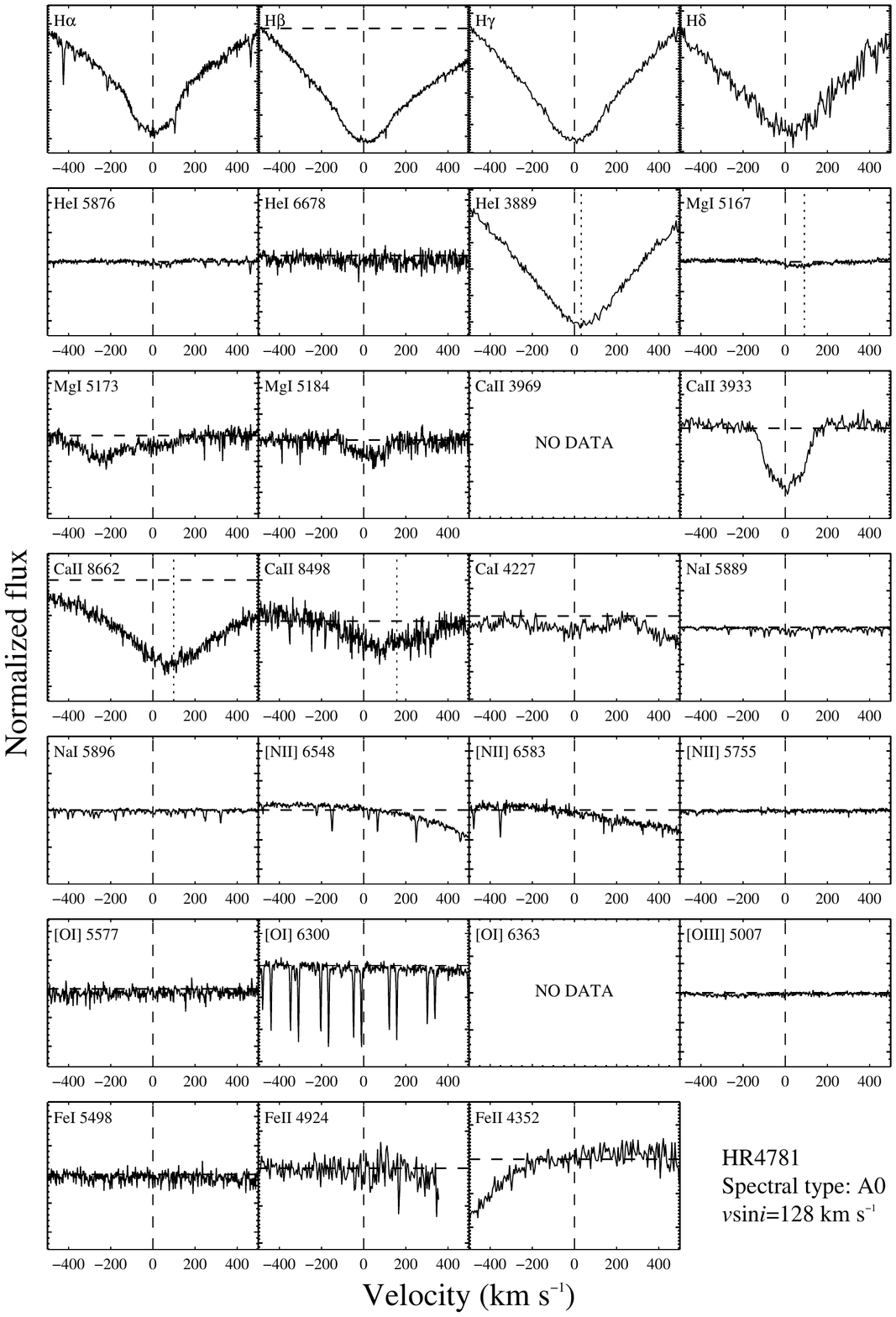}
     \caption{Extracted line profiles for the spectroscopic standard HR 4781.}
  %\end{center}
\end{figure*}

\begin{figure*}\label{fig:fig5b}
  %\begin{center}
     \includegraphics[scale=.93,trim= 20mm 35mm 5mm 5mm]{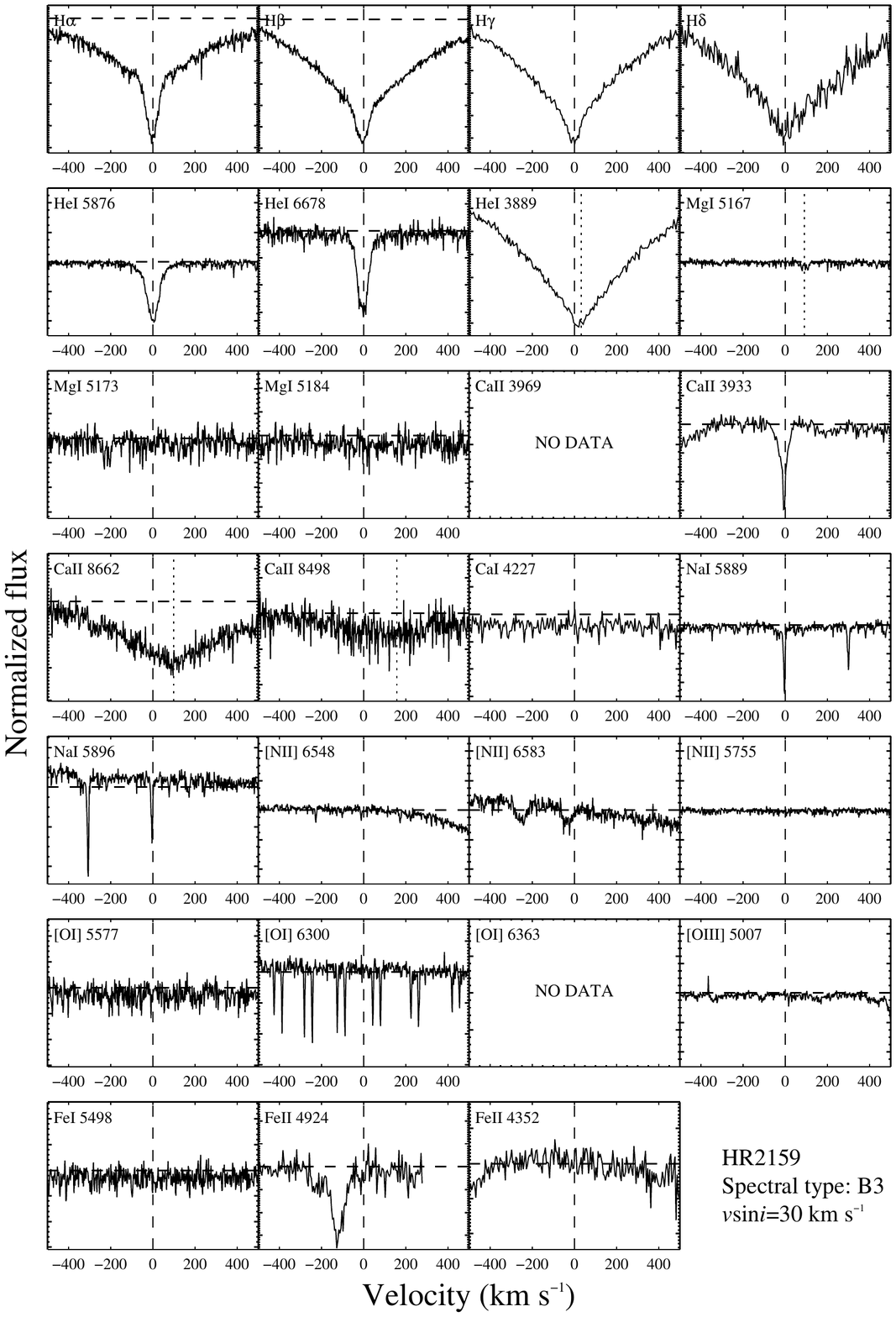}
     \caption{Extracted line profiles for the spectroscopic standard HR 2159.}
  %\end{center}
\end{figure*}

In order to accurately classify the morphologies, each profile was compared with a rotationally
broadened spectroscopic standard of a similar effective temperature. Objects with unknown $v$sin$i$
values did not have their comparison spectra broadened. For most objects, this allowed weak emission and absorption
features to be identified in lines with significant photospheric components. This comparison
is most important for the Balmer lines since most of our objects have strong photospheric Balmer line absorption. An example of this
effect is shown in the top panel of \autoref{fig:fig4} where the core of the H$\gamma$ profile of HD
190073 is clearly filled in by emission. A second example is shown in the bottom panel of
\autoref{fig:fig4} for a more complex line morphology at H$\beta$ for HD 290770. In this case the
blue--shifted absorption at $\sim$-100---200 \kms can clearly be seen superimposed on the broad
emission profile. The comparison with the standard makes it obvious that this is non--photospheric
absorption. To be clear, we note that the profiles shown in \autoref{fig:fig2} and
\autoref{fig:fig3} show only the continuum normalized profiles--no photospheric subtraction was done
to make these plots even though photospheric comparisons were made while classifying the profiles.
The photospheric comparisons are done by eye and, in general, it is obvious when a non--photospheric
contribution to the line profile is present. Thus no quantitative comparison of the photospheric and circumstellar
components is performed. The approximate comparisons prevent gross misclassification of line profiles, i.e.
identifying circumstellar absorption when the absorption actually forms in the photosphere. Thus objects
will often show line profiles that appear to show structure but are in fact due entirely to photospheric
absorption. Examples of an A0 and B3 spectroscopic standard are shown in \autoref{fig:fig5a} and
\autoref{fig:fig5b}.

In some objects there is significant contamination of the He I 3889 \AA, Mg I 5167 \AA, Ca II 3969
\AA, and Ca II 8662 \AA\hspace{0pt} spectral lines by neighboring features. This is especially
severe at Mg I 5167 \AA\ (e.g. Z CMa and DW CMa), where we tentatively identify the nearby Fe I
5168.89 \AA\ line (+81.8 km s$^{-1}$), and Ca II 3969 \AA\ and He I 3889 \AA\ which are often
overwhelmed by the nearby H$\epsilon$ and H$\zeta$ lines, respectively. For example, in
\autoref{fig:fig2} the He I 3889 \AA\ line appears to show a PC morphology. However, this profile
more likely due to the nearby H$\zeta$ line. Thus this line is listed as ``contaminated'' since the
true profile is obscured by the H$\zeta$ line. The N II 6548 \AA\ line lies in the far blue
wing of the H$\alpha$ and often appears to show structure when in fact the line emission and shape
are due entirely to H$\alpha$ (e.g., \autoref{fig:fig2}).  If there are other lines of the same
element that show a non--featureless line profile, e.g. the Mg I 5173 or 5184 \AA\ lines in the case
of Mg I 5167 \AA, the line is categorized as 'C' for 'contaminated'.  If the other lines of the same
element are featureless, we assume a featureless spectrum for the feature in question. The rest
velocities of these contaminating lines are marked with a dotted line in the individual profiles
plots. With the exception of Ca II 8662 \AA, the wavelength regions containing these diagnostics are
mostly featureless in the whole sample. Contaminated profiles comprise 16\% of the He I 3889 \AA\
profiles, 3\% of the Mg I 5167 \AA\ profiles, 7\% of the Ca II K profiles, and 9\% of the Ca II 8662
\AA\ profiles. Thus the contamination is not problematic concerning the classification of most
profile morphologies. 

Examples of the extracted profiles and their categorization are shown in \autoref{fig:fig2}
and \autoref{fig:fig3}. Plots of the individual line profiles for all objects are given in Appendix
A. The profile classification is shown in the upper right corner of each plot window; the spectral
feature being shown is listed in the upper left. Spectral lines for which there is no data are
labeled as such in the plots. These lines either fell in between orders on the detector or the S/N
was too low for analysis. Features with rest wavelengths below 5000 \AA\hspace{0pt} are plotted
using a 5 km s$^{-1}$ (2 pixel) bin window in order to enhance the S/N. The bottom--right panel of
each figure shows a bar plot of the number of line profiles in each morphology category. 
We note that only the PC and IPC bars are shown in color. The bar showing the count for pure
red-- or blue--shifted absorption (i.e., the A profiles) is shown in black even though the
subdivision of the A profiles into red-- and blue--shifted absorption is included in the statistics.

\section{Morphology Statistics}
\label{sec:sec5}

\subsection{Pre--main sequence evolutionary groups}
\label{subsec:sec51}

HAEBES display a wide variety of SEDs indicating that, as a group, they represent objects at
different phases of pre--main sequence evolution \citep{meeus01,malfait98,hillenbrand92}. An accurate
calculation of mass flow statistics would only include objects that are currently in the same
pre--main sequence phase as one another. However, this is challenging due to the relatively small number of HAEBES
each specific evolutionary phase would contain. Studies of this sort will
necessarily suffer from smaller sample sizes in each group thus increasing the uncertainties associated with the
mass flow statistics. 

In order to eliminate any objects from our sample that no longer show evidence of inner disk
material, and thus should not show evidence of mass accretion and mass loss, we employ the
color--color classification of \citet{hillenbrand92}. The $J-H$ vs $H-K$ classification is a proxy
for the full SED analysis presented in \citet{hillenbrand92}. Objects with $J-H$$<$0.3 and
$H-K$$<$0.3, the Group III objects from \citet{hillenbrand92}, likely have insufficient material in
their inner disks to be actively accreting and ejecting material. These objects should not be
included in any comparisons to CTTSs since all CTTSs show strong evidence for accretion and mass
loss. 

\begin{figure}\label{fig:fig5}
  %\begin{center}
     \includegraphics[scale=.56,trim=25mm 0mm 10mm 135mm]{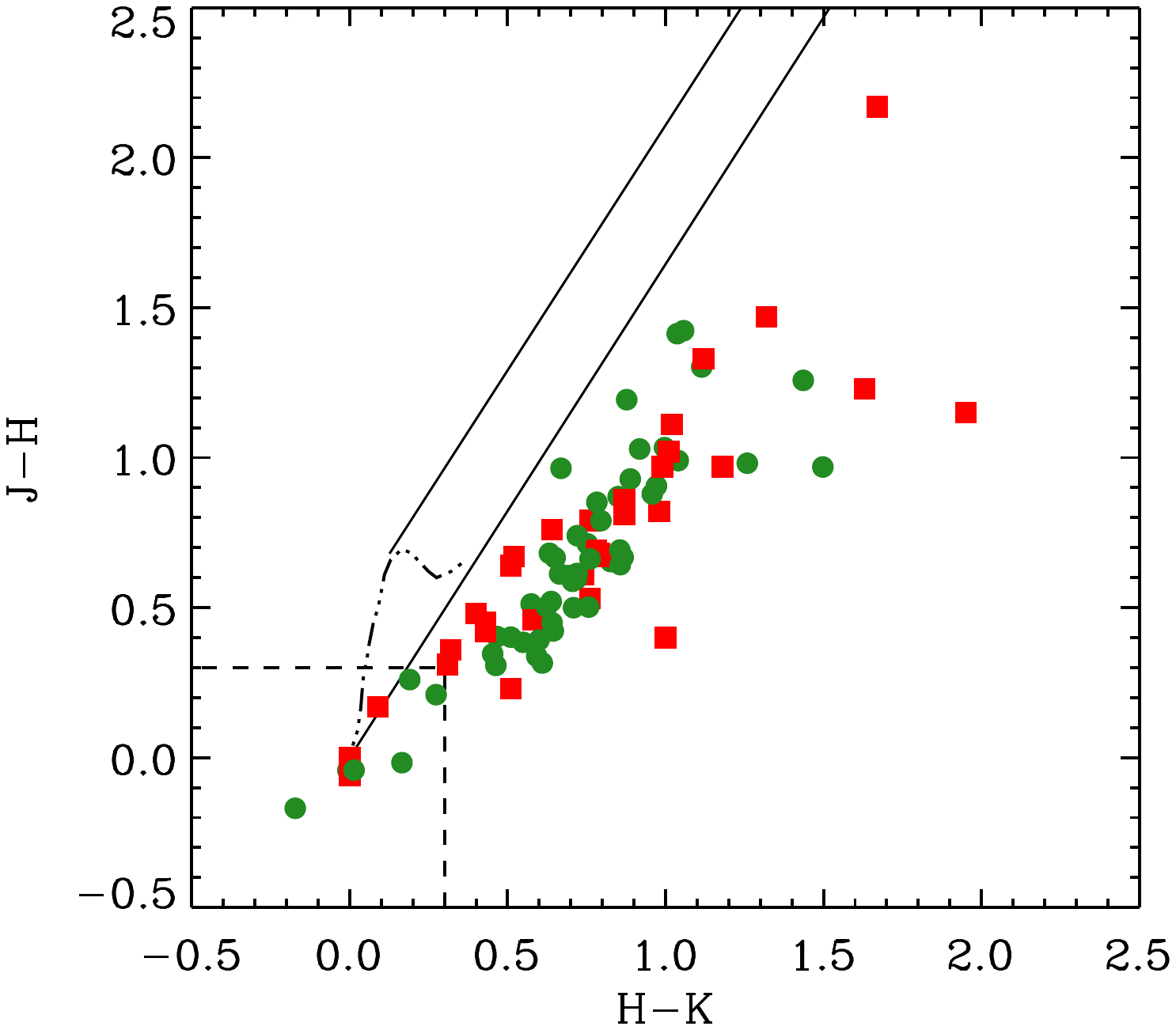}
     \caption{$J-H$ and $H-K$ colors for our sample. Reddening
vectors are plotted as solid lines. The approximate location of the main sequence is indicated by
the dashed--dotted line. The area within the dashed lines contains objects that are most
likely evolved HAEBES without any significant IR excess. Objects with de--reddened colors are
plotted as green circles; objects without reddening corrections are plotted as red squares. There are 
9 objects contained in the box that we exclude from further analysis.}
  %\end{center}
\end{figure}

We have collected $J$, $H$, and $K$ data from the literature for our observed sample
in order to identify any potential Group III objects. Optical extinctions are also collected
for some objects. The $J-H$ and $H-K$ colors for our sample are plotted in \autoref{fig:fig5}.
Objects with de--reddened colors are plotted as green circles while objects without known $A_V$
values are plotted as red squares. Reddening vectors for the standard $R$=3.1 extinction law are
plotted as solid black lines. An approximation of the main sequence for dwarf spectral types B8
through M6 is represented by the dashed--dotted line. Main sequence objects are required to remain
within the boundaries of the reddening vectors. It can be seen that most of our objects lie well to
the right of the main sequence region, confirming the existence of significant excess emission. The
Group III objects from \citet{hillenbrand92} all exist in region $J-H$$<$0.3, $H-K$$<$0.3. We use
the box defined by the intersection of these boundaries as our criteria for identifying Group III
objects in our sample.  The box is plotted in \autoref{fig:fig5} using dashed lines. There are 9
objects (AE Lep, HD 141569, HD 203024, HD 35929, HD 36408, HD 50083, IL Cep, MWC 610, and V361 Cep)
that lie within the Group III box. These objects are marked with a $\dagger$ in \autoref{tab:tab4} and
will be excluded from the analysis presented in the rest of the paper. This reduces the sample to 78
objects.

\subsection{Calculating the mass flow incidence}
\label{sec:sec52}

In order to exclude objects that are not unambiguously experiencing some sort of outflow or
accretion, we do not include pure emission lines (double or single--peaked) as evidence of mass
flows. While emission lines in many species can form in a wind or accretion flow, they can also form
in static geometries such as extended stellar and disk atmospheres or remnant circumstellar
envelopes. On the other hand, doppler shifted absorption is an unambiguous sign of mass motion. The
CTTS comparison statistics that are discussed in \autoref{sec:sec55} are calculated using the same
criteria. Thus we only count red and blue--shifted absorption below the local continuum (e.g. the
blue--shifted absorption seen at H$\beta$ in the bottom panel of \autoref{fig:fig4}) as indications
of mass flows when calculating the final statistics in \autoref{tab:tab3} and \autoref{tab:tab4}.
For objects with multiple observations ($\sim$20\% of the sample, see \autoref{tab:tab1}) we search
for absorption signatures in all of the exposures. Thus the computed statistics include objects
that, for example, show blue--shifted absorption in a line for one exposure but do not show the same
morphology in a separate exposure. This enables us to account for variability, in a limited way, in
at least some objects.   

The morphology statistics for the mass flow diagnostics are presented in \autoref{tab:tab3} and a
summary of each object's profile classifications is given in \autoref{tab:tab4}. In
\autoref{tab:tab3} the line diagnostic with the highest occurrence of each profile type is
highlighted in bold font. For each line the total number of object profiles examined is give in
column 2.  The number of each type of morphology is listed and the fraction of the total number of
profiles is given in parentheses. In column 8 of \autoref{tab:tab4} the total number of line
diagnostics examined is given for each object. The number of contaminated profiles for each object
is not shown although they are included in the total number of lines in column 8. It is immediately
obvious from \autoref{tab:tab3} that most of the diagnostics show featureless profiles. In fact,
H$\alpha$, H$\beta$, and Ca II $\lambda$8498 are the only lines for which a majority of the sample
displays some form of activity. The lines with the highest incidence of circumstellar features
displaying PC and IPC morphologies are H$\beta$ (30\%) and He I $\lambda$5876 \AA\ (14\%),
respectively. The Na I D lines show the highest rate (12\%) of pure absorption profiles. After the
Balmer lines, the next best spectral line for observing a PC morphology is Fe II $\lambda$4924 which
displays PC characteristics in 11\% of the object profiles.

The last row in \autoref{tab:tab4} displays the final mass flow statistics for our sample. The
uncertainties are the 68\% confidence intervals calculated using Wilson's score test for large
samples \citep{wilson27}. These intervals are very close to the values obtained assuming Poisson
statistics in the distribution of observed incidence. However, the Wilson score test results in a
weighted average towards the center of the distribution.  We note that the given incidence is not
the adjusted fraction given by the score test and thus the confidence intervals may be asymmetric
around the value. The occurrence of both inflow and outflow signatures in our sample are fairly low:
only 26\% of our objects have at least one profile displaying red--shifted absorption; only 37\%
have at least one profile displaying blue--shifted absorption.

\subsection{Comparison to previous HAEBE studies}
\label{subsec:sec53}

A comparison of our mass flow statistics to previous HAEBE studies is presented in
\autoref{tab:tab5}.  We find similar rates of blue--absorption in our sample compared to both
\citet[][FM84]{fink84} and \citet[][HP92]{hp92}. Both \citet[][V03]{viera03} and
\citet[][R96]{reipurth96} find rates of blue--shifted absorption that are significantly lower than
ours (as well as FM84's and HP92's). We attribute this large discrepancy to the fact that they only
observe H$\alpha$ which tends to mask sub--continuum absorption due to strong emission components
from multiple sources (e.g. emission from accretion, stellar winds, nebular background). We note,
however, that using solely H$\alpha$ in our sample results in a blue--shifted absorption fraction of
26\%$^{+5}_{-4}$, a value in between the low rates of \citet{viera03} and \citet{reipurth96} and our
final fraction (37\%). This may be due to our classification of profiles as P--Cygni if the blue feature is
clearly absorption and not a second peak, even if the absorption does not extend below the local
emission continuum (e.g. \autoref{fig:fig4}). Our results, and those of FM84 and HP92, show that in
order to recover accurate outflow statistics diagnostics other than H$\alpha$ should be examined.
The line with the largest occurrence of blue--absorption in our sample is H$\beta$ (30\%), although
the incidence in H$\alpha$ is similar at 26\%. In fact, all objects in our sample that show P--Cygni
type profiles in lines other than H$\alpha$ and H$\beta$ also show P--Cygni morphologies in either
H$\alpha$ or H$\beta$.  Thus it appears H$\beta$ is a very strong stand--alone diagnostic of
outflows in HAEBES.

\capstartfalse
\begin{deluxetable*}{lccccccc}
%\rotate
%\linespacing{1}
\tablecaption{Line morphology statistics \label{tab:tab3}}
\tablewidth{0pt}
\tablehead{\colhead{}&\colhead{}&\multicolumn{6}{c}{Profile morphology}\\
           \cline{3-8}\\
\colhead{Line ID} & \colhead{Total} &\colhead{PC}&
\colhead{IPC}&
\colhead{DP}&\colhead{E}&
\colhead{A}&\colhead{F}\\
\colhead{(1)}&\colhead{(2)}&\colhead{(3)}&\colhead{(4)}&\colhead{(5)}&\colhead{(6)}&\colhead{(7)}&
\colhead{(8)}}
\tabletypesize{\scriptsize}
\startdata
H$\alpha$          & 78 & 20(.26) & 5(.06) & 34(\textbf{.44}) & 17(.22) & 0(.00) & 2(.03) \\
H$\beta$           & 74 & 22(\textbf{.30}) & 4(.05) & 15(.20) & 18(.24) & 5(.07) & 10(.14) \\
H$\gamma$          & 61 & 12(.20) & 0(.00) & 3(.05) & 14(.23) & 3(.05) & 29(.48) \\
H$\delta$          & 51 &  5(.10) & 0(.00) & 1(.02) & 8(.16) & 5(.10) & 32(.63) \\
He I $\lambda$3889 & 49 &  0(.00) & 0(.00) & 0(.00) & 1(.02) & 0(.00) & 40(.82) \\
He I $\lambda$5876 & 78 &  1(.01) & 11(\textbf{.14})& 1(.01) & 11(.14) & 1(.01) & 53(.68) \\
He I $\lambda$6678 & 71 &  1(.01) & 1(.01) & 0(.00) & 2(.03) &  1(.01) &  66(.93) \\
Mg I $\lambda$5167 & 71 &  0(.00) & 0(.00) & 0(.00) & 0(.00) &  0(.00) &  69(.97) \\
Mg I $\lambda$5173 & 71 &  0(.00) & 0(.00) & 0(.00) &  3(.04) &  0(.00) &  68(.96) \\
Mg I $\lambda$5184 & 71 &  1(.01) & 0(.00) & 0(.00) &  5(.07) &  0(.00) &  65(.92) \\
Ca II $\lambda$3933 & 57 &  2(.03) & 1(.01) &  0(.00)  &  1(.01) &  4(.05) &  65(.89)\\
Ca II $\lambda$3969 & 42 &  0(.00) & 0(.00) &  0(.00)  &  3(.07) &  1(.02) &  35(.83)\\
Ca II $\lambda$4227 & 71 &  1(.01) & 0(.00) &  0(.00)  &  0(.00) &  0(.00) & 70(.99)\\
Ca II $\lambda$8498 & 27 & 2(.07) & 1(.04) &  2(.07) &   9(.33) &  0(.00) &  13(.48) \\
Ca II $\lambda$8662 & 70 & 2(.03) & 2(.03) &  3(.04) &  22(.31) &  0(.00) &  35(.50) \\
Na I $\lambda$5889  & 78 & 4(.05) & 1(.01) &  0(.00) &  19(.24) &  9(\textbf{.12}) & 45(.58) \\
Na I $\lambda$5896  & 78 & 4(.05) & 1(.01) &  0(.00) &  18(.23) &  8(.10) & 47(.60) \\
$[$N II$]$ $\lambda$5755& 78 & 0(.00) & 0(.00) &   0(.00) &    3(.04) &  0(.00)  & 75(.96) \\
$[$N II$]$ $\lambda$6548& 78 & 0(.00) & 0(.00) &   0(.00) &    2(.03) &  0(.00)  & 76(.97) \\
$[$N II$]$ $\lambda$6583& 40 & 0(.00) & 0(.00) &   0(.00) &    6(.15) &  0(.00)  & 34(.85) \\
$[$O I$]$ $\lambda$5577 & 78 & 0(.00) & 0(.00) &   0(.00) &    6(.08) &  0(.00)  & 72(.92) \\
$[$O I$]$ $\lambda$6300 & 77 & 0(.00) & 0(.00) &   0(.00) &   43(\textbf{.56}) &  0(.00) &  34(.44) \\
$[$O I$]$ $\lambda$6363 & 64 & 0(.00) & 0(.00) & 0(.00) & 20(.31) & 0(.00) & 44(.69) \\
$[$O III$]$ $\lambda$5007& 71 & 0(.00) &  0(.00) & 0(.00) & 3(.04) & 0(.00) &  68(.96) \\
Fe I $\lambda$5498  & 78 & 0(.00) &  0(.00) &  0(.00) &  1(.01) & 0(.00) &  77(.99) \\
Fe II $\lambda$4924  & 64 & 7(.11) & 2(.03)  &  1(.02) &  7(.11) & 3(.05) &  44(.69) \\
Fe II $\lambda$4352  & 55 & 3(.05) & 0(.00) &  0(.00) &  3(.05) &  0(.00) &  49(.89) \\
\enddata
\end{deluxetable*}
\capstarttrue

\capstartfalse
\begin{deluxetable*}{lccccccccc}
%\rotate
%\linespacing{1}
\tablecaption{Object profile classification summary \label{tab:tab4}}
\tablewidth{0pt}
\tablehead{\colhead{Object ID} & \colhead{$N_{PC}^a$}&\colhead{$N_{IPC}$}&\colhead{$N_{DP}$}&
\colhead{$N_{E}$}&\colhead{$N_{A}$}&\colhead{$N_{F}$}&\colhead{$N_{tot}^b$}&
\colhead{Red abs.}&\colhead{Blue abs.}\\
\colhead{(1)}&\colhead{(2)}&\colhead{(3)}&\colhead{(4)}&\colhead{(5)}&\colhead{(6)}&\colhead{(7)}&
\colhead{(8)}&\colhead{(9)}&\colhead{(10)}}
\tabletypesize{\scriptsize}
\startdata
AB Aur & 4 & 0 & 0 & 5 & 1 & 16 & 26 & No & Yes\\
AE Lep$^\dagger$ & 0 & 0 & 0 & 5 & 0 & 7 & 12  & No & No\\
BD+61 154 & 6 & 0 & 0 & 9 & 0 & 10 & 26  & No & Yes\\
BF Ori & 0 & 2 & 1 & 0 & 8 & 15 & 27  & Yes & No\\
BH Cep & 1 & 0 & 0 & 0 & 3 & 22 & 26  & No & Yes\\
CQ Tau & 0 & 0 & 1 & 2 & 0 & 24 & 27 & No & No\\
DW CMa & 5 & 0 & 0 & 7 & 0 & 12 & 25 & No & Yes\\
GSC 04794--08227 & 0 & 0 & 0 & 6 & 0 & 20 & 26  & No & No\\
HD 141569$^\dagger$ & 0 & 0 & 2 & 1 &	0&	27& 30 &	No&	No\\
HD 142666&0&1&1&1&	0&	25& 28 &	Yes&	No\\
HD 163296&1&1&0&3&	1&	21& 27 &	Yes&	Yes\\
HD 169142&0&0&0&2&	0&	23& 25 &	No&	No\\
HD 190073&2&0&0&12&	0&	14& 29 &	No&	Yes\\
HD 203024$^\dagger$&0&0&0&1&	0&	26& 27 &	No&	No\\
HD 244314&2&0&0&7&	0&	20& 29 &	No&	Yes\\
HD 244604&2&0&0&5&	0&	19& 26 &	No&	Yes\\
HD 245185&0&0&1&3&	2&	20& 26 &	Yes&	No\\
HD 249879&2&0&0&4&	0&	18& 24 &	No&	Yes\\
HD 250550&2&0&0&4&	0&	6 & 12 &	No&	Yes\\
HD 287823&1&0&0&1&	0&	24& 26 &	No&	Yes\\
HD 290409&0&0&1&	1&	0&	24& 26 &	No&	No\\
HD 290500&0&0&1&	0&	0&	25& 26 &	No&	No\\
HD 290764&0&0&0&	2&	0&	24& 26 &	No&	No\\
HD 290770&1&0&0&	5&	0&	20& 26 &	No&	Yes\\
HD 34282&0&0&1&0&	0&	25& 26 &	No&	No\\
HD 35187&0&2&0&1&	0&	24& 27 &	Yes&	No\\
HD 35929$^\dagger$&0&0&1&1&	0&	24& 26 &	No&	No\\
HD 36408$^\dagger$&1&0&0&5&	0&	19& 25 &	No&	Yes\\
HD 37357&2&0&0&1&	0&	23& 26 &	No&	Yes\\
HD 37411&0&0&1&1&	0&	10& 12 &	No&	No\\
HD 38120&0&1&1&2&	0&	8& 12 &	Yes&	No\\
HD 50083$^\dagger$&0&0&3&1&	0&	21& 26 &	No&	No\\
HK Ori&	0&0&2&3&	0&	21& 26 &	No&	No\\
IL Cep$^\dagger$&	0&0&2&0&	0&	24& 26 &	No&	No\\
IP Per&	0&0&0&4&	0&	22& 26&	No&	No\\
IRAS 05044-0325&	0&	0& 	2&	1&	0&22& 25&	No&	No\\
IRAS 06071+2925&	2&	0&	0&	5&	0&	18& 26 &	No&	Yes\\
IRAS 07061-0414&	0&	0&	0&	0&	0&	25& 25 &	No&	No\\
IRAS 17481-1415&	0&	0&	0&	2&	0&	18& 21 &	No&	No\\
IRAS 18306-0500&	0&	0&	0&	0&	0&	23& 23 &	No&	No\\
IRAS 18454+0250&	0&	0&	0&	1&	0&	19& 20 &	No&	No\\
LkH$\alpha$ 134&	2&	0&	0&	6&	0&	18& 26 &	No&	Yes\\
LkH$\alpha$ 208&	0&	0&	1&	2&	0&	24& 27 &	No&	No\\
LkH$\alpha$ 233&	2&	0&	0&	4&	0&	17& 23 &	No&	Yes\\
LkH$\alpha$ 257&	0&	0&	2&	0&	0&	20& 22 &	No&	No\\
LkH$\alpha$ 324&	0&	0&	2&	0&	0&	21& 23 &	No&	No\\
LkH$\alpha$ 339&	0&	0&	0&	3&	0&	19& 22 &	No&	No\\
MWC 1080&	9&	0&	0&	4&	0&	10& 25 &	No&	Yes\\
MWC 120&	0&	3&	3&	7&	0&	12& 26 &	Yes&	No\\
MWC 137&	0&	0&	0&	13&	0&	10& 24 &	No&	No\\
MWC 147&	0&	0&	3&	3&	0&	16& 22&	No&	No\\
MWC 300&	0&	0&	2&	12&	0&	12& 26 &	No&	No\\
MWC 361&	0&	0&	3&	1&	0&	21& 26 &	No&	No\\
MWC 480&	6&	0&	1&	1&	3&	13& 25 &	No&	Yes\\
MWC 610$^\dagger$&	0&	0&	2&	0&	0&	27& 29 &	No&	No\\
MWC 614&	0&	0&	0&	2&	0&	24& 26 &	No&	No\\
MWC 758&	2&	0&	0&	4&	0&	20& 26 &	No&	Yes\\
MWC 778&	0&	0&	2&	3&	0&	8& 13 &	No&	No\\
MWC 863&	2&	0&	0&	3&	0&	20& 25 &	No&	Yes\\
NZ Ser&	        0&	0&	0&	7&	0&	13& 21 &	No&	No\\
R Mon&	        1&	0&	1&	7&	0&	13& 22 &	No&	Yes\\
RR Tau&	        0&	1&	2&	3&	0&	21& 27 &	Yes&	No\\
SV Cep&	        0&	0&	1&	1&	0&	23& 25 &	No&	No\\
T Ori&	        0&	1&	2&	5&	0&	19& 27 &	Yes&	No\\
UX Ori&	        0&	5&	0&	3&	0&	15& 23 &	Yes&	No\\
UY Ori&	        0&	0&	1&	2&	2&	7& 12 &	No&	Yes\\
V1185 Tau&	1&	0&	0&	2&	0&	24& 27 &	No&	Yes\\
V1578 Cyg&	7&	0&	0&	2&	2&	14& 26 &	No&	Yes\\
V1685 Cyg&	1&	1&	4&	3&	0&	20& 29 &	Yes&	Yes\\
V1686 Cyg&	0&	0&	2&	2&	0&	16& 19 &	No&	No\\
V1787 Ori&	0&	0&	1&	1&	0&	21& 23 &	No&	No\\
V1818 Ori&	0&	0&	5&	1&	2&	16& 24 &	Yes&	No\\
V346 Ori&	0&	1&	0&	0&	1&	24& 26 &	Yes&	No\\
V351 Ori&	0&	2&	0&	0&	2&	22& 26 &	Yes&	No\\
V361 Cep$^\dagger$&	0&	0&	1&	3&	0&	21& 26 &	No&	No\\
V373 Cep&	3&	0&	0&	3&	0&	15& 22 &	No&	Yes\\
V374 Cep&	3&	0&	1&	0&	2&	22& 29 &	No&	Yes\\
V380 Ori&	0&	0&	0&	8&	0&	5& 14 &	No&	No\\
V586 Ori&	0&	1&	1&	6&	2&	16& 26 &	Yes&	No\\
V590 Mon&	0&	0&	2&	4&	0&	20& 26 &	No&	No\\
V599 Ori&	0&	1&	0&	2&	0&	15& 18 &	Yes&	No\\
V718 Sco&	0&	1&	0&	0&	1&	21& 23 &	Yes&	No\\
V791 Mon&	2&	0&	0&	6&	0&	5& 13 &	No&	Yes\\
VV Ser &	0&	2&	3&	3&	0&	17& 25 &	Yes&	No\\
WW Vul &	0&	2&	1&	3&	2&	20& 28 &	Yes&	No\\
XY Per &	0&	1&	1&	0&	5&	18& 26 &	Yes&	No\\
Z CMa  &	13&	0&      0&	5&	1&	7& 27 &	No&	Yes\\
\hline\\
\textbf{Total fraction$^{c}$:}& 0.36 & 0.23 & 0.45 & 0.83 & 0.22 & \nodata & \nodata & 
\textbf{0.26$^{+0.05}_{-0.05}$}&\textbf{0.37$^{+0.06}_{-0.05}$} \\
\enddata

\tablenotetext{a}{Total number of line profiles that show the specified morphology.}
\tablenotetext{b}{Total number of line profiles examined for each object.}
\tablenotetext{c}{Total fraction of objects that display the specified profile or absorption type.} 

\end{deluxetable*}
\capstarttrue     

\capstartfalse
\begin{deluxetable}{lccccc}
%\rotate
%\linespacing{1}
\tablecaption{Comparison to previous studies \label{tab:tab5}}
\tablewidth{0pt}
\tablehead{
\colhead{Study} & \colhead{$N_{tot}$} & \colhead{$N_{RA}$}&
\colhead{\% RA}&\colhead{$N_{BA}$}&\colhead{\% BA}\\
\colhead{(1)}&\colhead{(2)}&\colhead{(3)}&\colhead{(4)}&\colhead{(5)}&\colhead{(6)}}
\tabletypesize{\normalsize}
\startdata
Present work & 78 & 20 & 26$^{+5}_{-5}$ & 29 & 37$^{+6}_{-5}$\\
FM84         & 46 & 2  &  4$^{+4}_{-2}$ & 16 & 35$^{+7}_{-7}$\\
HP92         & 32 & 2  &  6$^{+6}_{-3}$ & 11 & 34$^{+9}_{-8}$\\
R96          & 18 & 1  &  6$^{+8}_{-3}$ &  2 & 11$^{+9}_{-5}$\\
V03          & 131&\nodata& \nodata     & 18 & 15$^{+4}_{-3}$\\
\enddata
\end{deluxetable}
\capstarttrue

The difference in red--absorption statistics between our study and the others listed in
\autoref{tab:tab5} is more pronounced. We find that 26\% of our objects show red--shifted absorption
in at least one line while the studies of FM84, HP92, and R96 show roughly a 4--6\% occurrence. The
discrepancy is easily explained for our study versus those of FM84 and R96: the occurrence of
red--shifted absorption at H$\alpha$ in our study is $\sim$6\%, comparable to that found by FM84 and
R96 both of which primarily examine H$\alpha$. FM84 also examine the Na I D lines but both their
sample and ours show a negligible occurrence of red--shifted absorption in those lines.

The discrepancy with HP92 is harder to explain. A large percentage (15\%) of our objects show
red--shifted absorption at He I $\lambda$5876. The rest of the red--shifted absorption detections
result mainly from H$\alpha$, H$\beta$, and Ca II IRT lines. HP92 also examine these lines but
they detect red--shifted absorption in only 2 out of 32 objects, a detection rate of $\sim$6\%.
This might be explained by the fact that HP92 only observe 6 objects at He I $\lambda$5876 and thus,
to obtain a detection rate at $\lambda$5876 similar to that of our sample, they would, on average,
detect red--shifted absorption in $\sim$1 object, which they do. The non--detections in the other
lines is more puzzling but may be accounted for by the lower resolution of their observations
(11--85 km s$^{-1}$). These medium--resolution observations may mask weak red--shifted absorption
signatures. We note that the red--shifted absorptions observed at H$\alpha$ and H$\beta$ in our
sample tend to appear as deeper absorptions superimposed on a photospheric absorption core. However,
when these lines are carefully compared to a rotationally broadened standard it is clear that the
red--shifted absorptions are non--photospheric and that the profile morphology is inverse P--Cygni.
It is not clear that HP92 compare their objects to photospheric standards. If not, this may also
contribute to the discrepancy. 

By comparison to the studies listed above, our analysis demonstrates the necessity of large sample,
high--spectral resolution studies of outflow and inflow diagnostics in HAEBES in order to gather
accurate statistics of mass flows around these objects. The H$\beta$ diagnostic shows the largest
percentage of P--Cygni profiles, although other lines (e.g. H$\alpha$, H$\gamma$, Fe I 4924) also
show non--negligible P--Cygni detection fractions. The He I $\lambda$5876 line is by far the most
sensitive line for detecting inverse P--Cygni morphologies. This is not surprising since the \hei
line is also a very efficient diagnostic for detecting mass flows \citep{EFHK,cauley14} and the
$\lambda$5876 transition lies immediately above the upper energy level of the $\lambda$10830
transition, preceding $\lambda$10830 in a recombination cascade sequence. The Na I D lines are also
relatively sensitive to outflows in HAEBES, although the lines more frequently display pure emission
profiles.  

\subsection{Comparison to \hei data}
\label{subsec:sec54}

The line profile statistics (\autoref{tab:tab6}) for both inflows and outflows from our
optical data set are similar to those from the \hei data set, although the fractions are
higher in the \hei sample.

We use contingency tests to compare the fractions of red and blue--shifted absorption between the
two studies. A contingency test of the measured fraction of a characteristic in two groups gives
the probability that the null hypothesis is true, i.e. that the intrinsic fraction is the same in
both groups \citep{feigelson12}. This probability is referred to here as the $p$--value. A high
$p$-value indicates a high probability of having obtained the difference in the measured fractions
by chance, i.e. the null hypothesis cannot be ruled out; a low $p$-value indicates a low probability
that the difference in the measured fractions was obtained accidentally, i.e. the difference is real
and the null hypothesis can probably be ruled out. A contingency test comparing the \hei sample to
the optical sample shows that neither the blue--shifted ($p=$0.85) nor the red--shifted
($p=$$\sim$1.0) absorption incidences are significantly
different. 

\capstartfalse
\begin{deluxetable}{rcc}
\tablecaption{Comparison to \hei HAEBE statistics \label{tab:tab6}}
\tablehead{\colhead{}&\colhead{Optical data}&\colhead{\hei data}\\
\colhead{} & \colhead{(1)} & \colhead{(2)}}
\tabletypesize{\normalsize}
\startdata
$N_{tot}$ & 78 & 48 \\
Red--shifted abs. & 26\%$^{+5}_{-5}$ & 27\%$^{+7}_{-6}$\\
Blue--shifted abs. & 37\%$^{+6}_{-5}$ & 40\%$^{+7}_{-7}$\\
\enddata
\end{deluxetable} 
\capstarttrue

Although the overall morphology statistics are similar between the optical and \hei data, the \hei
line is much more efficient at tracing mass flows around HAEBES than any \textit{single} optical
diagnostic. In addition, \hei is generally less contaminated by sources other than nearby
circumstellar material due to the high excitation temperatures needed to populate the lower energy
level \citep[$\sim$20,000 K;][]{dupree05}. This is not the case for other popular diagnostics such as
H$\alpha$, H$\beta$, and the Na I D lines which can frequently display strong emission signatures
from chromospheres, nebular background, and interstellar absorption in the case of Na I D. The
tendency of the \hei line to remain in its lower energy level until a photon is absorbed accounts
for the large opacity needed to generate the blue--shifted absorption commonly seen in the data from
\citetalias{cauley14}. 

\begin{figure}\label{fig:fig6}
  %\begin{center}
     \includegraphics[scale=.57,trim=25mm 0mm 10mm 120mm]{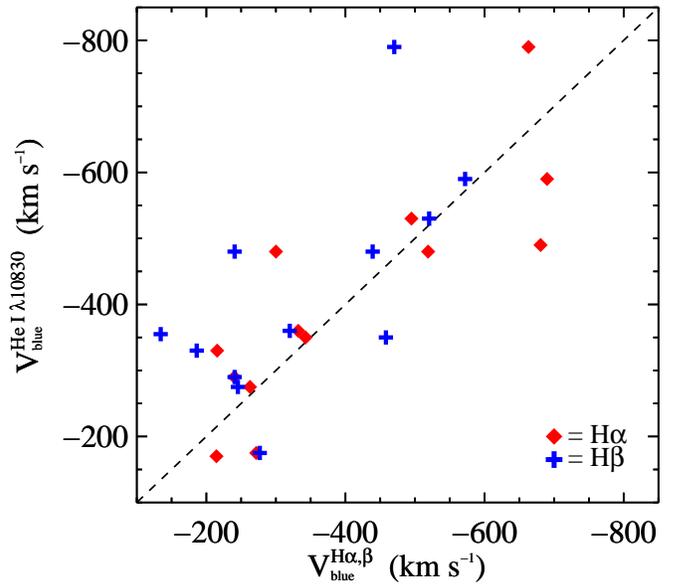}
     \caption{Maximum blue--shifted 
absorption velocities seen in \hei vs H$\alpha$ and H$\beta$ for objects with blue shifted
absorption at both \hei and either H$\alpha$ or H$\beta$. The \hei and the Balmer lines
generally trace the same velocity of outflowing material, although there is significant scatter. The
H$\beta$ line velocities match the \hei velocities less well than the H$\alpha$ velocities. This may
be an optical depth effect: the optical depth at H$\beta$ may simply be too low to trace the same
wind extent as \hei and H$\alpha$, resulting, in general, in smaller observed velocities at H$\beta$. 
Note that the data for the optical lines and the He I $\lambda$10830 lines are not simultaneous.}
  %\end{center}
\end{figure}

\autoref{fig:fig6} compares the maximum absorption velocities of \hei and H$\alpha$ or H$\beta$ for
objects that display blue--shifted absorption in both \hei and H$\alpha$ (red diamonds) and H$\beta$
(blue crosses). This figure shows that \hei traces approximately the same outflow velocities as
H$\alpha$ and H$\beta$, although the velocities in H$\beta$ are generally smaller when compared to
He I $\lambda$10830. This could be due to the larger optical depth of \hei and H$\alpha$ relative to
H$\beta$, allowing photons to be absorbed up to larger fractions of the terminal wind velocity. The
number of objects with both optical and \hei data that also show blue--shifted absorption is small
and thus it is not possible to make a strong case for choosing one of the optical diagnostics over
\hei based on a velocity comparison. Each line appears to trace similar material in the wind. Thus
for future studies concerned with detecting the presence of outflows and accretion around HAEBES,
\hei could serve as a primary diagnostic due to its high efficiency in detecting mass flows. In
addition, it is relatively free of photospheric and interstellar contaminants making the
circumstellar nature of the line morphologies unambiguous. In terms of mass flow statistics and
outflow velocities, it appears that optical data does not provide any additional information over
that provided by the \hei line.

\subsection{Comparison to CTTSs}
\label{sec:sec55}

To our knowledge, the study performed by \citet[][hereafter
AB00]{alencar00}\defcitealias{alencar00}{AB00} is the only current high--resolution, moderate sample
size investigation of CTTSs that examines many different spectral line diagnostics. We take the
statistics from \citetalias{alencar00} to be representative of CTTSs in general.
\citetalias{alencar00} examine the following accretion and outflow diagnostics: H$\alpha$, H$\beta$,
H$\gamma$, H$\delta$, He I $\lambda$5876, Na I D, Ca II K, Ca II $\lambda$8662, Ca II $\lambda$8498,
Fe I $\lambda$6192, $\lambda$5497, and Fe II $\lambda$4924 and $\lambda$4352. Our study is well
suited for comparison with \citetalias{alencar00} since we examine all of these except for Fe I
$\lambda$6192. The resolving power of their observations ($R\sim$48,000) is also similar to ours
increasing the chance that no weak features will be missed in one data set but not in the other. We
will also compare certain profile statistics from our sample to those of \citet{beristain01} and
\citet{batalha96}, studies which analyzed a smaller number of mass flow diagnostics in similar
sample sizes of CTTSs to that of \citetalias{alencar00}.

\citetalias{alencar00} find high fractions of both red--shifted absorption (12 objects,
40\%$^{+9}_{-8}$) and blue--shifted absorption (24 objects, 80\%$^{+6}_{-8}$) in at least one line
for the CTTSs in their sample. Our sample displays much lower incidences of both red and
blue--shifted absorption: 26\% and 37\%, respectively. Contingency tests comparing our study to the
of \citet{alencar00} are shown in \autoref{tab:tab7}. The red--shifted absorption comparison between
our final sample and that of \citetalias{alencar00} suggests a marginal difference ($p$=0.16) in the
incidence of red--shifted absorption. The blue--shifted absorption statistics are significantly
different for both sample comparisons, with $p<$0.001.

\capstartfalse
\begin{deluxetable}{llc}
%\linespacing{1}
\tablecaption{Optical contingency table results \label{tab:tab7}}
\tablewidth{0pt}
\tablehead{\colhead{Feature}&\colhead{Groups}&\colhead{$p$--value}}
\tabletypesize{\normalsize}
\startdata
Red absorption & HAEBES vs CTTSs & 0.160 \\
Blue absorption& HAEBES vs CTTSs & 9x10$^{-4}$ \\
\enddata
\end{deluxetable}
\capstarttrue

\begin{figure*}\label{fig:fig7}
  %\begin{center}
     \includegraphics[scale=.70,angle=180,trim=10mm 25mm 10mm 20mm]{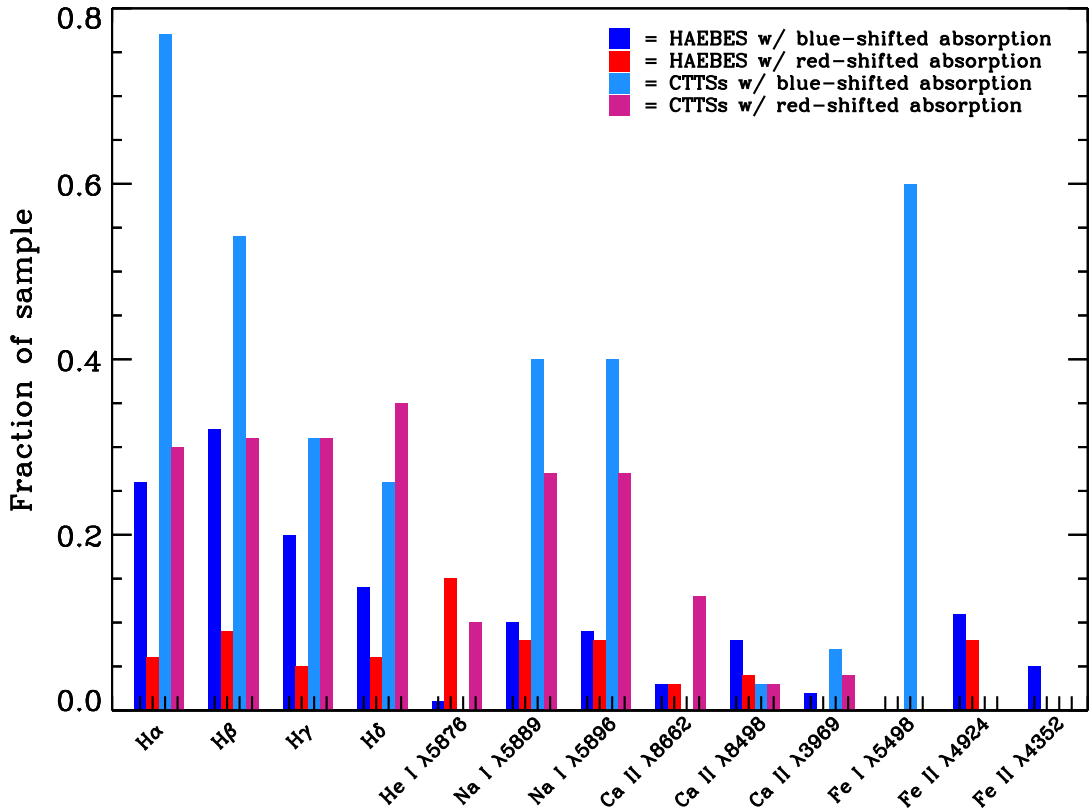}
     \caption{Line-by-line mass flow fraction comparison of CTTS and HAEBE samples. The fractions
are listed as percentages in \autoref{tab:tab8}. The CTTS data is
from \citetalias{alencar00}. The CTTSs, in general, show a higher incidence of both blue and
red--shifted absorption in most line diagnostics compared to HAEBES. The exceptions are at He I
$\lambda$5876 and the Fe II lines.}
  %d{figure}{center}
\end{figure*}

In addition to the total mass flow statistics, there are differences in the absorption
statistics in the lines that are common to both our study and that of \citetalias{alencar00}. This
comparison is presented in \autoref{tab:tab8} and is shown graphically in \autoref{fig:fig7}. The
most common line displaying blue--shifted absorption in their data is H$\alpha$ followed by
H$\beta$, H$\gamma$, H$\delta$ and the Na I D lines. A similar trend is observed in our data except
that H$\beta$ shows a slightly higher incidence of blue--shifted absorption than H$\alpha$.  The
most common red--shifted absorptions features appear at He I $\lambda$5876 and Na I D in their
sample. The highest red--shifted absorption occurrence in our data is also at He I $\lambda$5876.
However, we observe only 1 object with red--shifted absorption at Na I D. We find a small incidence
of inverse P--Cygni profiles in H$\alpha$ and H$\beta$ (6\% and 5\%, respectively) while
\citetalias{alencar00} observe a larger incidence of red--shifted absorption components at H$\alpha$
(23\%) and H$\beta$ (23\%).  \citetalias{alencar00} find Fe II $\lambda$4924 in emission in 40\% of
their sample. Our data show Fe II $\lambda$4924 emission in only 9\% of the sample. We also observe
both blue and red-shifted absorption at Fe II $\lambda$4352 in a non--negligible fraction (13\%) of
our sample.  \citetalias{alencar00} do not fit for the absorption components at Fe II $\lambda$4352,
likely due to strong nearby photospheric features. By comparing the CTTS $\lambda$4352 lines in
their sample to that of V410 Tau, a WTTS also from their sample, we searched for red and blue--shifted 
absorption in the line. We find no profiles that are not consistent with purely photospheric absorption. Thus we
estimate the mass flow incidence in this line to be $\sim$0 in the \citetalias{alencar00} CTTS
sample. 

The largest incidence differences between our sample and \citetalias{alencar00}, outside of the
Balmer lines, occur in the Na I D lines, He I $\lambda$5876, Fe I $\lambda$5498, and Fe II
$\lambda$4924. At He I $\lambda$5876 we find a higher occurrence of red--shfited absorption (15\%)
than \citetalias{alencar00} (10\%). Our observed fraction of He I $\lambda$5876 red--shifted
absorption is more similar to that found by \citet{beristain01} (16\%) for a sample of 31 CTTSs. We
find significantly lower fractions of red--shifted absorption in both Na I D lines (8\%) compared to
\citetalias{alencar00} (27\%). We find a red--shifted absorption fraction of 8\% at Fe II
$\lambda$4924.  \citetalias{alencar00} do not observe any red--shifted absorption at Fe II
$\lambda$4924.  Similar discrepancies are observed in the blue--shifted absorption statistics. At Fe
I $\lambda$5498, we find no evidence of blue--shifted absorption in \textit{any} objects while
\citetalias{alencar00} report a detected fraction of 60\%. Our detected blue--shifted absorption
fraction at Na I D ($\sim$10\%) is also much lower than \citetalias[]['s]{alencar00} (40\%).
Finally, the blue and red--shifted absorption fractions at H$\alpha$, H$\beta$, H$\gamma$, and
H$\delta$ are significantly higher in \citetalias{alencar00}'s sample of CTTSs than in our HAEBE
sample. 

Although we do not focus on the pure emission profiles, we note that our detected fraction of
emission at He I $\lambda$5876 is low (14\%) compared to that of \citet{beristain01} who detect
emission in 100\% of their 31 CTTSs. In \S 6 we will comment on possible causes for this differing
line profile statistic. The possible physical causes of the absorption statistics detailed above
will be outlined in the next section.

\capstartfalse
\begin{deluxetable*}{llccccccccccccc}
%\linespacing{1}
\tablecaption{Comparison to line absorption statistics from \citet{alencar00} \label{tab:tab8}}
\tablewidth{0pt}
\tablehead{ \colhead{} & \colhead{} & \colhead{} & \colhead{} & \colhead{}
& \colhead{}
& \colhead{He I} & \colhead{Na I} & \colhead{Na I}
& \colhead{Ca II} & \colhead{Ca II} & \colhead{Ca II}
& \colhead{Fe I} & \colhead{Fe II} & \colhead{Fe II}\\
\colhead{} & \colhead{} & \colhead{H$\alpha$} & \colhead{H$\beta$} & \colhead{H$\gamma$} &
\colhead{H$\delta$}
& \colhead{$\lambda$5876} & \colhead{$\lambda$5889} & \colhead{$\lambda$5896}
& \colhead{$\lambda$8662} & \colhead{$\lambda$8498} & \colhead{$\lambda$3969}
& \colhead{$\lambda$5498} & \colhead{$\lambda$4924} & \colhead{$\lambda$4352}\\
\colhead{} & \colhead{} & \colhead{(1)} & \colhead{(2)} & \colhead{(3)} & \colhead{(4)} &
\colhead{(5)} & \colhead{(6)} &
\colhead{(7)} & \colhead{(8)} & \colhead{(9)} & \colhead{(10)} & \colhead{(11)} & \colhead{(12)} &
\colhead{(13)}}
\tabletypesize{\scriptsize}
\startdata
\citetalias{alencar00} & $N_{RA}$ & 9 & 8 & 8 & 8 & 3 & 8 & 8 & 4 & 1 & 1 & 0 & 0 & 0 \\
                       & \% RA    & 30& 31& 31& 35& 10& 27& 27& 13& 3 & 4 & 0 & 0 & 0 \\
                       & $N_{BA}$ & 23& 14& 8 & 6 & 0 & 12& 12& 0 & 1 & 2 & 18& 0 & 0 \\
                       & \% BA    & 77& 54& 31& 26& 0 & 40& 40& 0 & 3 & 7 & 60& 0 & 0 \\
                       &          &   &   &   &   &   &   &   &   &   &   &   &   &   \\
This study             & $N_{RA}$ & 5 & 7 & 3 & 3 & 12& 6 & 6 & 2 & 1 & 0 & 0 & 5 & 0 \\
                       & \% RA    & 6 & 9 & 5 & 6 & 15& 8 & 8 & 3 & 4 & 0 & 0 & 8 & 0 \\
                       & $N_{BA}$ & 21& 24& 12& 7 & 1 & 8 & 7 & 2 & 2 & 1 & 0 & 7 & 3 \\
                       & \% BA    & 24& 32& 20& 14& 1 & 10 & 9 & 3 & 8 & 2 & 0 & 11 & 5 \\
\enddata
\end{deluxetable*}
\capstarttrue

\section{Discussion} 

The optical line statistics presented in \S 5 indicate that there are differences in the
characteristics of, and/or the physical mechanisms producing, the mass flows in CTTSs vs HAEBES.
Below we discuss possible scenarios capable of producing the trends and differences seen in the
measured statistics. 

\subsection{Fe II: Higher ionization states tracing mass flows in HAEBES} 

\autoref{tab:tab8} shows clear evidence that HAEBES, in general, show significantly more mass flow
activity in the Fe II lines than CTTS indicating an absorbing region that is hotter or is subject to
higher amounts of ionizing radition. Fe I $\lambda$5498 shows a 60\% incidence of blue--shifted
absorption in \citetalias{alencar00}'s sample but is not detected in absorption in any of our
HAEBES. The reverse is true for the Fe II $\lambda$4924 and $\lambda$4352 lines: we see non--zero
fractions of both lines exhibiting blue--shifted absorption while \citetalias{alencar00} do not
detect absorption in either of these lines in any objects. All of the objects in which we detect
blue--shifted absorption in Fe II are B--type or early A--type objects. This behavior was also noted
by \citet{hp92} concerning the relative strengths of Fe I and Fe II emission lines in HAEBES versus
CTTSs. The first ionization potential of Fe is 7.9 eV which corresponds to a wavelength of 1570 \AA.
The ionizing radiation flux for a 10,000 K blackbody, a similar temperature to $T_{eff}$ for an A0
star, is $\sim$10$^{6}$ times higher at 1570 \AA\hspace{0pt} than for a 3,800 K blackbody, a typical
$T_{eff}$ for a CTTS of spectral type M0. In addition, CTTSs display large UV excess fluxes due to
accretion which can be 10$^4$--10$^5$ times higher than the stellar photosphere at typical
temperatures used to model the excess continuum. Thus the actual flux ratio at 1570 \AA\hspace{0pt}
is likely much lower than 10$^6$ but still favors HAEBES over CTTSs. The larger number of ionizing
photons emitted by B and early A stars compared to CTTSs could account for the observed disparities
in the Fe II line profile statistics observed here. 

This trend also holds for another low ionization potential metal, Na I, which is ionized by photons
with energies higher than  5.1 eV, or $\lambda$$\lesssim$2431 \AA. The overall mass flow incidence
in our sample is much lower at Na I D than in \citetalias{alencar00}'s sample. However, we do
observe Na I outflow signatures in a majority of our objects that also display blue--shifted
absorption at Fe II, indicating that significant amounts of neutral material are simultaneously
present in these outflows. Although it is unclear to what extent the temperature of the outflow
determines the ionization state of the wind compared to the ionizing flux from the star, Na I is
likely not present in the regions of the outflow being sampled by Fe II. The observed statistics for
Ca II are not significant enough to warrant a comparison to those from \citetalias{alencar00}.  It
appears that the differences in Fe I and Fe II and Na I morphology statistics between HAEBES and
CTTSs suggest more highly ionized environments around HAEBES, likely due to the higher number of
ionizing photons emitted by HAEBES.

\subsection{Infall velocities}  

\begin{figure}\label{fig:fig8}
  %\begin{center}
     \includegraphics[scale=.55,trim=25mm 0mm 10mm 135mm]{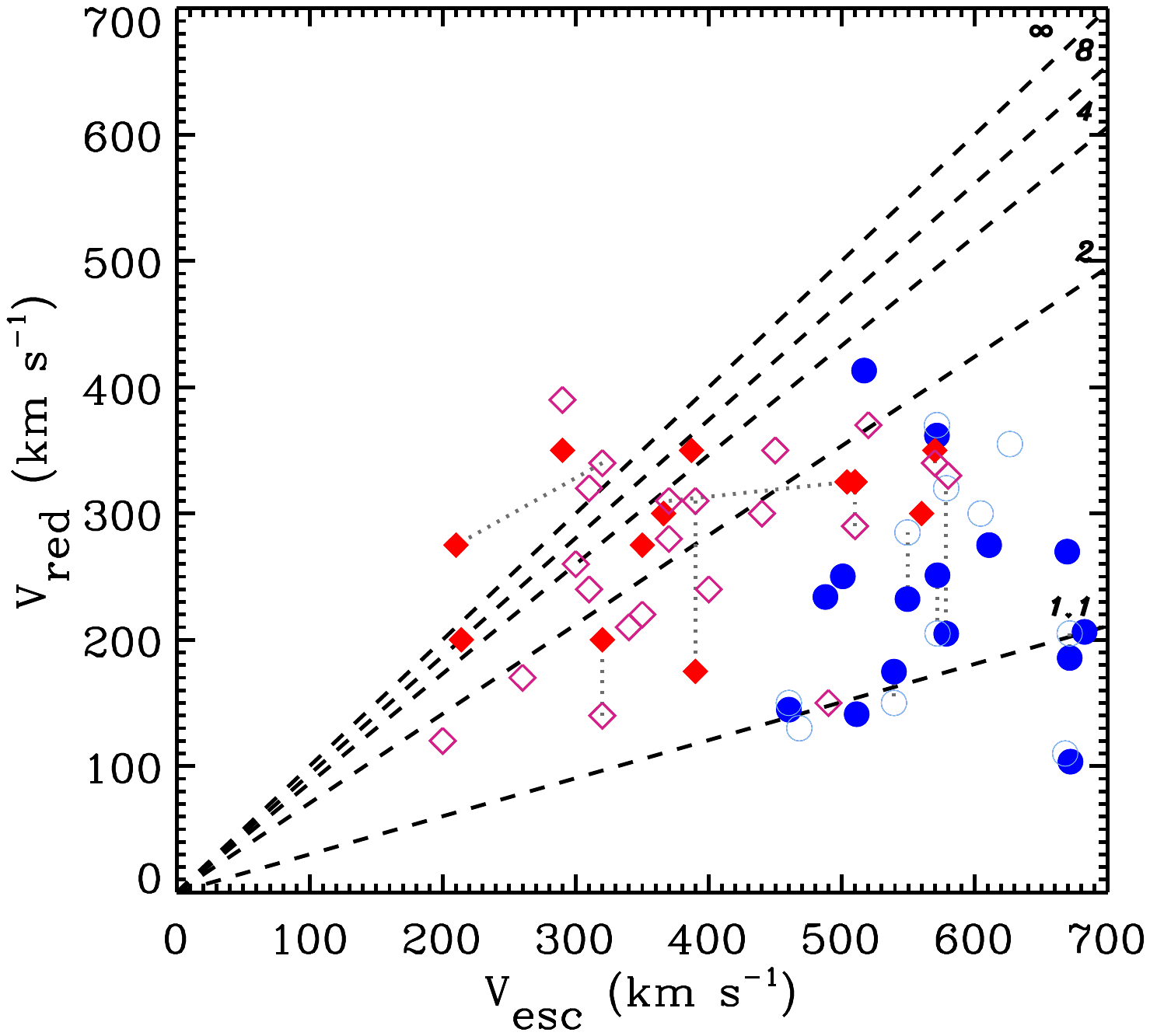}
     \caption{Observed maximum red--shifted absorption velocities versus stellar escape velocity for
the HAEBES (filled blue circles) from this study and CTTSs from \citet{edwards94} (filled red
diamonds). The unfilled plot symbols are the HAEBE (light blue) and CTTS (light red) objects from
the \hei samples of \citetalias{cauley14} and \citet{fischer08}, respectively. Objects common to
both the optical and \hei samples are connected by a dotted line. The dashed lines show, for a given
$V_{esc}$, the maximum velocity obtained if infall begins from the distance labeled next to the
line (in stellar radii). The HAEBES, on average, show
maximum red--shifted absorption velocities that are smaller fractions of their stellar escape
velocities, indicating that the infalling material originates closer to the stellar surface.}
\end{figure}

In \citetalias{cauley14} we showed that, on average, the maximum red--shifted absorption velocities
in HAEBES are smaller fractions of the stellar escape velocity than in CTTSs (see Figures 5 and 6
from \citetalias{cauley14}). The comparison in \citetalias{cauley14} was performed using only the
line velocities measured from \hei data. Here we present the same comparison in \autoref{fig:fig8}
but incorporating all of the optical diagnostics displaying red--shifted absorption for each object.
For HAEBES that show red--shifted absorption in more than one diagnostic, their average value is
used. We note that, as in \citetalias{cauley14}, \autoref{fig:fig8} is identical in format to Figure
5 from \citet{fischer08} in order to facilitate easy comparison with the red--shifted \hei
absorption velocities presented in that study. The diagonal lines mark the velocity ratios that
correspond to the infall distance from the stellar surface, given in stellar radii, marked above
each line. In other words, the dotted line shows the maximum free--fall velocity that can be
obtained if infall begins at the specified distance from the star. We estimate the red--shifted
absorption velocities of 13 CTTSs from the published optical profiles of \citet{edwards94}.
\citet{edwards94} observed Balmer line profiles as well as Na I D and He $\lambda$5876. Most of the
red--shifted absorption profiles in our sample are seen in the same lines so the comparison should
not be subject to lines tracing mass flows with significantly different physical conditions. 

\begin{figure}\label{fig:fig9}
     \includegraphics[scale=.60,trim=70mm 20mm 20mm 20mm]{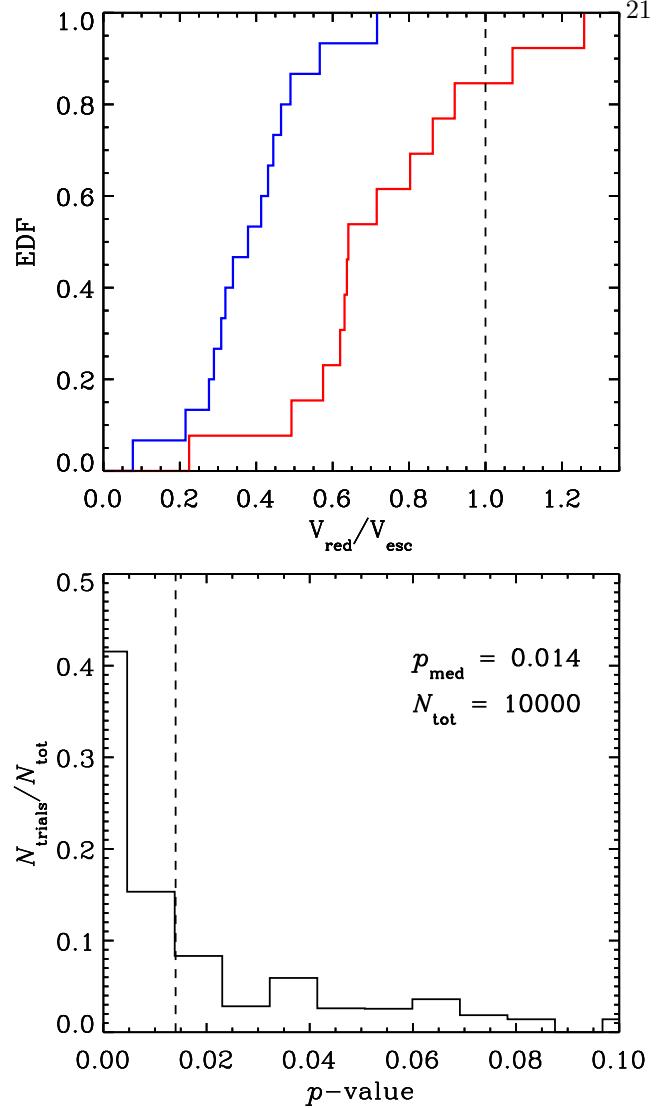}
     \caption{Histogram showing the distribution of KS test
$p$--values (bin width of 0.0005) for ten thousand trials (bottom panel) and EDFs for the HAEBE and
CTTS samples (top panel). The median value of $p_{KS}$ is marked by the vertical dashed line in the
bottom panel. In the top panel, the HAEBE EDF is shown in blue; the CTTS EDF is shown in red. A 
majority of the tests (74\%) return $p$$<$0.05 indicating a significant difference between the 
two EDFs.}
\end{figure}

In \autoref{fig:fig8} the CTTSs are plotted as solid red diamonds; the HAEBES are represented by the
filled blue circles. For comparison we have overplotted the HAEBES \hei velocities from
\citetalias{cauley14} as light blue unfilled circles and the \hei CTTS velocities from
\citet{fischer08} as unfilled light red diamonds. Objects that are common to both the optical and
\hei samples have their velocities, as measured at \hei and in the optical diagnostics, connected by
a dotted line\footnote{Two CTTSs are plotted with two different estimates of their escape velocities
due to different $M_*$ and $R_*$ estimates from the studies that the spectral line velocities are
taken from.}. The mean velocity ratio ($V_{red}/V_{esc}$) for the HAEBE sample is 0.40; for the
CTTSs it is 0.78. A two--sided KS--test performed on the sample returns a $p$--value of
2.2x10$^{-5}$ indicating a significant difference between the two populations. 

In order to test the effect of uncertainties in determining $V_{esc}$ on the KS statistic, we generated ten
thousand random velocity distributions and performed a KS test on each one. We follow the same
procedure for assigning uncertainties to the stellar mass and radius (50\% and 25\%, respectively)
as performed in \citetalias{cauley14}. In addition, we let the measured red--shifted absorption velocity vary
by 50\%. This provides some accounting of accretion variability. The result of the KS--tests
is shown in the bottom panel of \autoref{fig:fig9} where the median $p$--value is seen to be
$\sim$0.014. The emipirical distribution function (EDF) for each sample is shown in the upper panel.
For each of the trials each object is assigned a random value within the limits of the uncertainty
in the object's escape velocity and red--shifted absoprtion velocity.  The low $p$--values suggest a
real difference in the distributions which in turn indicates that accretion flows in HAEBES begin
deeper in the system's gravitational potential, i.e. closer to the star. This supports the similar
result found in \citetalias{cauley14}, although the $p$-values are, on average, higher in the
optical case due to the red-shifted absorption velocity being allowed to vary in the simulation.

\autoref{fig:fig8} and \autoref{fig:fig9} provide additional support for our suggestion in
\citetalias{cauley14} that HAEBES have smaller magnetospheres than CTTSs. The evidence in the
optical is perhaps even more convincing due to the larger number of red--shifted absorption lines
used in the comparison, although there is a large spread in the measured line absorption velocities
for some individual objects. However, the lack of large $p$--values in our KS test simulation
indicates that the velocity ratio difference between HAEBES and CTTSs is independent of the
particular line velocity chosen for the comparison. The statistical tests also support the idea that
the red--shifted absorption in HAEBES originates deeper within the gravitational potential of the
central star. If the relative distances are similar at which the magnetospheres in HAEBES and CTTSs
are dominating the kinematics of the accretion flow, the velocity ratios of the HAEBE and CTTS
samples should be comparable. Since they are not, there must be some difference in the formation
mechanisms of the absorption signatures. We attribute this difference to smaller magnetospheres in
HAEBES compared to CTTSs. We note that smaller corotation radii, and thus smaller magnetospheres,
are suggested for most HAEBES compared to CTTSs due to their large $v$sin$i$ values \citep[see \S 2
of \citetalias{cauley14};][]{muzz04}. Thus, this result is expected from magnetospheric accretion
physics.            

\subsection{HAe stars vs HBe stars}

\begin{figure}\label{fig:fig10}
  %\begin{center}
     \includegraphics[scale=.58,angle=180,trim=35mm 45mm 55mm 30mm]{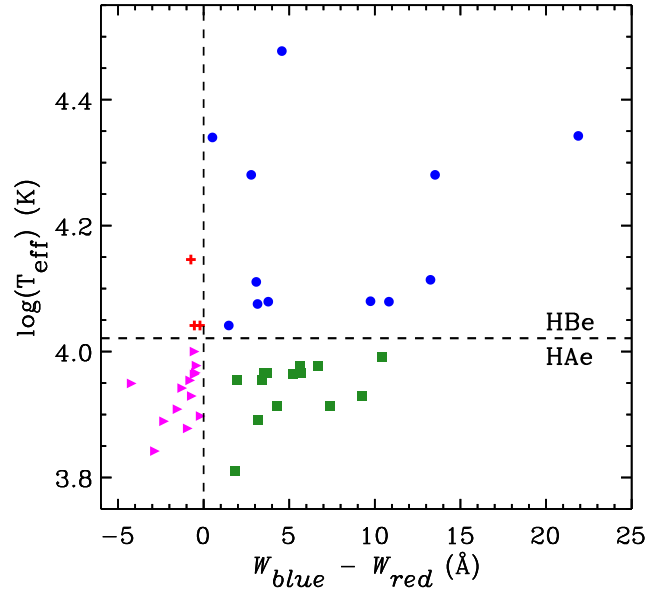}
     \caption{The $T_{eff}$ vs blue equivalent width minus
red equivalent width for all of the objects in our sample that have absorption profiles. Objects
with more than one line in a quadrant are shown using the average of the equivalent width
differences for those lines. Colors are used for clarity in distinguishing quadrants. The
trend found for the \hei data from \citetalias{cauley14} holds for the optical sample: there is a 
lack of red--shifted absorption profiles in HBe stars. The strongest blue--shifted absorption 
profiles are found in the HBe group.}
  %d{figure}{center}
\end{figure}

In \citetalias{cauley14} we demonstrated the lack of red--shifted absorption at \hei in HBe stars
(Figure 7 of \citetalias{cauley14}). This same trend holds for our larger optical sample. This is
shown in \autoref{fig:fig10}, which plots the difference between the blue ($v<0$) and red ($v>0$)
line equivalent widths versus $T_{eff}$, for the objects with at least one profile displaying evidence of blue
or red--shifted absorption. If an object shows more than one profile in absorption, we average the
equivalent width difference. The equivalent width difference is essentially a measure of profile
type: red--shifted absorption results in a negative equivalent width difference while blue--shifted
absorption results in a positive equivalent width difference. Objects in each quadrant are
color--coded for clarity.  There is an absence of objects in the upper-left quadrant. Although V1818
Ori shows evidence of red--shifted absorption in the Na I D lines, we do not calculate an equivalent
width for either line due to the strong interstellar absorption features superimposed on the line. Thus
there are only 3 HBe objects with red--shifted absorption shown in \autoref{fig:fig10} and they all lie very
close to the vertical dashed line. The IPC and
red absorption He I $\lambda$5876 profiles generally have small equivalent width differences (0.0
\AA$>$$W_b$-$W_r$$>$-2.0 \AA) and comprise most of the red--shifted absorption objects (red crosses and
magenta right--facing triangles). 

This result is also born out in the overall profile statistics: 15 of the 19 red--shifted absorption
objects are HAe stars; 4 are HBe stars, with 2 of those having spectral types of B9. There are 45
HAe stars in the final sample and 33 HBe stars. A contingency test comparing HAe and HBe
red--shifted absorption returns $p$=0.04 providing support for a real difference in the morphology
statistics. A test comparing the blue--shifted absorption reveals no difference between HAe and HBe
objects ($p$=0.48). A similar trend is seen in the \hei data but the larger optical sample has added
significance to the red--shifted absorption comparison.

The similarity in the optical and \hei HAe versus HBe comparisons suggests that the conclusions
reached in \citetalias{cauley14} are not altered by the analysis presented here. HAe stars exhibit
red--shifted absorption more frequently than HBe stars indicating that mass infall, most likely
magnetically channeled, is more common in HAe stars. Outflows seem to be present in similar
fractions of HAe and HBe objects suggesting that the wind mechanism is equally as efficient in both
groups of objects. 

\subsection{Simultaneous blue and red--shifted absorption and accretion driven outflows}

In \citetalias{cauley14} we noted the lack of simultaneous blue and red--shifted absorption in
the \hei profiles. To clarify, simultaneous absorption signatures in the current, optical context refer to detected
absorption in \textit{any} diagnostics from the same exposure, or exposures taken very close in time
to one another. This includes simultaneous red and blue--shifted detections in the same line, such
as that seen in \hei for the CTTSs from \citet{EFHK}, or detections in two different lines as seen in \citet{alencar00}. 
This lack of simultaneous absorption contrasted sharply with the large detected fraction of simultaneous 
absorption signatures in CTTSs presented by \citet{EFHK}. 

The optical data continues this trend: we observe only 1 object (HD
163296) that shows clear red and blue--shifted absorptions signatures in at least one line from the
same exposure. Based on the published centroid velocities of the detected absorption features from
\citetalias{alencar00}, we count 17 out of 30 CTTSs, or 57\%, showing red and blue--shifted
absorption in at least one line from nearly simultaneous exposures. The lack of simultaneous
absorption features observed in our optical sample supports the arguments presented in
\citetalias{cauley14}, namely that accretion driven outflows in HAEBES may be more difficult to
produce. The strong optical outflow signatures observed in both HAe and HBe objects that
do not also exhibit red--shifted absorption, combined with the conclusion from \citetalias{cauley14}
that the \hei PC morphologies are likely accretion--driven, suggests that accretion in some HAEBES
may proceed through a non--magnetic mechanism such as a boundary layer.  This would explain the
relative lack of simultaneous red and blue--shifted absorption signatures in HAEBES compared to
CTTSs. 

It is still unclear whether or not the accretion flows detected in the red--shifted absorption
objects can drive outflows. Outflow signatures in these objects are not observed (except for HD
163296) suggesting that if the accretion is related to driving outflows, the outflow signatures are
too weak to be detected or are launched in a geometry that requires a specific viewing angle to
observe (e.g. a highly collimated jet emerging from the polar region). This seems unlikely since
these simultaneous signatures are almost entirely absent from our sample, suggesting that they are
present in very few objects. Indeed, jets have been detected in a relatively fewer number of HAEBES
\citep[e.g.][]{grady04,ellerbroek14} compared to CTTSs hinting that they do not form as
efficiently in HAEBES.  

One potential explanation is that, in general, the accreting material as traced by red--shifted absorption is
not energetic enough to provide the energy necessary to drive an outflow. Support for this idea is
seen in \autoref{fig:fig8} from this study and Figure 5 from \citetalias{cauley14} where the
red--shifted absorption velocities of HAEBES are found to be smaller fractions of their stellar
escape velocities. We also find support for this scenario in Figure 7 from \citetalias{cauley14}
which demonstrates the relative lack of objects with low accretion rates that also exhibit
blue--shifted absorption profiles. We note that this distinction is not clear in our optical data,
i.e. the separation of profile type with accretion rate is not distinct. We have fewer estimates of
$\dot{M}$ in hand, however, for the red--shifted absorption objects from the optical sample and thus
the comparison is less significant. This could be improved and further tested by obtaining
simultaneous line profile morphologies and accretion rate estimates for a large sample of HAEBES. 

The idea of accretion powered stellar winds was first proposed for CTTSs by \citet{matt05}. They
suggest that a portion of the deposited accretion energy will heat the corona (via conduction or
magnetosonic wave dissipation) and assist in launching stellar winds.  While it is unclear whether
or not this process is occurring in CTTSs, magnetically channeled accretion, involving smaller
magnetospheres in HAEBES compared to CTTSs, will provide a smaller fraction of the
necessary energy for launching an outflow from the stellar surface of HAEBES than from the surface
of CTTSs. Since the outflows observed in CTTSs are generally accepted to be driven largely by
accretion \citep[e.g.][]{hartigan95,EFHK}, the amount of energy deposited by the accretion flow onto
the stellar surface could strongly influence the strength of the outflow. For HAEBES, if we imagine
that all of the kinetic energy of the accretion flow is converted into kinetic energy in a wind
driven from very near the stellar surface (i.e. a stellar wind), these outflows will be incapable of escaping the
gravitational potential of the star due to the small ($\sim$0.5; \autoref{fig:fig8}) fractions of
the stellar escape velocity that the accretion flows attain during their infall.  Obviously this
is an ideal scenario, but the energy budget argument holds if a large portion of the
energy in the wind is provided by the accretion flow, which it seems to be in CTTSs.  To summarize,
the lower infall velocities relative to the stellar escape velocities in HAEBES may indicate that
these accretion flows cannot efficiently drive mass outflows from the stellar surface.

If magnetically controlled mass infall in HAEBES is incapable of driving outflows, another form of
accretion may be partly responsible for driving the outflows from objects with large $\dot{M}$.  We
suggested in \citetalias{cauley14} that boundary layer accretion in HAEBES could provide the
necessary energy, although it is unclear exactly how the energy is tranferred to the outflow.  One
possible scenario discussed in \citetalias{cauley14} for launching outflows from the boundary layer
is the \citet{shu88} X--celerator mechanism. The high $v$sin$i$ values of HAEBES are ideal for
producing this type of outflow. Better $\dot{M}$ estimates for a larger number of HAEBES will help
elucidate the relationship between profile type and the role of the accreting material in producing
the outflow signatures.

\subsection{He I $\lambda$5876 emission lines}

\citet{beristain01} examine the He I $\lambda$5876 in a sample of 31 CTTSs. They find an emission
component (broad, narrow, or both) in all of the objects in their sample. They interpret the
broad component emission as arising in a hot wind as well as in the accretion flow
onto the star. The narrow emission component is interpreted as forming in the post--shock gas at the
base of an accretion column at the surface of the star. Our sample contains 13 HAEBES that display
emission profiles at He I $\lambda$5876. None of these objects show the narrow component that is
commonly seen in CTTSs, as well as WTTSs. This may be due to the much higher average $v$sin$i$
values of HAEBES compared to CTTSs. However, 11 of the 13 objects show broad emission profiles
similar in width to those observed by \citet{beristain01} and 8 objects show unambiguous
blue--shifted absorption in a separate but simultaneous diagnostic. Of these 8 objects, five show
unambiguous evidence of mass outflow at He I $\lambda$10830, although these observations are not
simultaneous to the He I $\lambda$5876 observations. Due to the prevalence of blue--shifted
absorption observed in the He I $\lambda$5876 emission objects, our data supports the suggestion by
\citet{beristain01} that the broad He I $\lambda$5876 emission component may partly arise in a hot
wind near the star.         

\section{Summary and Conclusions}

We have presented a large sample census of mass flows around Herbig Ae/Be stars using a wide variety
of mass flow diagnostics and determined the fraction of objects that display unambiguous (i.e. red
and blue--shifted absorption) mass flow signatures. We find that the Na I D interstellar absorption line centers 
are good approximations, to within 8.1$^{+3.6}_{-3.6}$ km s$^{-1}$, to the center-of-mass radial velocity in most 
HAEBE systems, confirming the conclusion reached by \citet{finkjank84}. Of the 78 HAEBES included in
the final analysis, 26\%$^{+5}_{-5}$ show direct signs of mass infall and
37\%$^{+6}_{-5}$ show direct evidence of mass outflow. These percentages are very similar to those
found in \citetalias{cauley14} using the He I $\lambda$10830 line.   

Overall, the optical data reinforce the conclusions outlined in \citetalias{cauley14}. More specifically,
we find lower rates of both red and blue--shifted absorption for HAEBES than is observed for
CTTSs, suggesting that, in general, the innermost environments of HAEBES 
are not simply scaled analogues of CTTSs. Similar to the \hei data, the HAEBE optical lines show
maximum red--shifted absorption velocities that are smaller fractions of the stellar escape
velocities than is found for CTTSs. This adds support to the argument presented in
\citetalias{cauley14} concerning the smaller size of HAEBE magnetospheres compared to those of
CTTSs. This has implications for the ability of HAEBE accretion flows to power strong winds: less energetic 
funnel flows provide less of the required energy for launching
outflows. This is supported by the lack of simultaneous blue and red--shifted absorption signatures
seen in both the optical sample and the \hei sample from \citetalias{cauley14}. This contrasts with
CTTSs for which the accretion process is believed to be the main energy source for outflows.

We also find a significant difference between the occurrence of red--shifted absorption in HAe stars
compared to HBe stars; both groups appear to exhibit outflow signatures at a similar rate. This
suggests that HBe stars accrete material from their disks differently than HAe stars, perhaps
through a boundary layer instead of along magnetic field lines. The prominence of blue--shifted absorption in
accreting objects in both subgroups suggests that the outflows in HAEBES are accretion driven.
However, the lack of simultaneous red and blue--shifted absorption in all but one HAEBE (HD 163296)
indicates that whatever geometry is producing the absorption signature is not capable of producing
the opposite absorption signature. In other words, boundary layer accretion, which should not
produce red--shifted absorption, may be capable of driving strong outflows which results in observed
blue--shifted absorption; magnetically controlled accretion from a small magnetosphere produces
red--shifted absorption but is perhaps not energetic enough to drive outflows from the stellar
surface, resulting in the lack of simultaneous red--shifted absorption in these profiles. Due to the
observed statistics, this suggests a general transition from magnetically controlled accretion in
HAe stars to boundary layer accretion in HBe stars.  

\acknowledgements

We thank the McDonald Observatory staff in West Texas for their hospitality during the observing
runs for this research. P. W. C. especially thanks David Doss for his help, without which the collection
of some of this data would not have been possible. P. W. C. would also like to acknowledge Pat Hartigan
for his comments and discussion throughout this project. We thank the referee for their helpful comments
which improved the quality of this manuscript. This research has made use of the SIMBAD
database and the NASA Astrophysics Data System. We acknowledge partial support for this research from 
the NSF through grant 1212122, from NASA Astrophysics Data Analysis Program through grant NNX13AF09G,
and from Space Telescope Science Institute through grant HST-60-12996.001, all made to Rice University.

\appendix

\section{Optical line profiles}

Below are the extracted profiles for a piece of the sample. The profiles for the entire sample will be
available with the published version. The format of each figure is the same as
\autoref{fig:fig2} and \autoref{fig:fig3}.

\begin{figure*}[hbt]\label{fig:fig12}
  %\begin{center}
     \includegraphics[scale=.93,trim= 20mm 35mm 5mm 5mm]{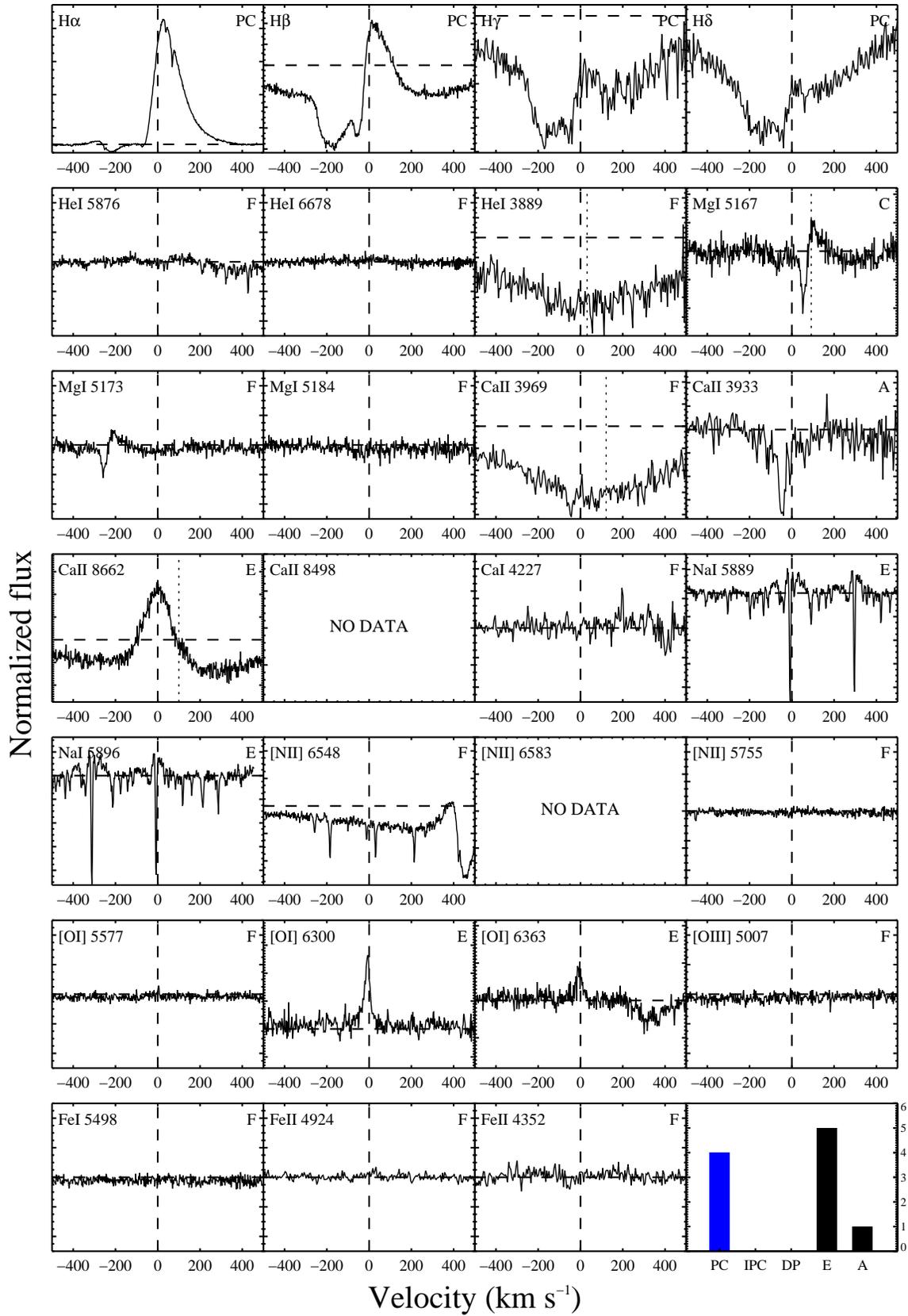}
     \caption{Extracted line profiles for AB Aur.}
  %\end{center}
\end{figure*}

\begin{figure*}[hbt]\label{fig:fig13}
  %\begin{center}
     \includegraphics[scale=.93,trim= 20mm 35mm 5mm 5mm]{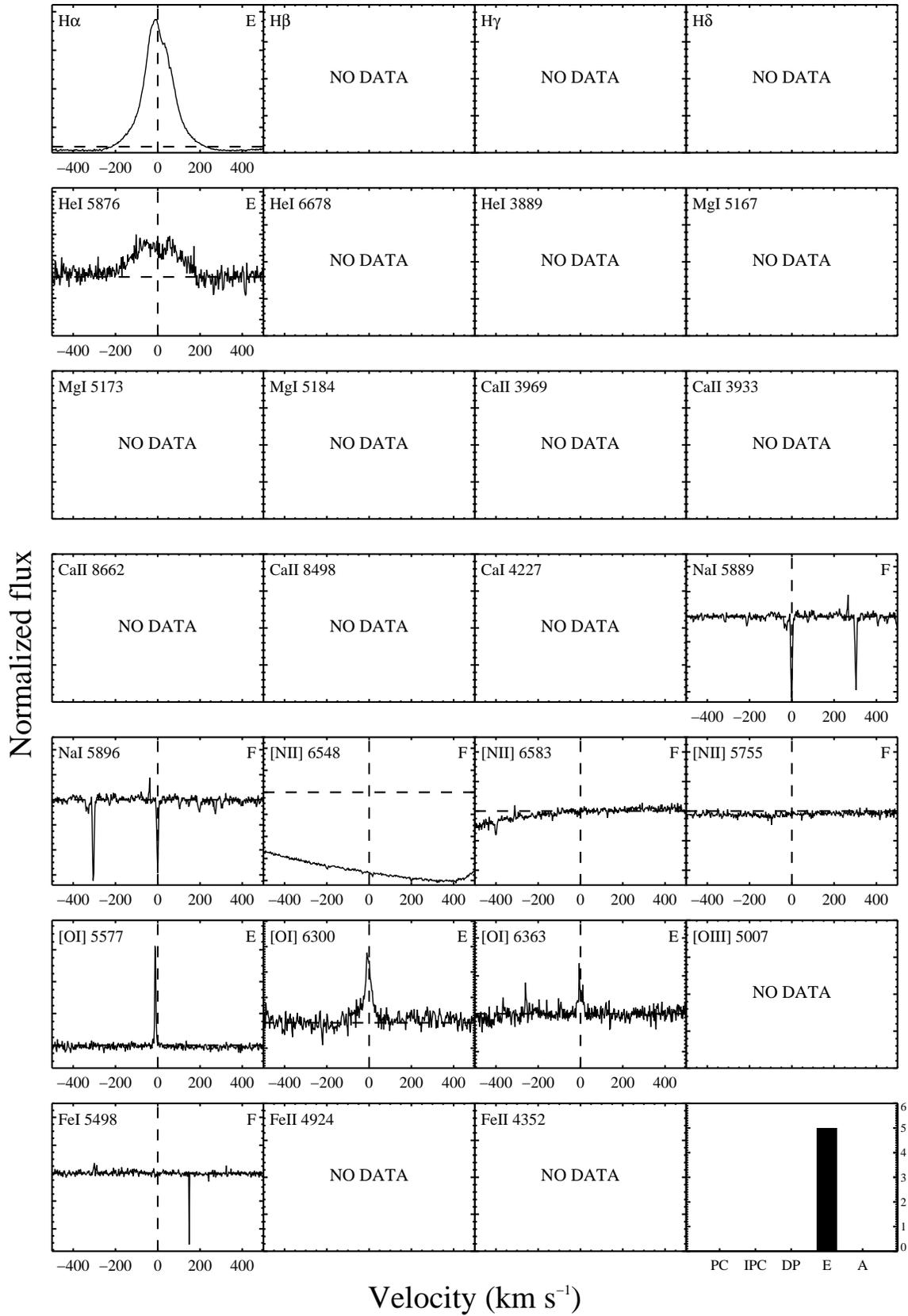}
     \caption{Extracted line profiles for AE Lep.}
  %\end{center}
\end{figure*}

\begin{figure*}\label{fig:fig14}
  %\begin{center}
     \includegraphics[scale=.93,trim= 20mm 35mm 5mm 5mm]{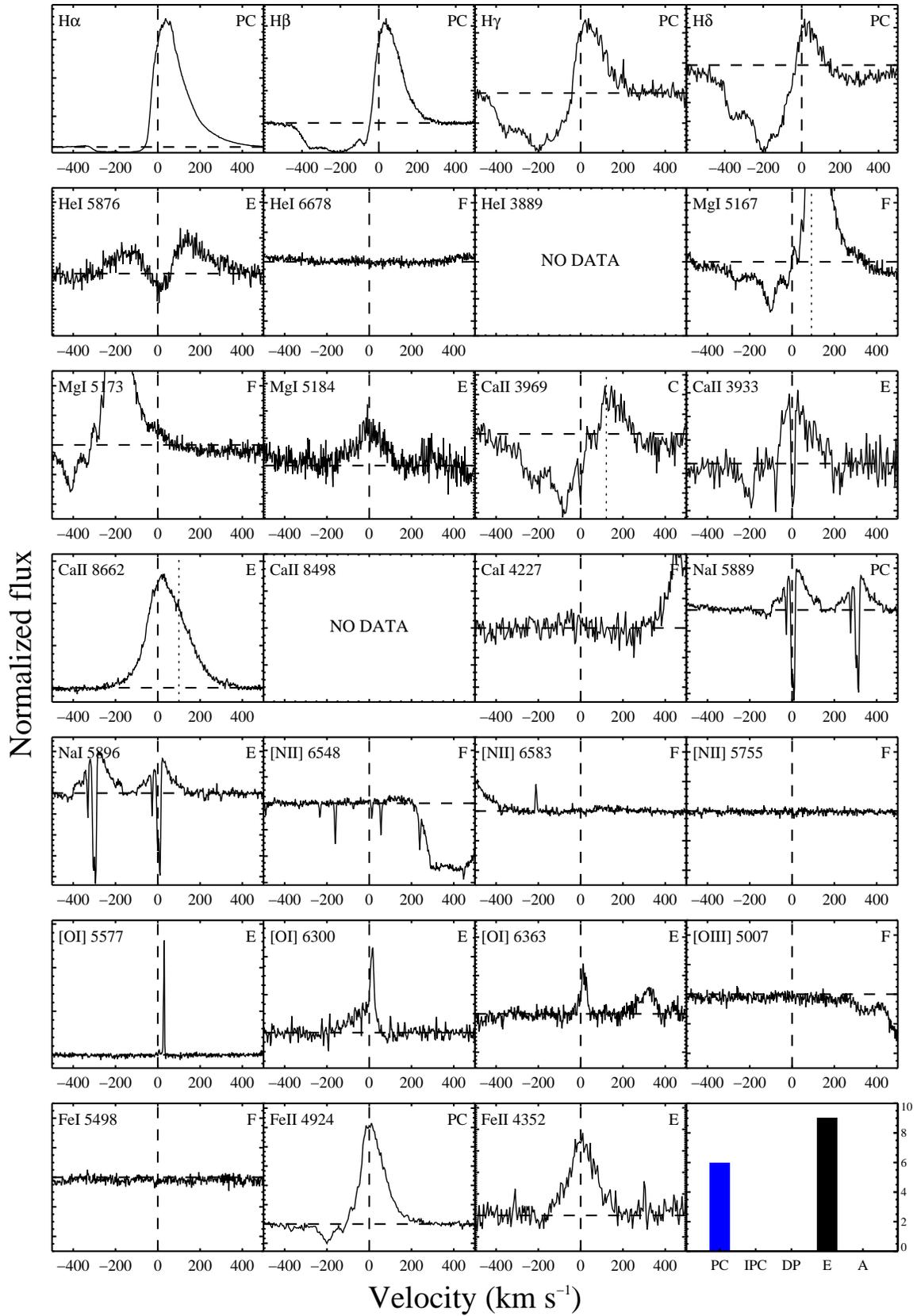}
     \caption{Extracted line profiles for BD+61 154.}
  %\end{center}
\end{figure*}

\begin{figure*}\label{fig:fig15}
  %\begin{center}
     \includegraphics[scale=.93,trim= 20mm 35mm 5mm 5mm]{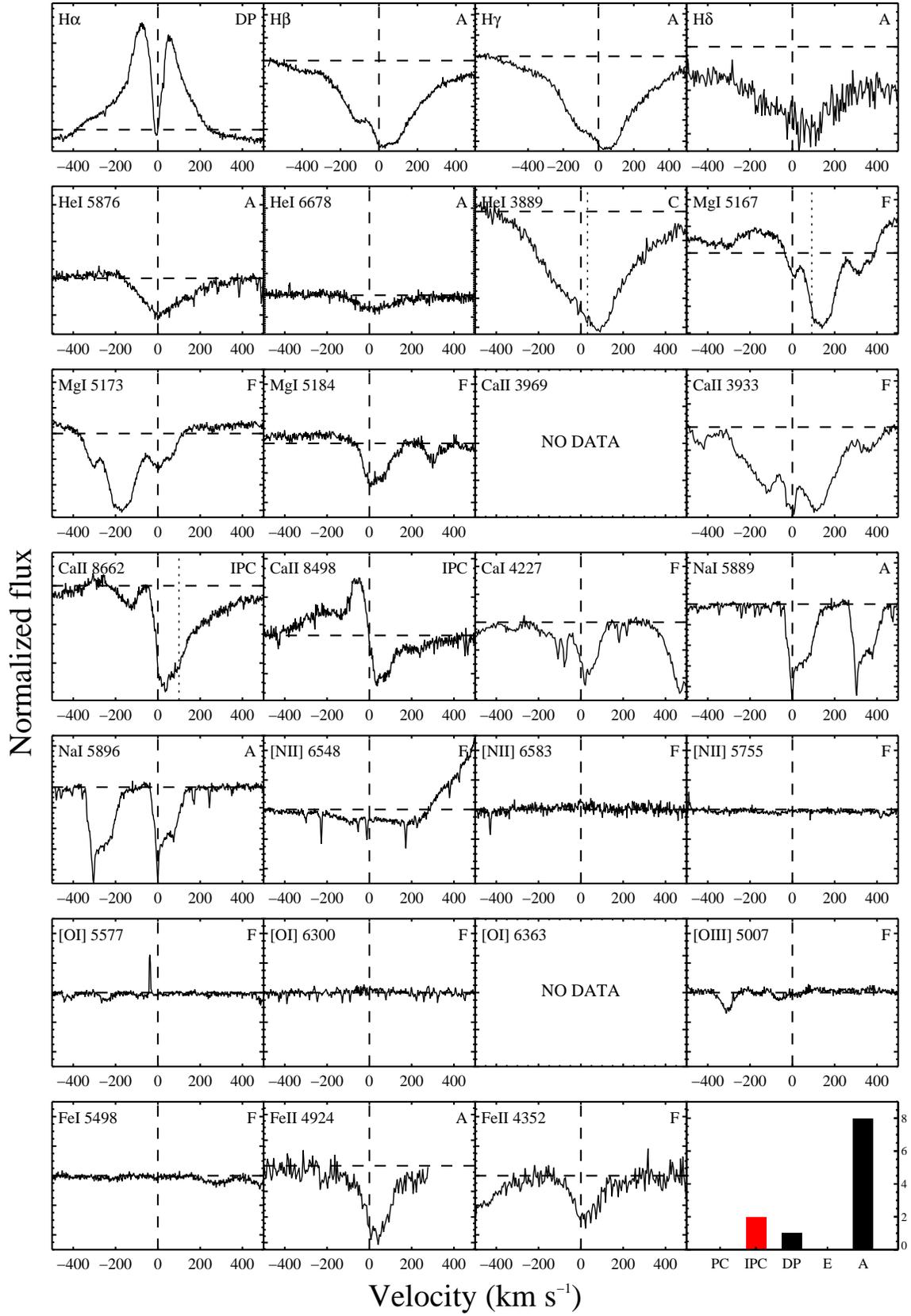}
     \caption{Extracted line profiles for BF Ori.}
  %\end{center}
\end{figure*}

\begin{figure*}\label{fig:fig16}
  %\begin{center}
     \includegraphics[scale=.93,trim= 20mm 35mm 5mm 5mm]{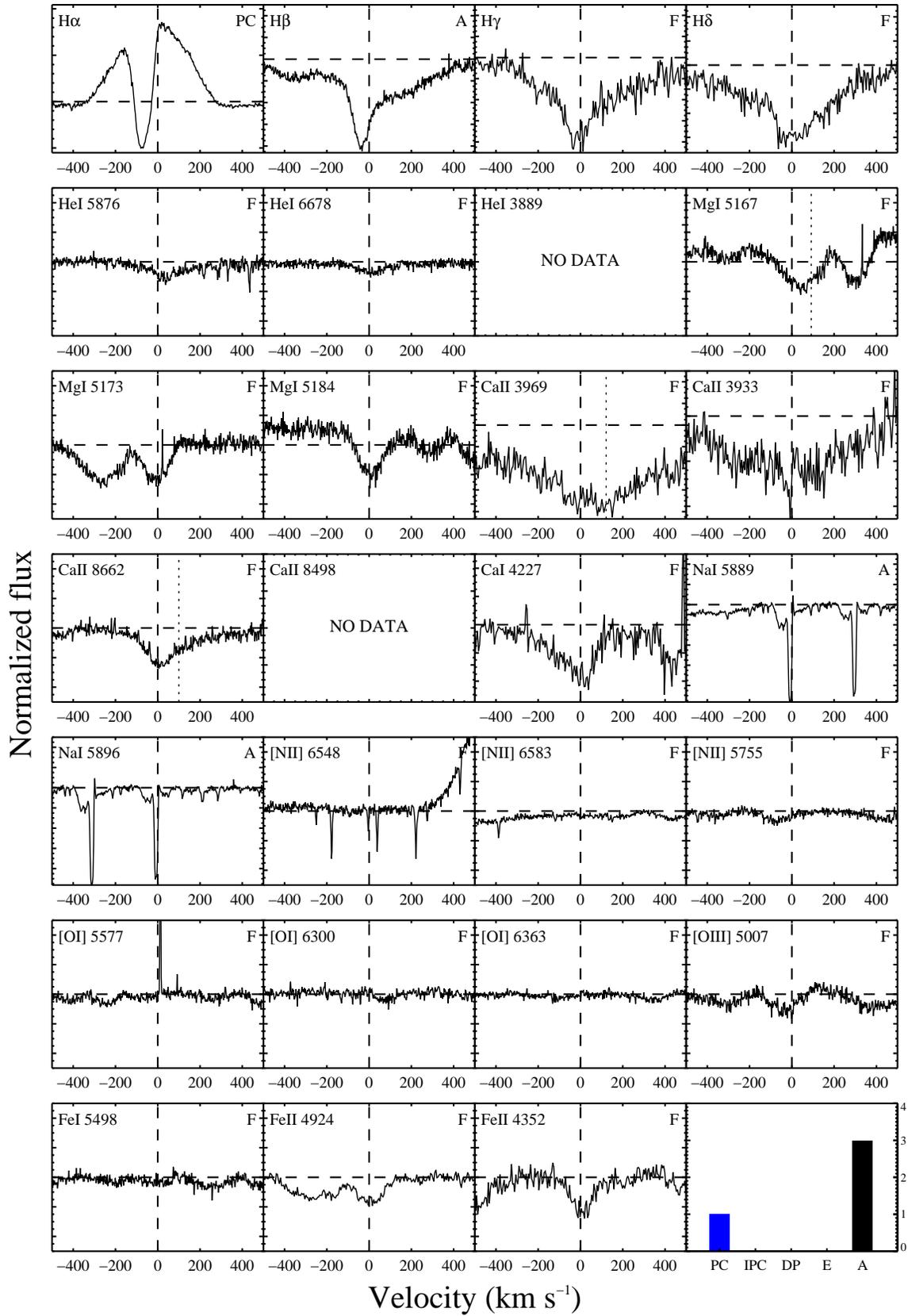}
     \caption{Extracted line profiles for BH Cep.}
  %\end{center}
\end{figure*}

\begin{figure*}\label{fig:fig17}
  %\begin{center}
     \includegraphics[scale=.93,trim= 20mm 35mm 5mm 5mm]{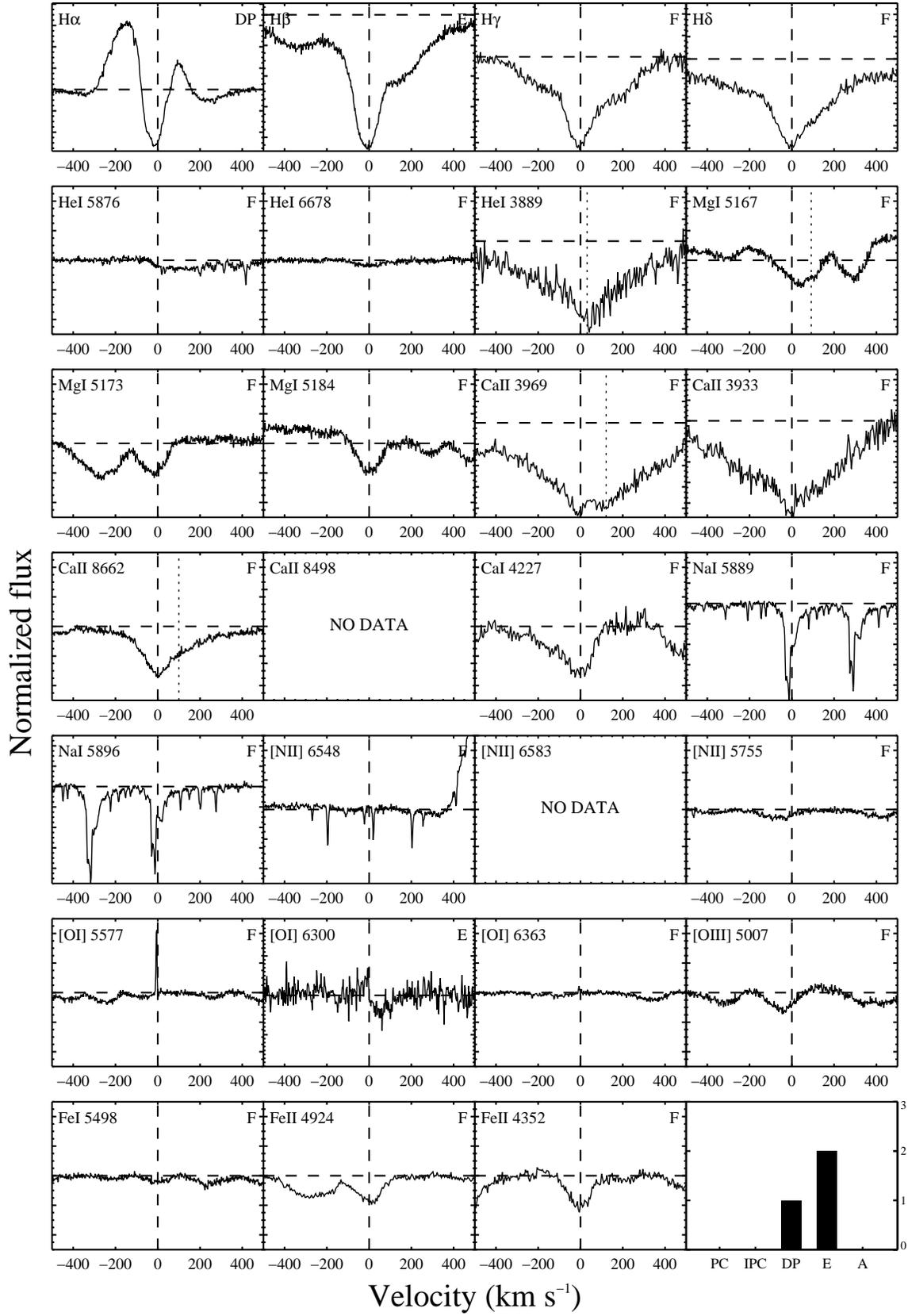}
     \caption{Extracted line profiles for CQ Tau.}
  %\end{center}
\end{figure*}

\begin{figure*}\label{fig:fig18}
  %\begin{center}
     \includegraphics[scale=.93,trim= 20mm 35mm 5mm 5mm]{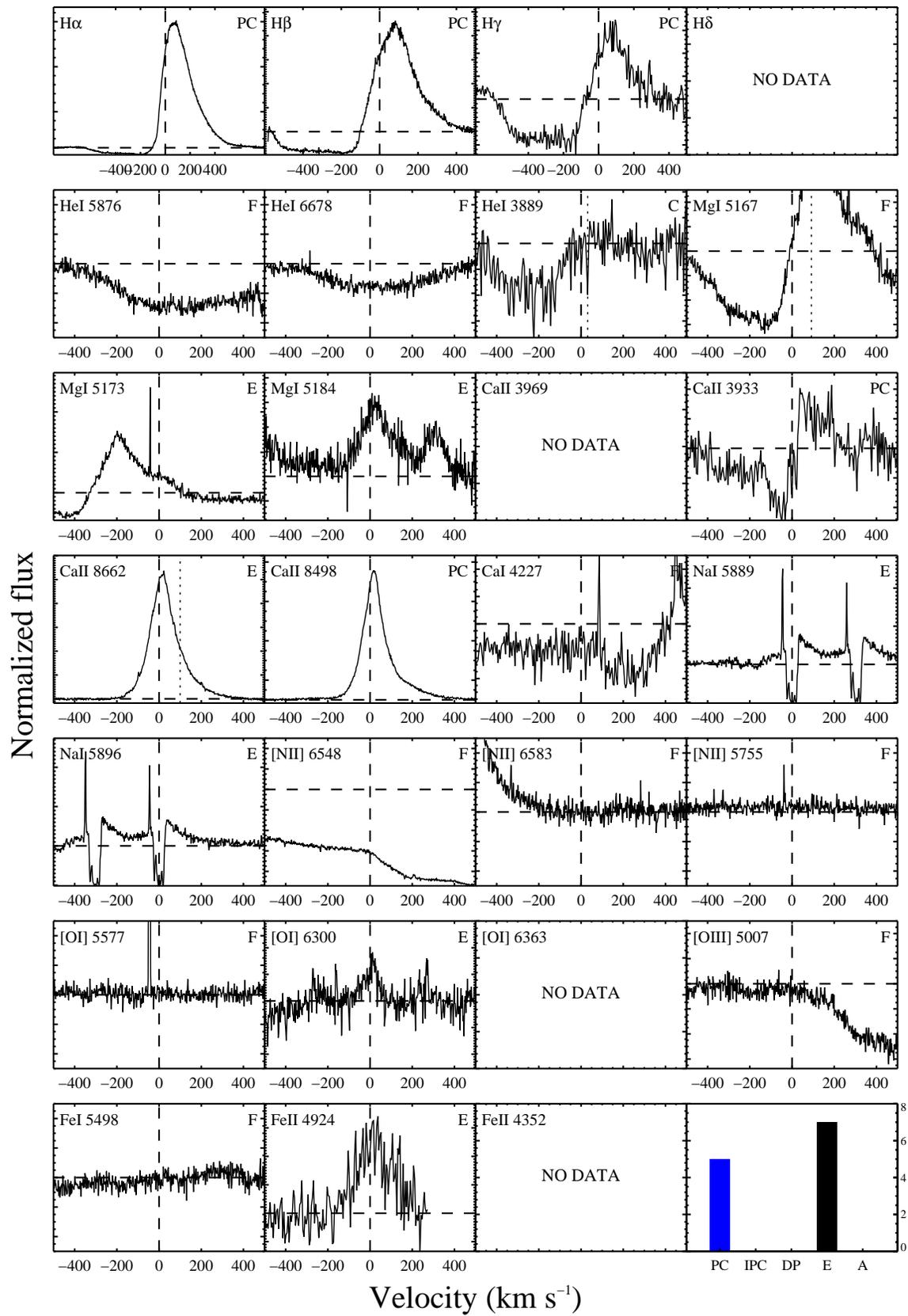}
     \caption{Extracted line profiles for DW CMa.}
  %\end{center}
\end{figure*} 

\begin{figure*}\label{fig:fig19}
  %\begin{center}
     \includegraphics[scale=.93,trim= 20mm 35mm 5mm 5mm]{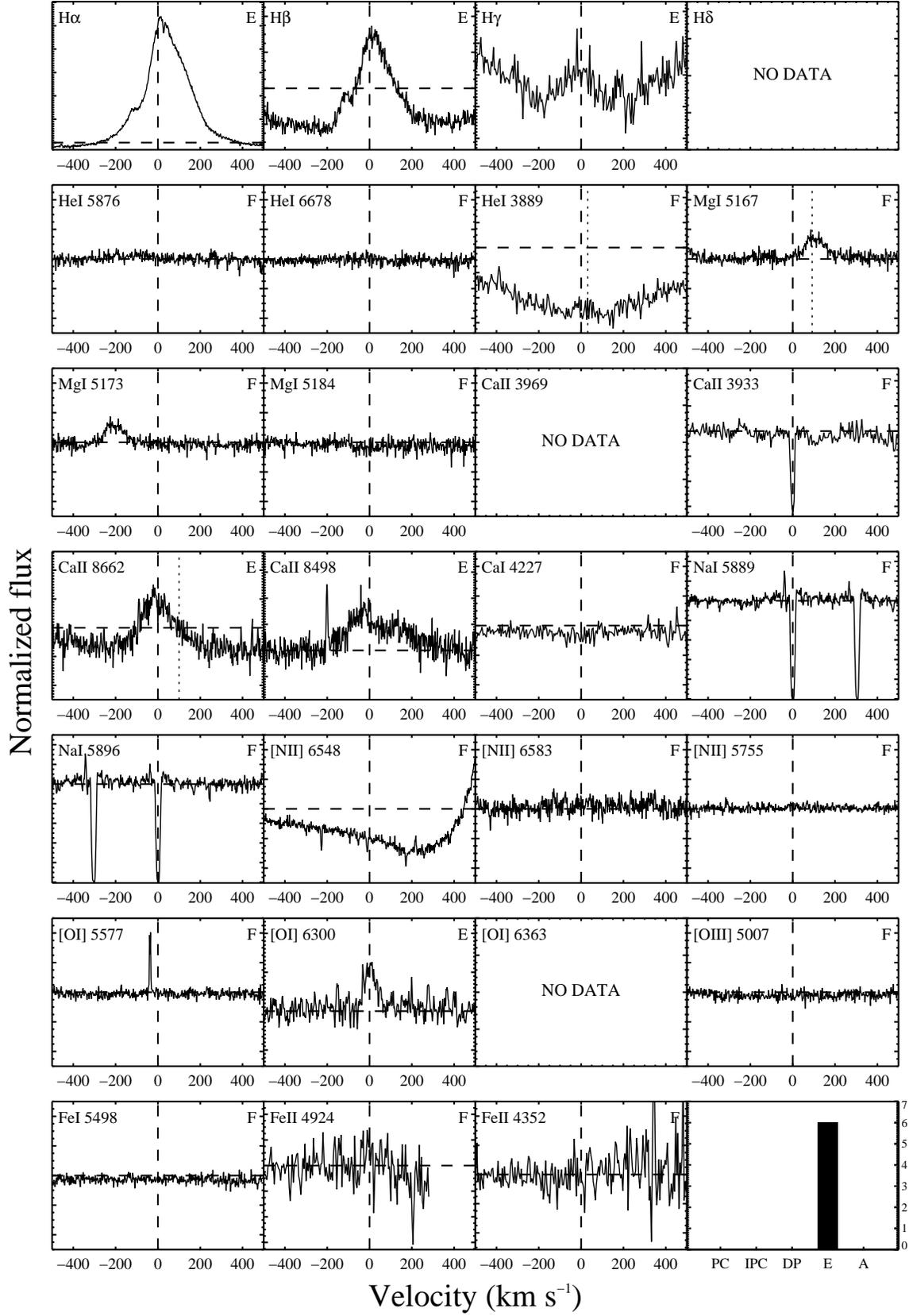}
     \caption{Extracted line profiles for GSC 04794--00827.}
  %\end{center}
\end{figure*}

\begin{figure*}\label{fig:fig20}
  %\begin{center}
     \includegraphics[scale=.93,trim= 20mm 35mm 5mm 5mm]{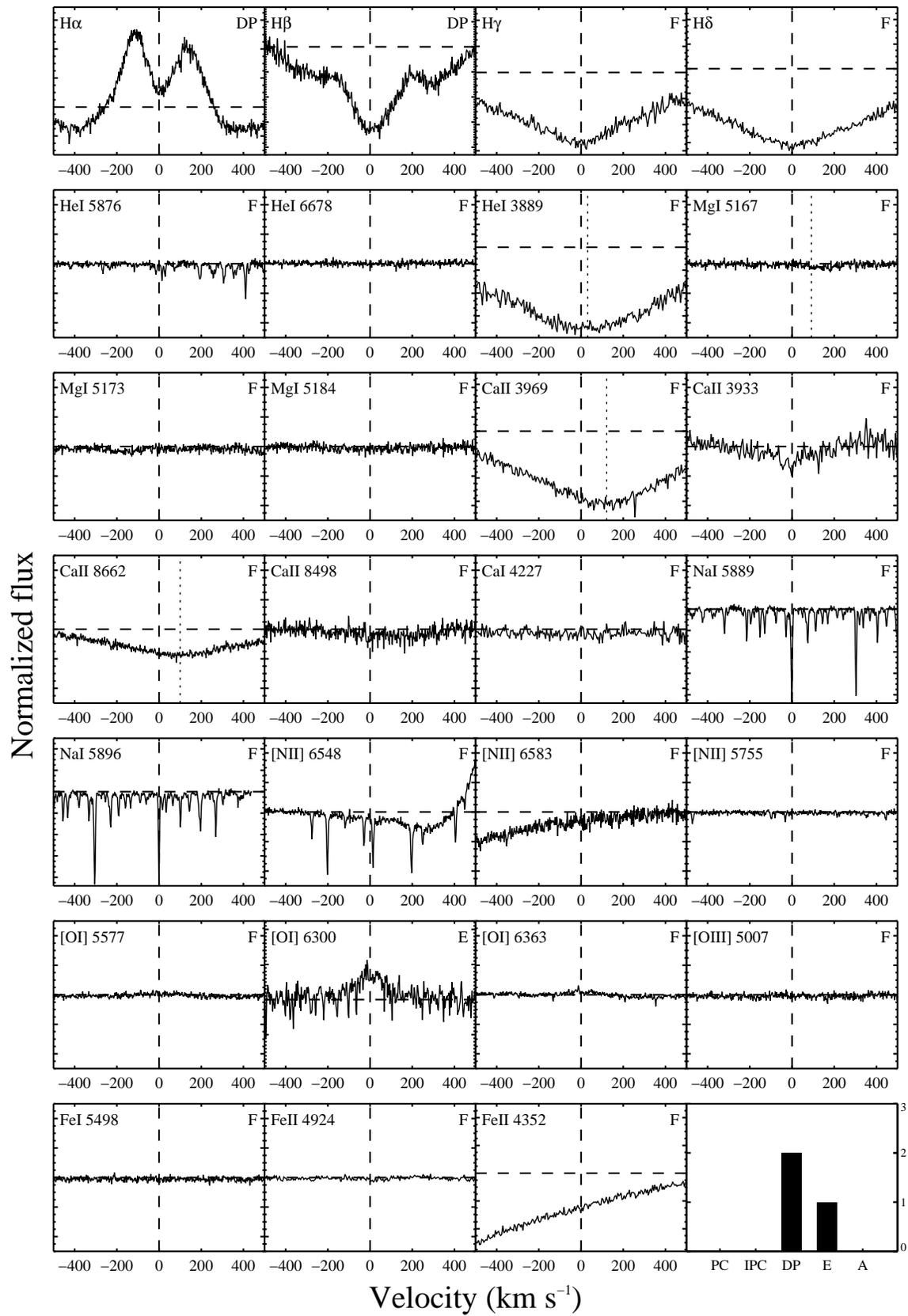}
     \caption{Extracted line profiles for HD 141569.}
  %\end{center}
\end{figure*}

\begin{figure*}\label{fig:fig21}
  %\begin{center}
     \includegraphics[scale=.93,trim= 20mm 35mm 5mm 5mm]{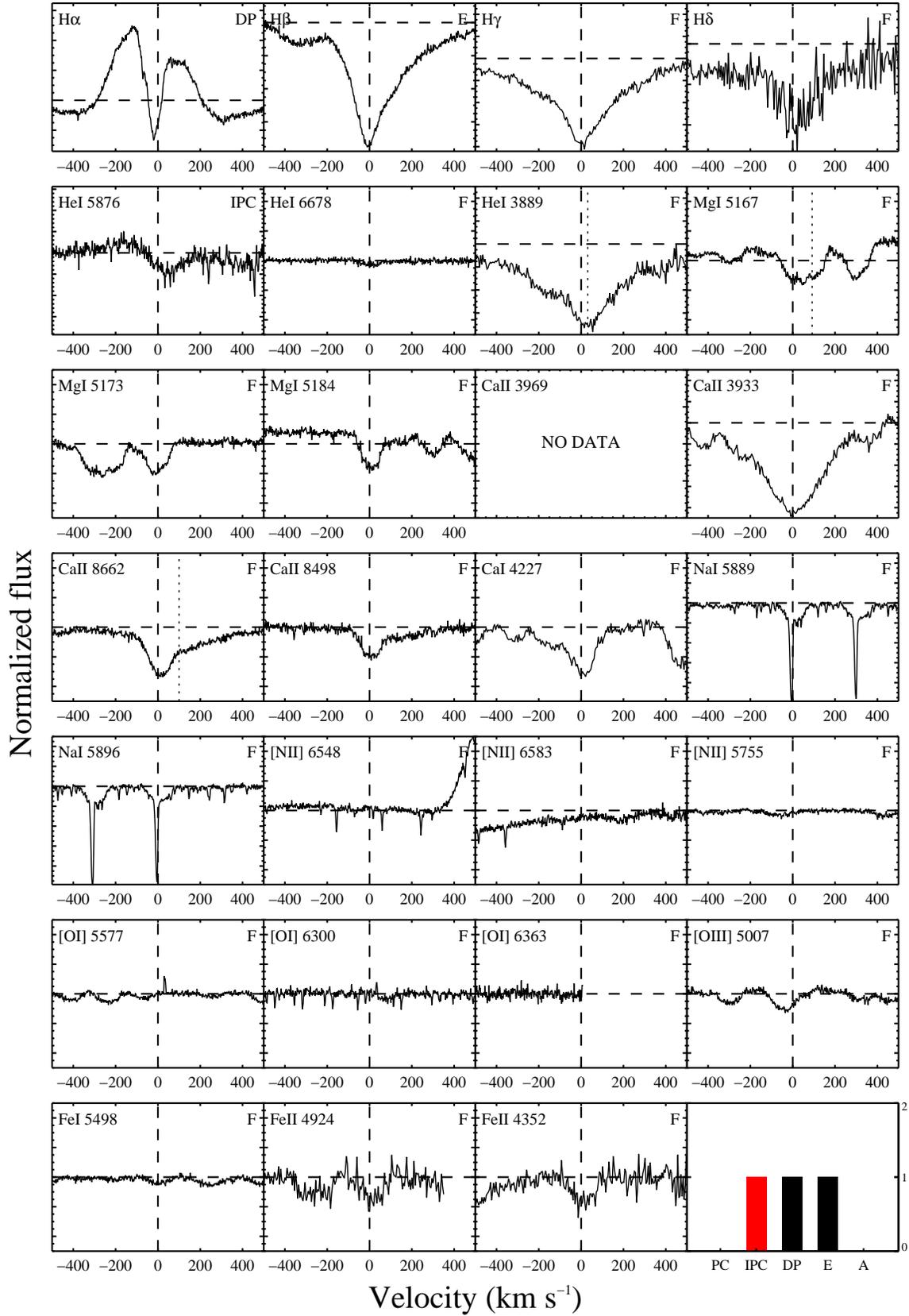}
     \caption{Extracted line profiles for HD 142666.}
  %\end{center}
\end{figure*}

\begin{figure*}\label{fig:fig22}
  %\begin{center}
     \includegraphics[scale=.93,trim= 20mm 35mm 5mm 5mm]{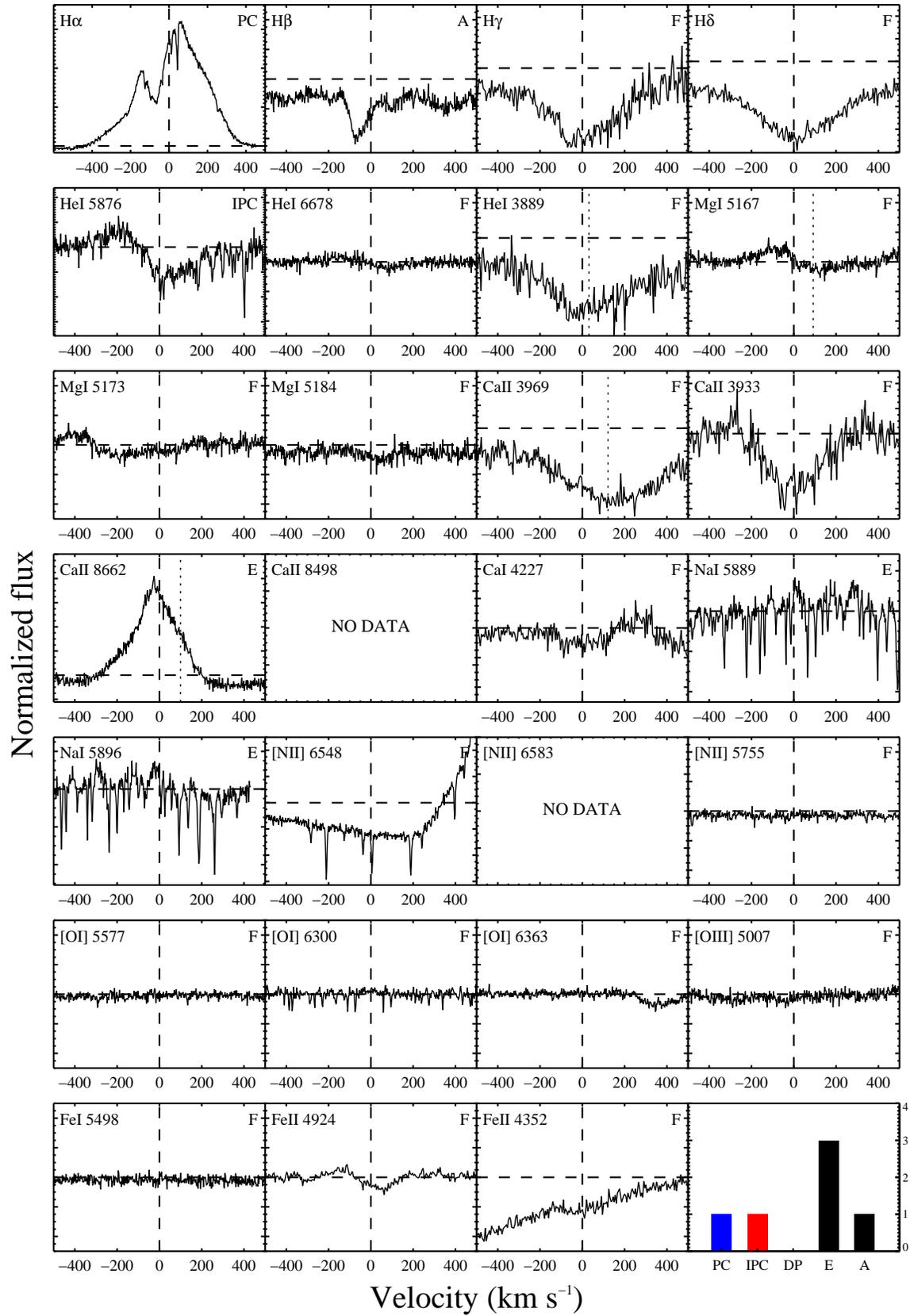}
     \caption{Extracted line profiles for HD 163296.}
  %\end{center}
\end{figure*}

\begin{figure*}\label{fig:fig23}
  %\begin{center}
     \includegraphics[scale=.93,trim= 20mm 35mm 5mm 5mm]{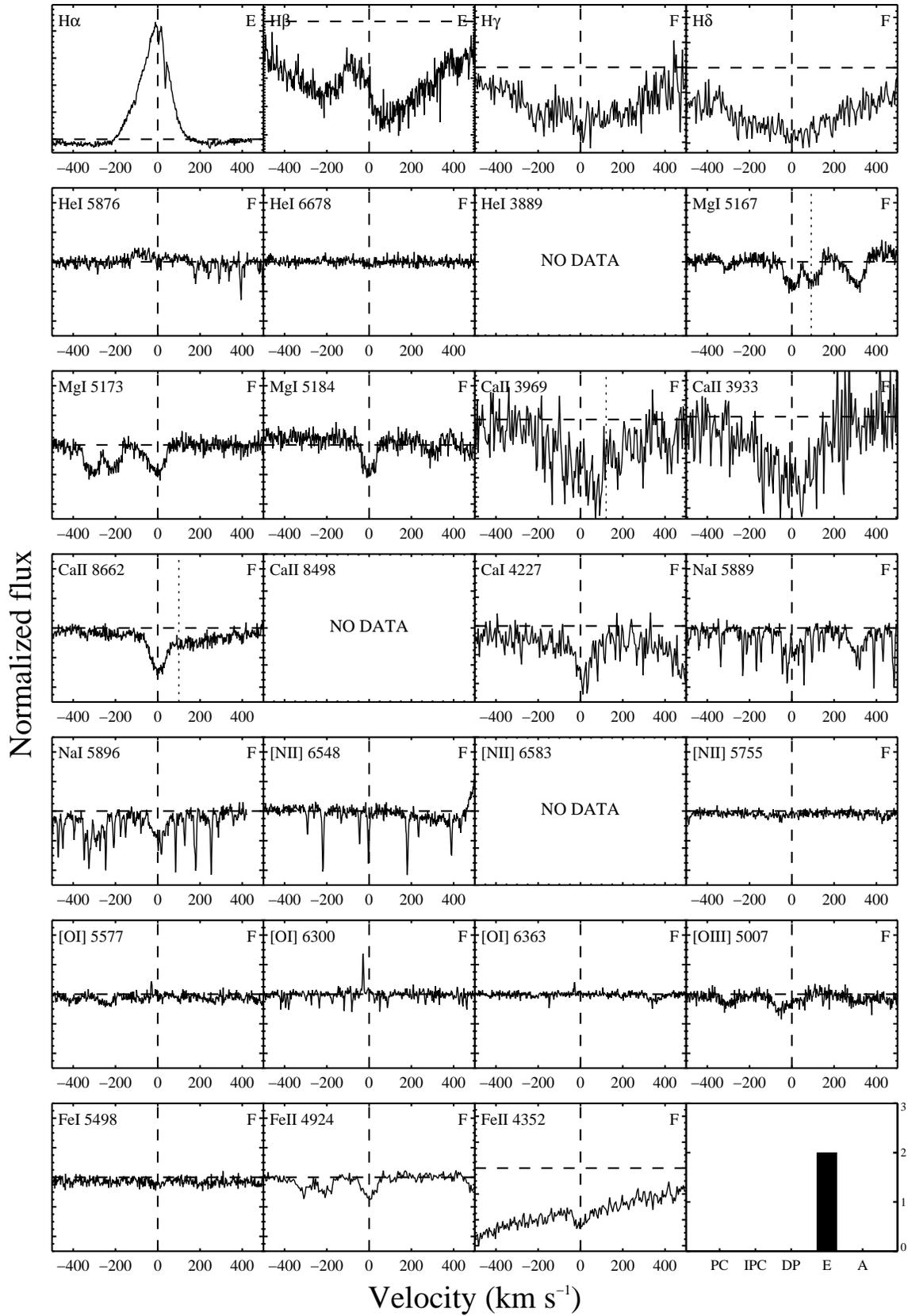}
     \caption{Extracted line profiles for HD 169142.}
  %\end{center}
\end{figure*}

\begin{figure*}\label{fig:fig24}
  %\begin{center}
     \includegraphics[scale=.93,trim= 20mm 35mm 5mm 5mm]{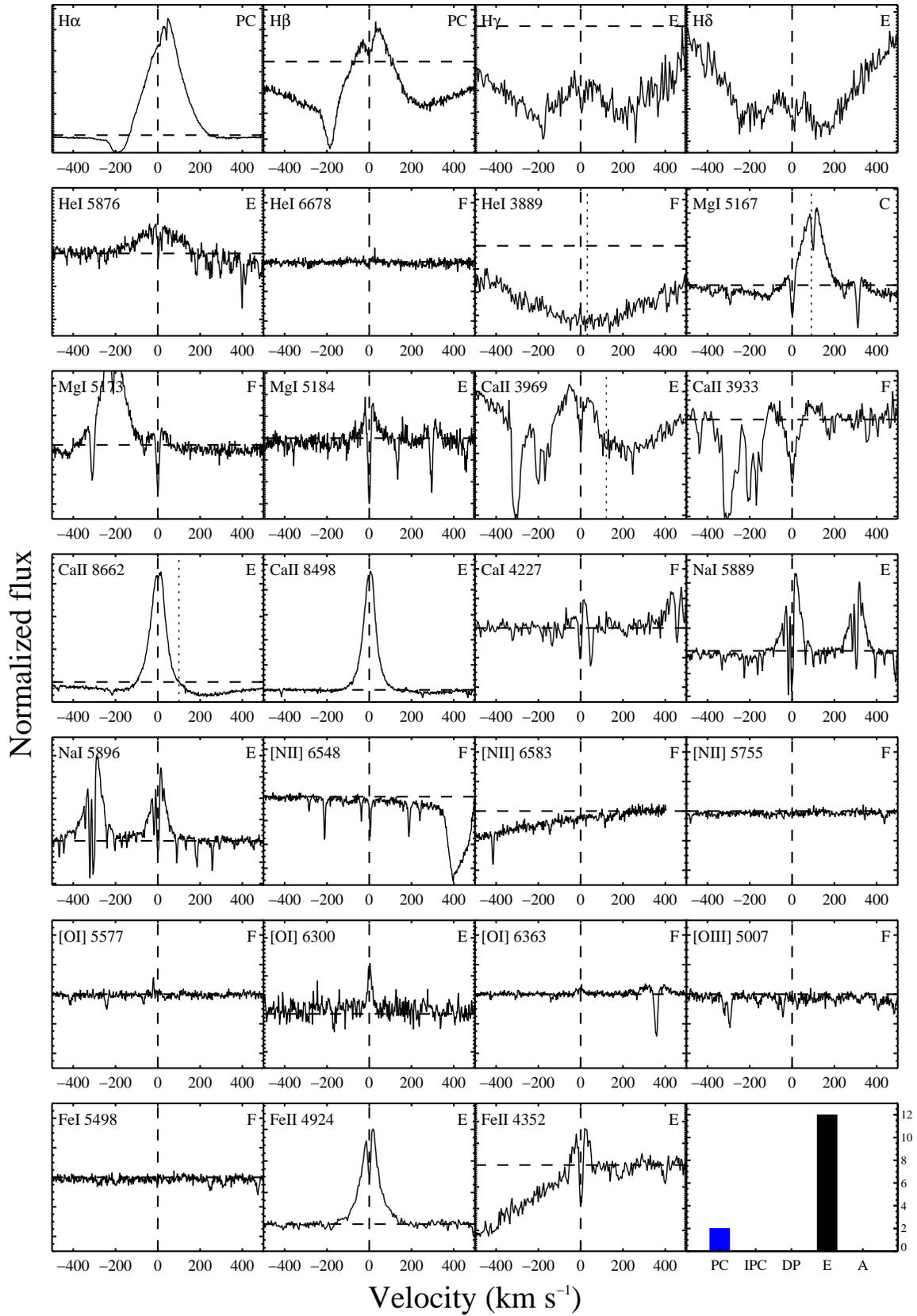}
     \caption{Extracted line profiles for HD 190073.}
  %\end{center}
\end{figure*}

\begin{figure*}\label{fig:fig25}
  %\begin{center}
     \includegraphics[scale=.93,trim= 20mm 35mm 5mm 5mm]{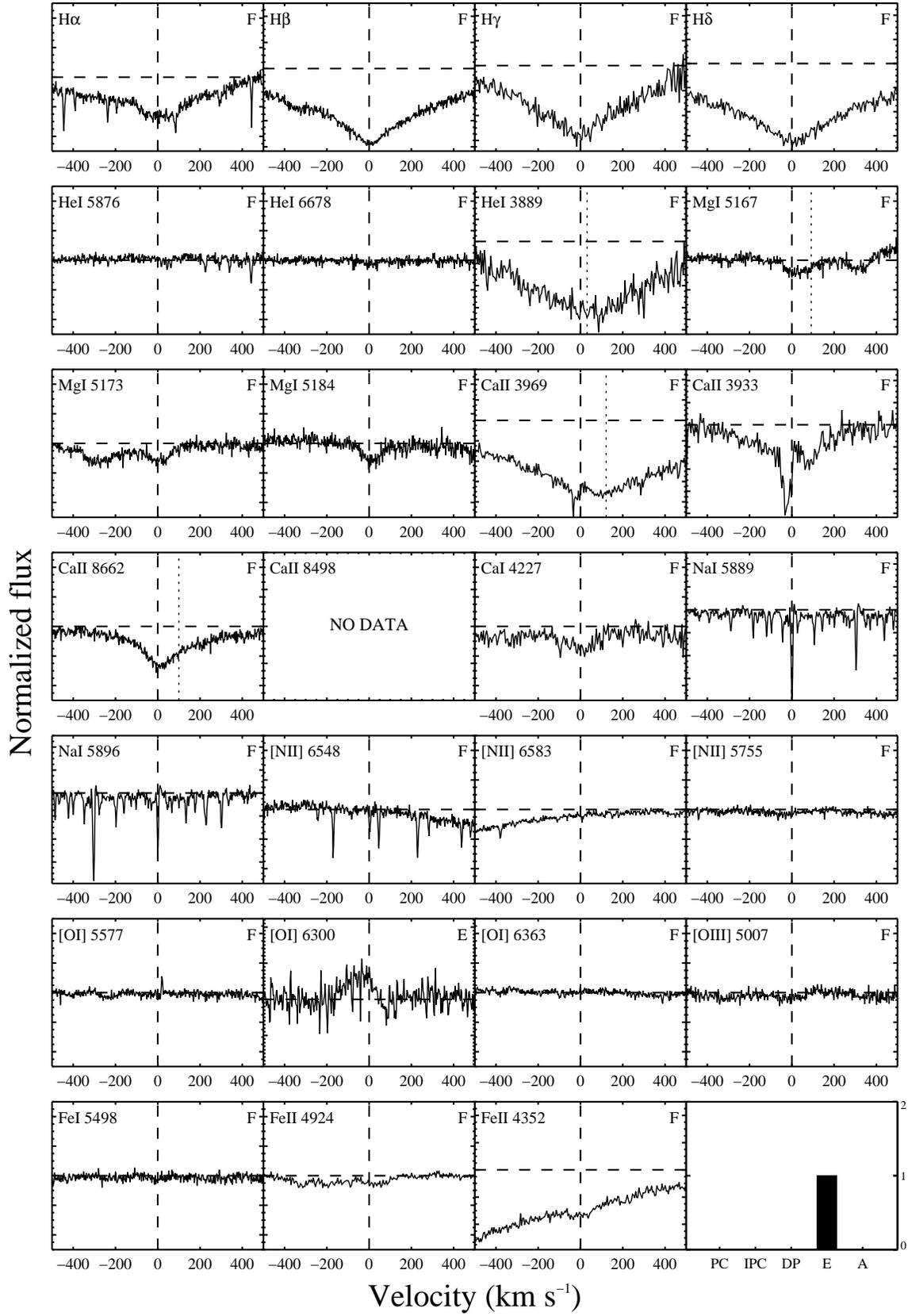}
     \caption{Extracted line profiles for HD 203024.}
  %\end{center}
\end{figure*}

\begin{figure*}\label{fig:fig26}
  %\begin{center}
     \includegraphics[scale=.93,trim= 20mm 35mm 5mm 5mm]{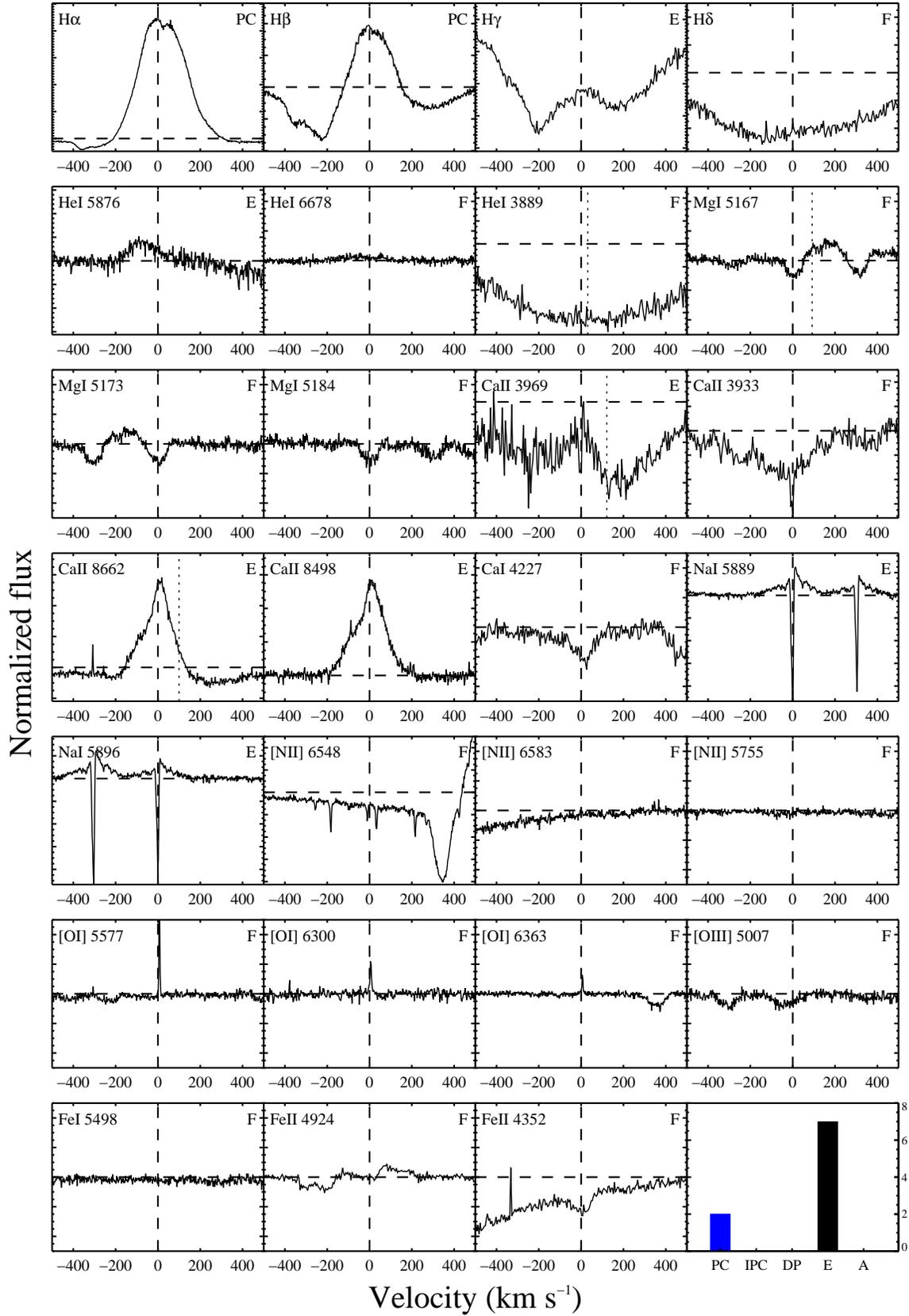}
     \caption{Extracted line profiles for HD 244314.}
  %\end{center}
\end{figure*}

\begin{figure*}\label{fig:fig27}
  %\begin{center}
     \includegraphics[scale=.93,trim= 20mm 35mm 5mm 5mm]{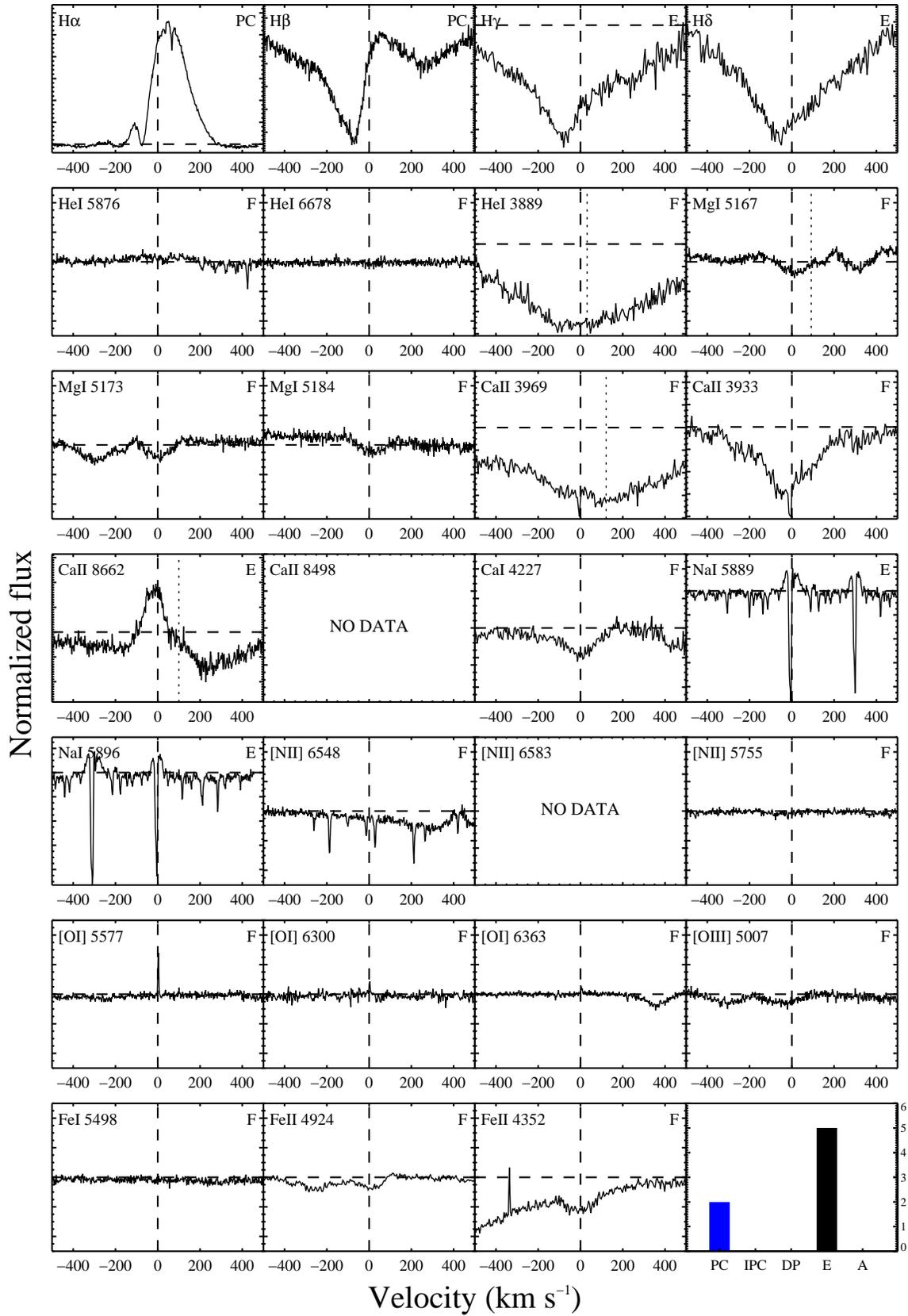}
     \caption{Extracted line profiles for HD 244604.}
  %\end{center}
\end{figure*}

\begin{figure*}\label{fig:fig28}
  %\begin{center}
     \includegraphics[scale=.93,trim= 20mm 35mm 5mm 5mm]{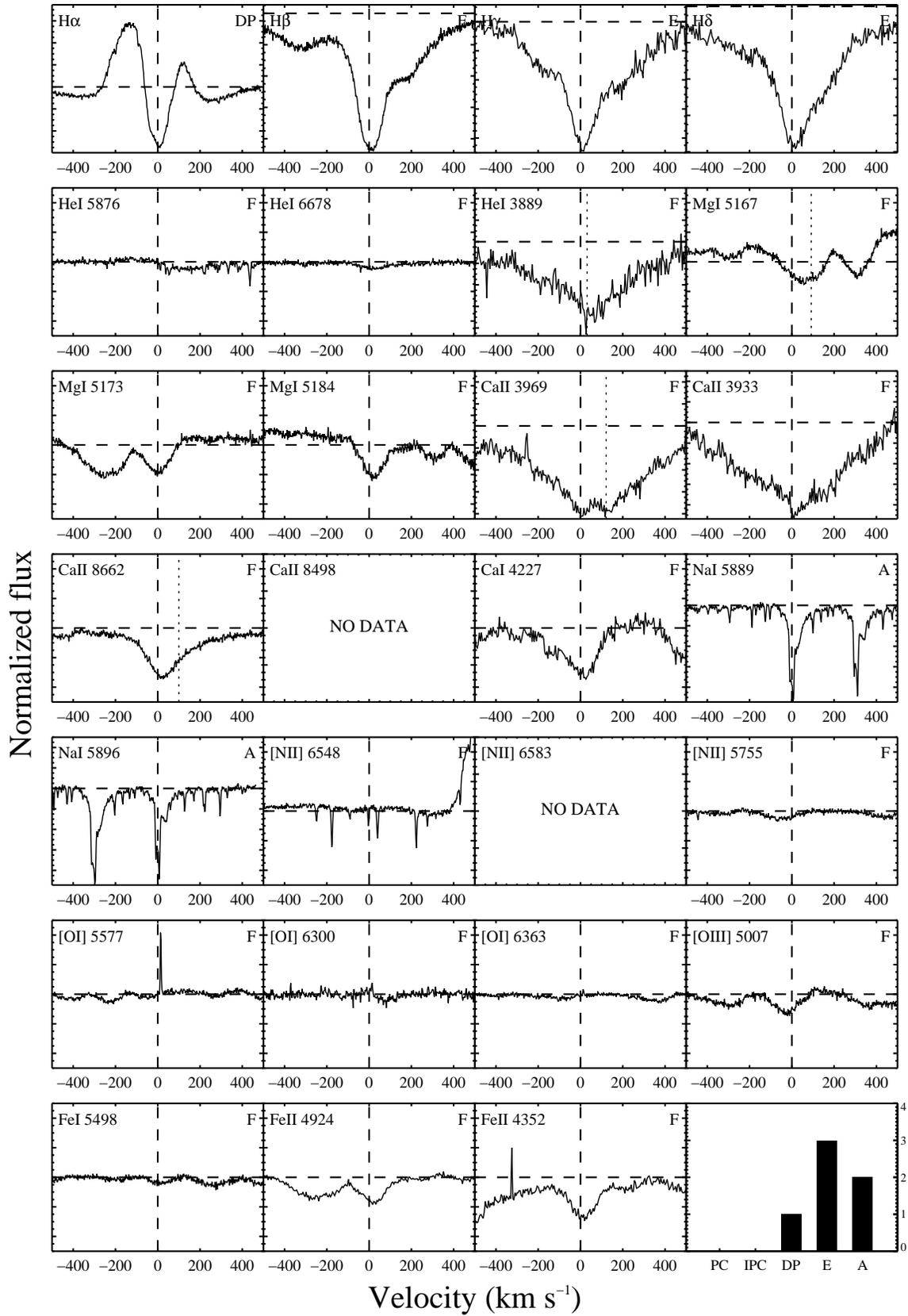}
     \caption{Extracted line profiles for HD 245185.} 
  %\end{center}
\end{figure*}

\begin{figure*}\label{fig:fig29}
  %\begin{center}
     \includegraphics[scale=.93,trim= 20mm 35mm 5mm 5mm]{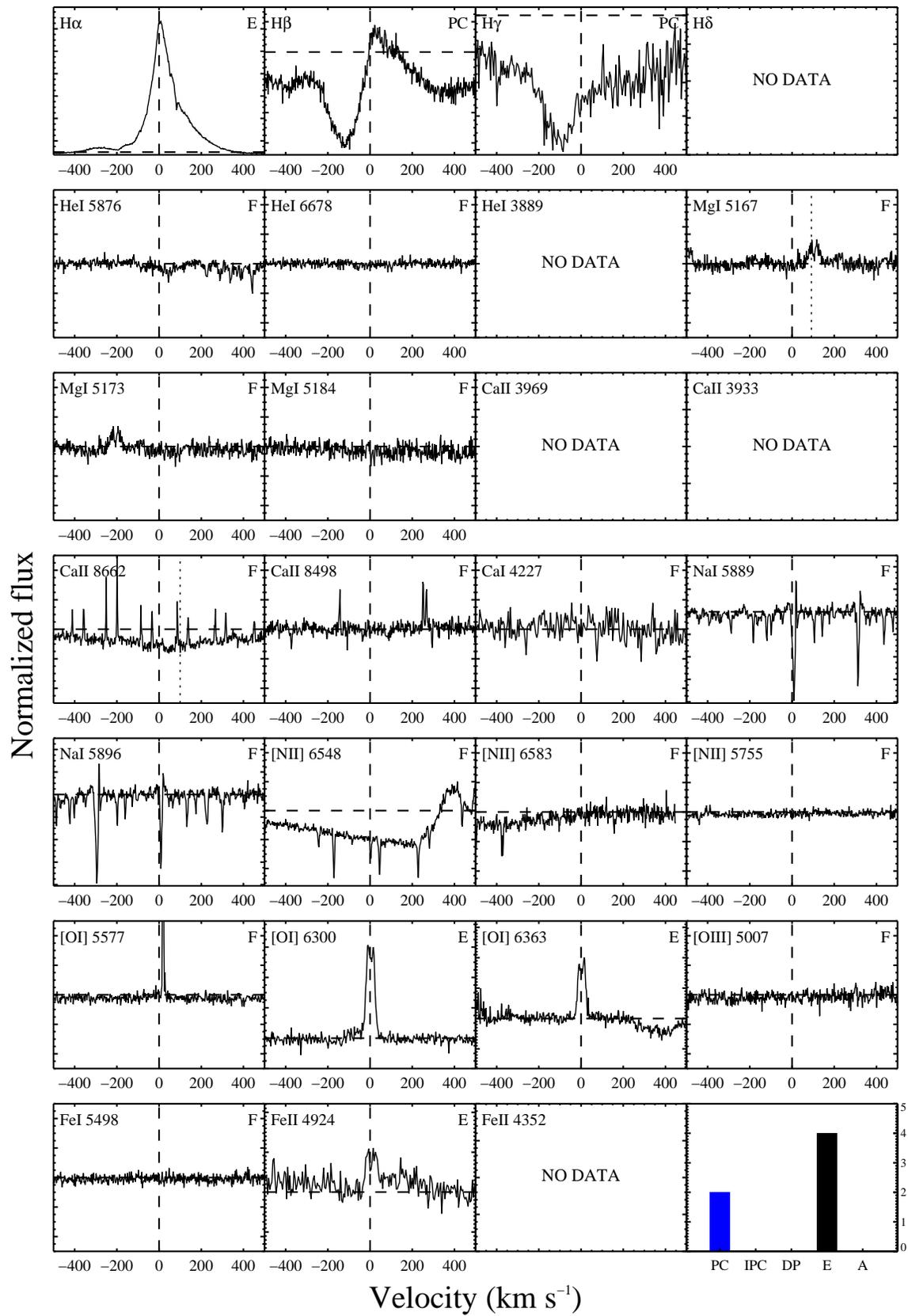}
     \caption{Extracted line profiles for HD 249879.}
  %\end{center}
\end{figure*}

\end{document}